\documentclass[reprint,superscriptaddress,
longbibliography,aps,prl,floatfix,footinbib]{revtex4-2}
\usepackage{graphicx}
\usepackage{dcolumn}
\usepackage{amsmath}
\usepackage{amssymb}
\usepackage{bm}
\usepackage{mathtools} 
\usepackage{mathrsfs}
\usepackage{bbm}
\usepackage[normalem]{ulem}
\usepackage{physics}
\usepackage[dvipsnames,x11names]{xcolor}
\usepackage{xr-hyper}
\usepackage{csquotes}
\usepackage[colorlinks=true,allcolors=blue,hypertexnames=true]{hyperref}
\usepackage{comment}
\excludecomment{temphide}
\usepackage[caption=false]{subfig} 
\usepackage{placeins}
\usepackage{booktabs}
\usepackage{orcidlink}

\newcommand{\otPrefactor}{\chi}

\newcommand{\qsspeed}{v_{\mathrm{QS}}}

\newcommand{\vloc}{\bar{v}}
\newcommand{\rcut}{r_\mathrm{cut}}
\newcommand{\cms}{c_\mathrm{MS}}

\newcommand{\kernel}{\mathcal{K}}
\newcommand{\zetac}{\zeta_{\mathrm{c}}}
\newcommand{\therm}{D_{\mathrm{T}}}

\newcommand{\Jv}{\vb{J}}

\newcommand{\qv}{{\vb{q}}}

\newcommand{\xv}{{\vb{x}}}
\newcommand{\uv}{{\vb{u}}}

\newcommand{\ofour}{\mathcal{O}\left(\nabla^4\right)}
\newcommand{\osix}{\mathcal{O}\left(\nabla^6\right)}
\newcommand{\Lv}{\boldsymbol{\Lambda}}
\newcommand{\Lvt}{\boldsymbol{\tilde{\Lambda}}}
\newcommand{\mob}{\mathcal{M}}

\newcommand{\epslvap}{\varepsilon_0}

\newcommand{\sigmaVL}{\sigma_\mathrm{V,L}}

\newcommand{\rhom}{\rho_\mathrm{m}}
\newcommand{\rhow}{\rho_\mathrm{w}}

\newcommand{\inlinesection}[1]{%
  \par\noindent\textit{#1}---\ignorespaces%
}

\begin{document}

\title{Bulk and microphase separation in chiral active systems}

\author{Sumeja Burekovi\'c\,\orcidlink{0009-0004-1600-2112}}
\email{sumeja.burekovic.phys@gmail.com}
\affiliation{Service de Physique de l'\'Etat Condens\'e, CEA, CNRS Universit\'e Paris-Saclay, 
CEA-Saclay, 91191 Gif-sur-Yvette, France}

\author{S. J. Kole,\orcidlink{0000-0002-6817-4643}}
\email{swapnilkole370@gmail.com}
\affiliation{Pritzker School of Molecular Engineering, University of Chicago, Chicago, Illinois 60637, USA}

\author{Ananyo Maitra,\orcidlink{0000-0003-3701-6981}}
\email{nyomaitra07@gmail.com}
\affiliation{Laboratoire de Physique Théorique et Modélisation, CNRS UMR 8089,
CY Cergy Paris Université, F-95032 Cergy-Pontoise Cedex, France}
\affiliation{Laboratoire Jean Perrin, UMR 8237 CNRS, Sorbonne Université, 75005 Paris, France}

\author{Cesare Nardini,\orcidlink{0000-0002-0466-1418}}
\email{cesare.nardini@gmail.com }
\affiliation{Service de Physique de l'\'Etat Condens\'e, CEA, CNRS Universit\'e Paris-Saclay, 
CEA-Saclay, 91191 Gif-sur-Yvette, France}
\affiliation{Sorbonne Universit\'e, CNRS, Laboratoire de Physique Th\'eorique de la Mati\`ere Condens\'ee, 
75005 Paris, France}

\begin{abstract}
Many active particles phase-separate due to quorum-sensing interactions, 
and their self-propulsion mechanisms often break chiral symmetry.
Using particle and continuum models, 
we uncover the role of chirality in inducing bulk or microphase separation, 
including a chiral phase formed of vapor bubbles. 
Analytical predictions for the emergence of these phases require a coarse-graining technique 
based on multiple-scale analysis. 
Further, introducing a minimal active field theory, we show that, in the bulk phase separation regime, 
chirality does not alter the diffusive~$t^{1/3}$ coarsening law nor the dynamical exponent 
associated with capillary waves, but induces traveling waves at the interface.
We finally demonstrate that, even in the absence of fluid flows, chirality can cause the breakup of elongated droplets,
resembling phenomena previously observed experimentally.
\end{abstract}
\maketitle
\inlinesection{Introduction}
Active particles are often chiral due to their shape or to
their self-propulsion mechanism~\cite{teeffelen-2008,lowen-2016, liebchen-2022, mecke-2024}. 
Two-dimensional examples include sperm cells or bacteria such as \textit{E.~coli} that swim in circles near 
surfaces~\cite{diLuzio-2005,riedel-2005,lauga-2006},
starfish embryos~\cite{tan-2022}, 
robots~\cite{yang-2020}, asymmetric colloids~\cite{kummel-2013}, 
and magnetic spinners~\cite{soni-2019,massana-cid-2021}. 
In turn, it is known that activity and chirality together give rise to distinctive nonequilibrium phenomena, 
including odd transport~\cite{hargus-2021,faedi-2026-arxiv,maire-2026-arxiv}, odd viscosity and 
elasticity~\cite{avron-1998,banerjee-2017,scheibner-2020,maitra-2020,fruchart-2023}, 
edge currents~\cite{zuiden-2016,soni-2019,yang-2020,massana-cid-2021,beppu-2021,yashunsky-2022,%
caporusso-2024,langford-2025-chiral, hargus-2025-prl, hargus-2025-pre,caprini-2025,caprini-2025-thermodynamics-arxiv,%
pisegna-2025-chiral-arxiv,alsallom-2026-arxiv,wang-2026-arxiv,metzger-2026-arxiv}, 
and rotating microflocks in two-dimensional systems~\cite{liebchen-2017}. 
A natural question suggests itself:  
how does chirality modify the known collective behaviors of active systems~\cite{maitra-2025}?  
In this Letter, we will provide a partial answer by examining the effect of chirality on phase separation.

Phase separation in active systems can be induced by several distinct microscopic mechanisms~\cite{cates-2025}. 
These include motility-induced phase separation (MIPS)~\cite{tailleur-2008,cates-2015} 
that operates even in the absence of interparticle 
attractive forces~\cite{fily-2012, redner-2013a, stenhammar-2014,stenhammar-2021,obyrne-2023-mips,caporusso-2024}; 
effective attraction induced by hydrodynamic interactions~\cite{bardfalvy2024collective,thutupalli2018flow}; 
demixing of two species of particles induced by their differential growth and death~\cite{hupe-2026}, 
or by coupling different particle species to unequal thermal 
baths~\cite{weber2016binary,grosberg2015nonequilibrium,ilker2020phase,smrek2017small,mccarthy2023demixing}. 
A particularly interesting case is when the effective attraction is induced by quorum-sensing (QS)
interactions~\cite{schnitzer-1993, miller-2001}, 
in which particles adapt their self-propulsion speed to the local density. 
QS interactions are present in a wide variety of systems, 
ranging from animate matter like bacteria~\cite{liu-2011,fu-2012,mukherjee-2019,curatolo-2020,%
dinelli-2023,ridgway-2023,li-2024,he-2025,dinelli-2026-random} 
and mussels~\cite{van-de-koppel-2008,liu-2013,desouza-2025} 
to light-activated colloids~\cite{bauerle-2018} and Quincke rods~\cite{lefranc-2025}.  
While a crude understanding of phase separation in active systems is often possible 
by an effective mapping to equilibrium, 
it is now clear that it also exhibits novel large-scale phenomenologies 
that are impossible when detailed balance is respected~\cite{cates-2025}. 
These include long-range correlations in the steady state~\cite{de2026generic}; 
microphase separation (coexisting domains that do not coarsen to system size) 
in the absence of long-range interactions~\cite{cates-2025,tjhung-2018,fausti-2024,%
li2021hierarchical,singh2019hydrodynamically,zwicker2017growth,caporusso2020micro,%
langford2026hexatic,shi-2020,redner-2013b,prymidis2015self}; 
bubbly phase separation~\cite{fausti-2024,tjhung-2018,redner-2013a,stenhammar-2014,%
bialke-2015,caporusso2020micro,langford2026hexatic}; 
network-like states that resemble active foams~\cite{fausti-2021,gulati2025active,%
toffenetti-2026-arxiv,redner-2013b,prymidis2016vapour,prymidis2015self}; 
and traveling bands~\cite{saha-2020,you2020nonreciprocity,frohoff2023non,frohoff2021suppression,brauns-2024}. 
However, in the absence of chirality, QS active particles with self-propulsion speed decreasing with the local density have a simpler phenomenology: 
they undergo bulk phase separation~\cite{cates-2015,solon-2018, solon-2018-njp,dinelli-2024}, 
and are amenable to analytical treatments that predict even quantitative features such as values of the 
binodals~\cite{solon-2018, solon-2018-njp,burekovic-2026} or surface tensions~\cite{burekovic-2026}. 
Because of its simplicity, QS interactions played a crucial theoretical role for understanding 
the emergence of MIPS in models in which particles interact via two-body 
forces~\cite{tailleur-2008,cates-2015,stenhammar-2013,speck-2014-weaknl-prl,vrugt-2023}. 

In chiral systems, active phase separation has been comparatively less investigated. 
Earlier works on particles endowed with two-body repulsive interactions have shown 
that chirality suppresses MIPS~\cite{liao-2018,bickmann-2022, kalz-2024, langford-2025-chiral} 
and induces microphase separation at low mean density~\cite{sese-sansa-2022,ma-2022,semwal-2024}, 
see also Ref.~\cite{pisegna-2025-chiral-arxiv} for the case of two particle species with nonreciprocal interactions.
However, surprisingly, the effect of chirality on QS active particles has not been addressed. 
Furthermore, while for achiral systems, significant understanding of the novel phenomenology 
that arises due to activity has been achieved by the study of continuum theories that extend Models A--J 
(in the Hohenberg-Halperin classification~\cite{hohenberg-1977,chaikin2000principles}) 
to describe active systems~\cite{cates-2025}, 
analogous studies for chiral materials have not been undertaken. 

In this Letter, we address both these issues. 
We introduce and study both a model of chiral QS active particles (cQSAPs) and chiral Active Model~B (cAMB), 
a generic continuum model that minimally extends Model~B to chiral, active phase-separating systems. 
We show that bulk phase-separation is transformed into microphase separation 
beyond a threshold in translational diffusivity~$\therm$ 
(for small~$\therm$, we predict that cQSAPs undergo bulk phase separation). 
Microphase separation can take the form of dense clusters at low mean density or of vapor bubbles at high mean density. 
Remarkably, predicting the emergence microphase separation in cQSAPs 
requires the use of a coarse-graining technique based on multiple-scale analysis 
that some of us recently introduced~\cite{burekovic-2026}; the widely employed diffusion-drift 
approximation~\cite{cates-2013, cates-2015, solon-2015, solon-2018, solon-2018-njp, dinelli-2024,dinelli-2026-long} 
fails to predict it. 
We further demonstrate that chirality does not alter the~$t^{1/3}$ law 
when the system coarsens toward bulk phase separation, 
and describe propagating capillary waves that emerge at the liquid-vapor interface.  
We finally uncover a generic mechanism by which elongated clusters break up 
into smaller droplets for sufficiently strong chirality. 
Such phenomena, which we show arise even in the absence of fluid flows, 
are reminiscent of experimental observations on magnetic colloidal spinners~\cite{soni-2019}.  
\par 
\inlinesection{Chiral QS particles}
We start by introducing cQSAPs, whose position and orientation are~$\xv_i$  and~$\uv_i$, $i=1,\ldots,N$.
As in other QS models, particles self-propel with a speed~$\qsspeed$ that depends on the local 
density~\cite{tailleur-2008,cates-2013, cates-2015, solon-2015, solon-2018, solon-2018-njp, %
martin-2021,dinelli-2024}. Their dynamics is given by  
\begin{subequations}\label{eq:chiral-qs_langevin}
\begin{align}
    \dot{\xv}_i  &=  \qsspeed[\rho_{\mathrm{d}}](\xv_i) \uv_i 
    + \sqrt{2\therm} \, \Lvt_i\,,
    \\[0.4em]
    \dot{\uv}_i  &=  -\frac{\uv_i}{\tau}
    + \omega \boldsymbol{\varepsilon}\cdot\uv_i  + \sqrt{\frac{2}{\tau}} \, 
   \Lv_i
   \,,  
\end{align}
\end{subequations}
where~$\omega$ is the angular frequency of the chiral rotation of the particles, 
and~$\boldsymbol{\varepsilon}$ is the 2D Levi-Civita tensor.  
In Eqs.~\eqref{eq:chiral-qs_langevin},~$\therm$ denotes the thermal diffusivity,~$\tau$ the persistence time, 
and $\Lv_i$,~$\Lvt_i$ are independent Gaussian white noise with zero mean and unit variance.  
The QS speed is defined as~$\qsspeed[\rho_{\mathrm{d}}] = \vloc(\kernel*\rho_{\mathrm{d}})$, 
where~$\kernel$ is an isotropic kernel that decays over a characteristic length scale~$\rcut$,  
$\rho_{\mathrm{d}}=\sum_i \delta(\xv-\xv_i)$ is the empirical density, 
and~$\vloc$ is a decreasing, purely local function of its argument. 
Particles slow down from a speed~$v_0$ (when isolated) to~$v_1$ (at high density) over a density range~$\rhow$, 
centered at the midpoint density~$\rhom$ (details reported in~\cite{si}).
In the absence of chirality, bulk phase separation emerges when~$\qsspeed$ 
decreases fast enough with density~\cite{tailleur-2008, cates-2015}.  
\begin{figure}[ht!]
    \centering
    \includegraphics[width=\linewidth]{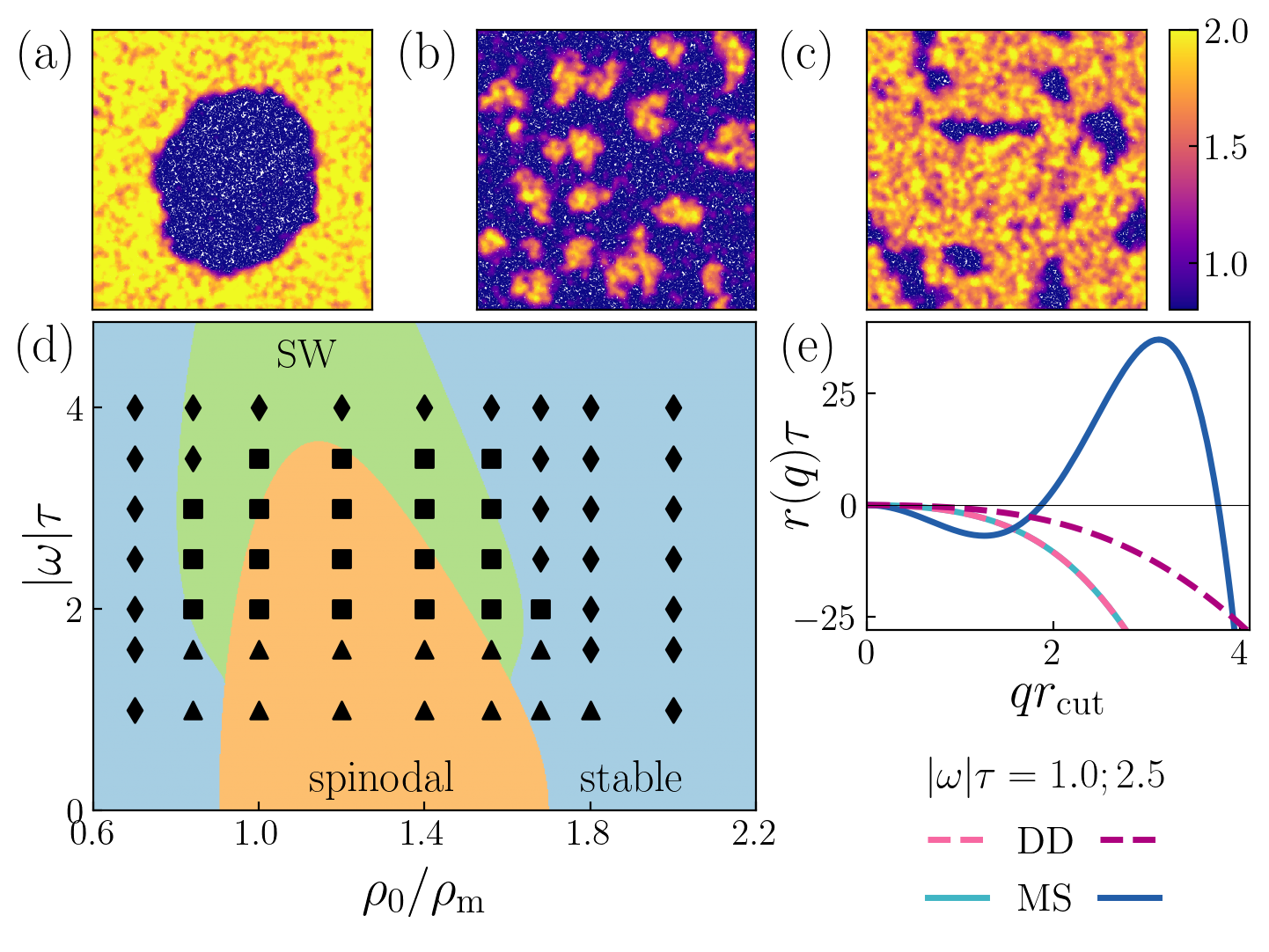}
    \caption{Phase diagram of cQSAPs for translational diffusivity~$\therm = 0.3$. 
    a)--c): Steady-state configurations ($L = 48\rcut$) 
    with the color bar denoting the local density in units of~$\rhom$. 
    d): Predictions of the linear stability from our MS theory: 
    linearly stable (blue); SW instability (green); spinodal instability (orange). 
    Phases observed from simulations of cQSAPs: 
    bulk phase separation (triangles); microphase separation (squares); 
    homogeneous phase (diamonds).    
    e) Growth rate~$r$ for~$\rho_0 = 0.84\rhom$ and two values of~$|\omega|\tau$, 
    shown for both DD and MS coarse-graining theories 
    (here, the MS growth rate for~$|\omega|\tau = 2.5$ is multiplied by a factor~$16$). 
    Other  parameters were set to 
    $v_0 = 5$, $v_1 = 1$, $\rhom = 50$, $\rcut = 1$, $\tau = 1$. 
    In a) $\rho_0 = 1.56\rhom$, $\omega=1.0$; 
    b) $\rho_0 = \rhom$, $\omega=2.5$; 
    c) $\rho_0 = 1.56\rhom$, $\omega=2.0$.}
    \label{fig:particle_phase-diagram}
\end{figure}
\par 
\inlinesection{Coarse-graining}
To describe the large-scale dynamics of cQSAPs analytically, we coarse-grain Eq.~\eqref{eq:chiral-qs_langevin} 
and obtain an evolution equation for the particle density~$\rho(\xv,t)$. 
For this, we extend the procedure introduced in Ref.~\cite{burekovic-2026} to cQSAPs.  
The method is based on a multiple-scales (MS) expansion 
in the parameter $\epslvap = v_0\tau/\xi$ (assumed small), 
which is the ratio of the persistence length~$v_0\tau$ and the interfacial width~$\xi$. 
Our procedure consistently retains all terms up to order~$\mathcal{O}\left(\epslvap^2\right)$, 
or equivalently,~$\ofour$. 
In Ref.~\cite{burekovic-2026}, some of us showed that, for the achiral case, 
the MS coarse-graining produces additional terms 
in~$\partial_t \rho$ at fourth order in gradients compared to the widely employed diffusion-drift (DD) 
approximation~\cite{cates-2013,cates-2015,solon-2015,solon-2018,solon-2018-njp,dinelli-2024,dinelli-2026-long}. 
Hence, it is natural to expect that this also happens for the chiral case. 
Indeed, the evolution equation obtained using \emph{both} techniques 
for chiral QS particles can be compactly written as
{\allowdisplaybreaks
\begin{align}
    & \partial_t \rho = 
    \therm \laplacian \rho
    + 
    \otPrefactor \left(\delta_{\alpha\beta} - \omega \tau \varepsilon_{\alpha\beta} \right)
    \nabla_\alpha \left[\qsspeed \nabla_\beta \left(\qsspeed  \rho\right) \right]
    \nonumber\\[0.4em] 
    & +  
    \cms \otPrefactor^2
    \therm \left[\left(1 - \omega^2 \tau^2\right)  \delta_{\alpha\beta} 
    - 2 \omega \tau \varepsilon_{\alpha\beta}
    \right] 
    \, \times 
    \nonumber\\[0.4em] 
    &\phantom{+} 
    \times \nabla_\alpha \left\{\qsspeed   
    \nabla_\beta \left[\laplacian \left(\qsspeed \rho\right)  
    - \qsspeed \laplacian \rho \right] \right\}
    \nonumber\\[0.4em] 
    &\phantom{} + \cms \otPrefactor^3 
    B_{\alpha\beta\mu\nu} 
    \nabla_\alpha \left(\qsspeed  \nabla_\beta \left\{\qsspeed 
    \nabla_\mu \left[\qsspeed  \nabla_\nu \left(\qsspeed \rho\right)
    \right] \right\} \right) 
    \nonumber
    \\[0.4em]
    &+ \osix 
    \,,
\label{eq:ms-result_dimensionful}
\end{align}}\noindent 
where Greek indices denote spatial variables,~$\qsspeed$ stands for~$\qsspeed[\rho]$, 
and the tensor~$B_{\alpha\beta\mu\nu}$ is expressed in terms of Kronecker deltas 
and~$\boldsymbol{\varepsilon}$ (cf.~\cite{si}). 
The bookkeeping parameter $\cms=1$ for the MS coarse-graining procedure 
and $\cms=0$ for the DD method. 
We also introduced the parameter~$\otPrefactor = \tau/\left(1 +\omega^2 \tau^2\right)$, 
which describes the reduction of the effective orientational correlation time due to 
chirality~\cite{teeffelen-2008,ebbens-2010,marine-2013,sevilla-2016,caprini-2019,langford-2025-chiral}.  
\par 
We start by analyzing the DD predictions obtained from Eq.~\eqref{eq:ms-result_dimensionful}. 
Since $\vloc'<0$, the homogeneous state~$\rho_0$ undergoes 
the MIPS spinodal instability when 
\begin{align}\label{eq:spinodal-DD}
    \vloc(\rho_0)\otPrefactor
    \left[\vloc'(\rho_0)\rho_0 \kernel_q + \vloc(\rho_0)\right] < - \therm
    \,,
\end{align}
where~$\kernel_q$ denotes the Fourier transform of~$\kernel$.
Eq.~\eqref{eq:spinodal-DD} closely resembles the standard criterion 
for the spinodal instability of QS particles~\cite{cates-2015}, 
with chirality entering only via~$\otPrefactor$. 

We next compute the interfacial tensions governing Ostwald ripening,~$\sigmaVL$, 
and the relaxation of capillary waves,~$\sigma_{\mathrm{cw}}$ 
as it is known that when either of these becomes negative, 
the system self-organizes into phase-separated morphologies 
that differ from equilibrium-like bulk phase separation~\cite{tjhung-2018,fausti-2021,cates-2025}. 
Following approaches developed for describing other active 
systems~\cite{tjhung-2018,fausti-2021,cates-2023,cates-2025,burekovic-2026}, we obtain~\cite{si}  
\begin{align} 
    \sigmaVL &= \mob_{\mathrm{V,L}} \sigma^0
    \,, \quad  
    \sigma_{\mathrm{cw}}
    = \frac{\mob_\mathrm{V} \sigma_{\mathrm{L}}^0 
    + \mob_\mathrm{L} \sigma_{\mathrm{V}}^0}{2}
    \,, 
    \label{eq:tensions_dd}
\end{align}
where $\sigma^0 > 0$ is the interfacial tension of 
active Model~B~\cite{wittkowski-2014,solon-2018-njp,omar-2023}, and $\mob_{\mathrm{V,L}}>0$. 
Within the DD approximation, Eq.~\eqref{eq:tensions_dd} shows that~$\sigmaVL$ 
and~$\sigma_{\mathrm{cw}}$ are all positive: 
Hence, the DD approximation predicts that cQSAPs are either homogeneous or bulk phase-separated, 
with chirality only modifying the phase diagram of QS particles 
through the quantitative dependence of spinodal and binodal values~\cite{si}. 

In contrast, MS theory leads to very different conclusions. 
By linearizing Eq.~\eqref{eq:ms-result_dimensionful} with $\cms = 1$ 
around a homogeneous state~$\rho_0$ and transforming to Fourier space, 
we obtain $\partial_t \delta \rho_\qv = r(q) \delta \rho_\qv$, 
where the growth rate~$r(q)$ is
\begin{align}
    &r(q) = 
    - \vloc(\rho_0)\otPrefactor
    \left[\vloc'(\rho_0)\rho_0 \kernel_q + \vloc(\rho_0)\right]  q^2 
    - \therm q^2 
    \nonumber
    \\[0.6em]
    &\phantom{={}} 
    + \cms  \therm \otPrefactor^2  \left(1 - \omega^2 \tau^2\right) \vloc(\rho_0)
    \vloc'(\rho_0) \rho_0 \kernel_q  q^4 
    \,.
    \label{eq:linear-stability_conv-ms}
\end{align}
It is then straightforward to show that the boundaries of the spinodal instability region 
are identical in the DD and MS theories. 
However, as shown in Fig.~\ref{fig:particle_phase-diagram}e, 
MS theory also predicts a short-wavelength (SW) instability characterized by 
$r''(q = 0) < 0$ and $r(q^*)>0$ for some $q^*>0$. 
(In contrast, spinodal instabilities satisfy $r''(q = 0) \geq 0$.) 
As shown in~\cite{si}, the presence of such SW instability requires intermediate values of translational diffusivity~$\therm$. 
We also provide a detailed comparison between the DD and MS predictions at linear level 
in terms of microscopic parameters.

In top-down descriptions of pattern formation, such as phase-field crystal models~\cite{emmerich2012phase} 
and the conserved Swift--Hohenberg equation~\cite{matthews-2000,elder2002modeling,thiele2013localized}, 
it is known that short-wavelength (SW) instabilities such as the one described above 
lead to either microphase-separated states or smectic patterns. 
We thus conclude that MS theory suggests their emergence in cQSAPs, 
while DD theory only predicts bulk phase separation and homogeneous states. 
We now test these predictions in large-scale simulations of cQSAPs.
\par 
\inlinesection{Particle simulations}
We perform particle simulations of Eqs.~\eqref{eq:chiral-qs_langevin} 
at fixed~$\rhow = 25$ and $\therm = 0.3$, varying the chirality~$\omega$ 
and the mean particle density~$\rho_0 = N/L^2$, 
starting from homogeneous initial conditions. 
For computational efficiency, 
we use a kernel with a compact support (denoted by~$\kernel^{\mathrm{c}}$ in the following). 
The phase diagram of cQSAPs obtained numerically 
and its comparison to the linear stability analysis predictions obtained from our MS theory 
are displayed in Fig.~\ref{fig:particle_phase-diagram}d.  
At low chirality~$|\omega|\tau$, the system undergoes bulk phase separation, 
consistent with the absence of a SW instability in the MS theory (snapshot Fig.~\ref{fig:particle_phase-diagram}a).
At intermediate chirality, where the MS theory predicts a SW instability, 
we observe microphase separation. 
Depending on the mean density~$\rho_0$, microphase separation consists of liquid clusters 
dispersed in a dilute environment, (at low~$\rho_0$), 
or of vapor bubbles (at high~$\rho_0$) dispersed in a dense environment.  
Upon increasing chirality, the typical cluster or bubble size decreases~\cite{si}.  
This persists up to values of~$|\omega|\tau$ where the spinodal instability disappears, 
beyond which the system becomes homogeneous. 
We note, however, that there is no reason to expect the disappearance 
of microphase separation to coincide exactly with the spinodal boundary.   
We also found no evidence of striped patterns in our simulations; 
however, we cannot rule out the possibility that they arise in a small region of the phase diagram. 
Importantly, our discovery of microphase separation in cQSAPs contrasts with achiral QS particles, 
which are known to undergo only bulk phase separation~\cite{cates-2013, cates-2015, solon-2015, solon-2018, %
solon-2018-njp,martin-2021,dinelli-2024}. 
\par 
That we obtain a remarkably correct prediction of both the emergence of microphase separation 
and the parameter range where it occurs
is surprising for several reasons. 
First, MS is derived as a perturbative expansion in~$\epslvap = v_0\tau/\xi$, 
which compares the particles' persistence length to the interfacial width~$\xi$. 
While for achiral QSAPs~$\xi$ diverges at the critical point---and, 
therefore, MS is quantitatively valid in its vicinity 
(outside the Ginzburg interval)~\cite{burekovic-2026}---in the presence of a SW instability, 
there is no clear parameter regime in which~$\epslvap$ can be made arbitrarily small 
when~$\therm$ is such that both bulk and microphase separation are present. 
Next, the predictions in Fig.~\ref{fig:particle_phase-diagram}d are based on a linear stability analysis, 
and the MS theory is only consistent to fourth order in gradients; 
we however expect the region of linear stability to be modified by~$\nabla^6$ terms. 
While our MS theory could be extended to include them,
which would also be needed for quantitatively predicting the cluster size in the microphase-separated regime,  
doing so is technically cumbersome and not attempted here.
\begin{figure}[ht!]
    \centering
    \includegraphics[width=\linewidth]{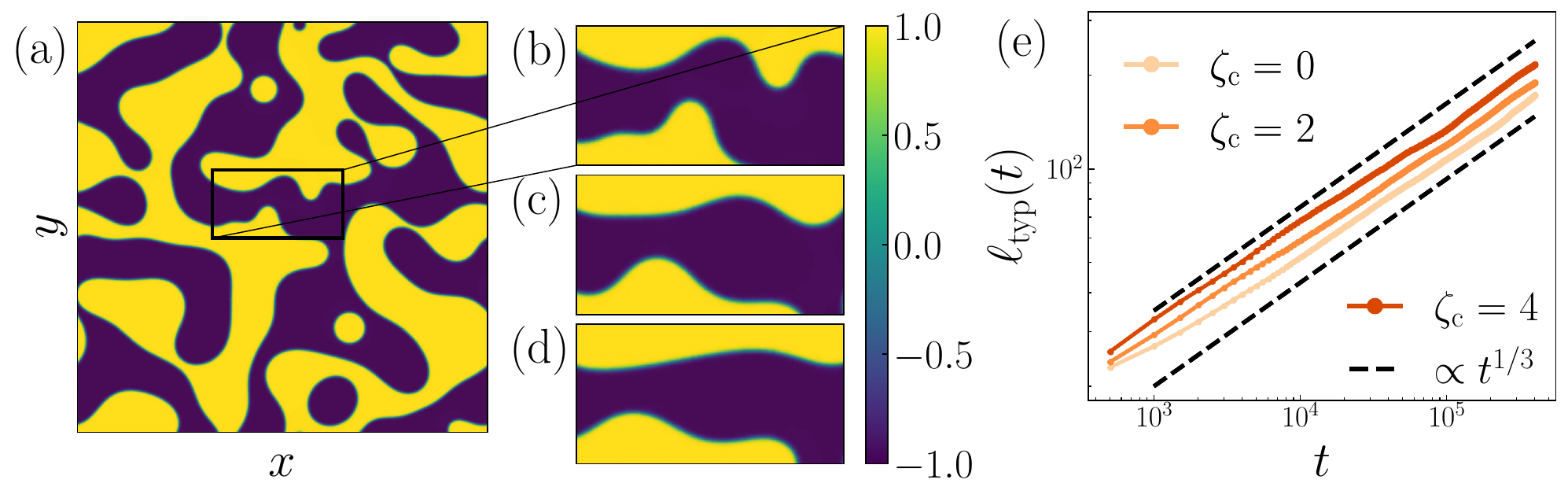}
    \caption{Coarsening of cAMB at fixed global density $\phi_0= 0$ 
    for system size~$1024\times1024$. 
    (a) Snapshot of the density field for $\zetac = 4$. 
    (b-d) Time evolution of the interface showing propagating waves. 
    (e) Characteristic length scale versus time for various~$\zetac$ compatible with the 
    diffusive $\ell_{typ}\sim t^{1/3}$ coarsening law as analytically predicted.}
    \label{fig:amb*_waves-coarsening-phi0-0.0}
\end{figure}
\par 
While the mechanism underlying microphase separation in cQSAPs 
does not require steric or hydrodynamic interactions, it does require sufficiently large translational  diffusivity~\cite{si}. 
When steric repulsion is present, an effective~$\therm$
is generated at the coarse-grained level by collisions alone~\cite{sese-sansa-2022,ma-2022}, 
which are absent in QS particles. 
As this effective~$\therm$ is not easily controlled in terms of microscopic parameters, 
a regime where microphase separation is absent for any value of~$\rho_0$ and~$\omega$ 
has not been previously identified in these models. 
Moreover, chiral microphase-separated states formed of vapor bubbles have hitherto 
only been observed in systems with odd~\cite{caprini-2025, caprini-2026-modelingchiral, marconi-2026, %
petrini-2026-arxiv, digregorio-2026} or hydrodynamic~\cite{shen-2023} interactions. 
These mechanisms are however distinct from the one described here as they require inertia; 
in contrast, the mechanism described in this Letter arises even in the non-inertial regime, 
which is the physically relevant one for describing~$\mu$m-sized active particles.  
\par 
Our MS theory yields explicit predictions for the phase diagram of cQSAPs, but they are strongly model-dependent.
To capture the generic large-scale phenomenology of chiral active systems, 
we now consider the bulk phase separation regime and construct a minimal chiral active continuum model 
that we term `chiral active model~B' (cAMB). 
This is defined in two spatial dimensions by 
\begin{subequations}\label{eq:amb*}
\begin{align}
    \partial_t\phi &= -\grad \vdot \left(\Jv +\sqrt{2D\mob} \, \boldsymbol{\Lambda}\right)
    \,,\\
    J_\alpha/\mob &= -\nabla_\alpha\mu
          +
          \left(\zetac \varepsilon_{\alpha\beta}\,
          + \zeta \delta_{\alpha\beta}
          \right) \,\laplacian \phi\,\nabla_\beta\phi
          \,,\label{eq:chiral-current}
    \\[0.4em]
    \mu &= \frac{\delta \mathcal{F}}{\delta\phi}
    +\lambda |\grad\phi|^2
    \label{eq:mu-amb*}
    \,,
\end{align}
\end{subequations}
where $\phi = \rho-\rho_{\mathrm{crit}}$ is the difference between the density 
and the critical density, and $\mathcal{F}=\int \dd\xv\,[f(\phi)+(K_0/2)|\grad\phi|^2]$. 
We choose for simplicity $f(\phi)=-a\phi^2/2+b\phi^4/4$, 
although our results could be extended to a generic double-well~$f(\phi)$.

 \begin{figure}[ht!]
    \centering
    \includegraphics[width=\linewidth]%
    {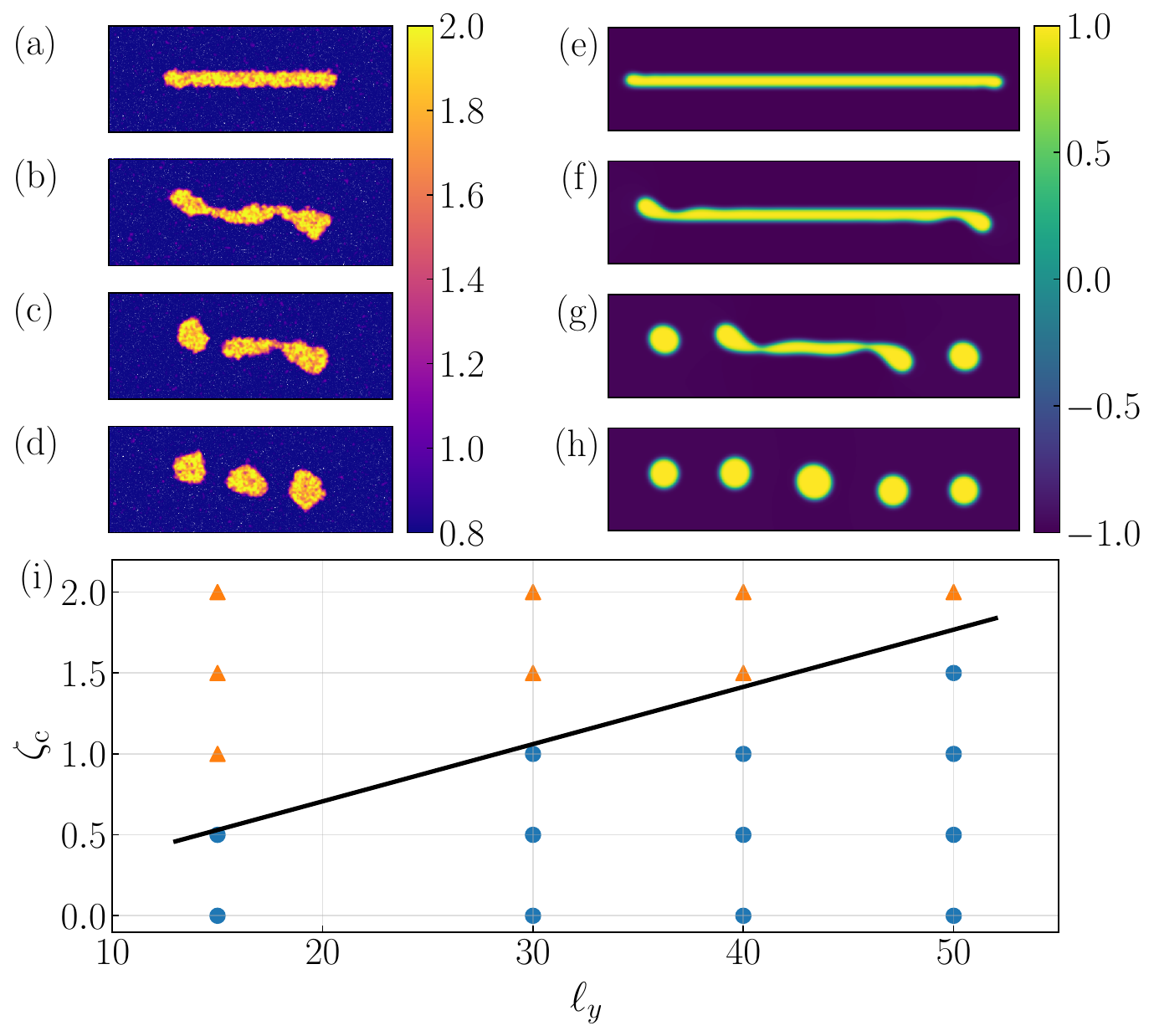}
    \caption{Simulation snapshots illustrating droplet splitting in particle simulations of cQSAPs (Top left) 
    and in cAMB (Top right), with time evolving from top to bottom. 
    Bottom:
    numerical test of our criterion in Eq.~\eqref{eq:droplet-splitting} for droplet splitting 
    (solid line, where $c_{\mathrm{split}} = 0.075$ was obtained from fitting). 
    Triangles indicate emergence of droplet splitting as observed in simulations, 
    while circles its absence. 
    The initial aspect ratio was fixed at $\Upsilon = 20$. 
    In particle simulations, the color encodes the local particle density in units of~$\rhom$. 
    Parameters used: (Left) $\omega = 1.5$, $\rho_0 = 38.4, \Upsilon = 11$ 
    and system size $128\rcut\times48\rcut$.
    (Right) $\zetac = 4.0$, $\ell_y = 15, \Upsilon = 31.25$, and system size~$512\times128$.} 
    \label{fig:amb*_droplet-split}
\end{figure}
cAMB is the minimal extension of Model~B~\cite{hohenberg-1977,chaikin2000principles} 
to describe chiral active systems undergoing phase separation. 
It includes all terms allowed by symmetry to leading order in a gradient and~$\phi$ expansion and, 
to order $\mathcal{O}\left(\nabla^4\phi^2\right)$, the term proportional to~$\zetac$ 
is the only possible chiral contribution~\cite{model-footnote}.
cAMB reduces to equilibrium Model~B~\cite{hohenberg-1977} 
when $\lambda=\zeta=\zetac=0$ and to Active Model~B+ 
that describes achiral active phase separation when $\zetac=0$~\cite{tjhung-2018,cates-2025}. 
In the following, we set $\lambda=\zeta=0$; 
this amounts to a minimal extension of Model~B to active chiral systems. 
In numerical simulations performed below, we fix $-a=b=1/4$, $K_0=1$, and $\mob=1$, 
and denote the global density by $\phi_0 = \int \dd \xv \, \phi/L^2$.
We mainly focus on the noiseless dynamics by setting~$D=0$.
\par 
Since the term proportional to~$\zetac$ in Eq.~\eqref{eq:amb*} vanishes in one-dimensional geometries, 
including circularly symmetric configurations, 
chirality cannot change the interfacial profile, binodals, or the Ostwald ripening process. 
This suggests that the coarsening law remains~$t^{1/3}$, 
as in equilibrium Model~B~\cite{bray-1994} (as long as $\lambda, \zeta = 0$). 
However, it has been debated whether the diffusive~$t^{1/3}$ coarsening law persists 
in the presence of activity even in the absence of chirality~\cite{stenhammar-2013,wittkowski-2014,%
pattanayak-2021,dikshit-2024-arxiv,yadav-2025,bhowmick2026critical}. 
To confirm that the coarsening in unchanged by chirality, we perform simulations of cAMB 
and measure the time evolution of the typical length-scale of domains~$\ell_{\mathrm{typ}}$. 
As expected, our results are compatible with $\ell_{\mathrm{typ}}\propto t^{1/3}$. 
Stronger finite-size effects
are observed (as expected~\cite{wittkowski-2014,cates-2025}) 
for non-bicontinuous systems ($\phi_0\neq0$)~\cite{si}. 
A study of the coarsening dynamics via cQSAPs simulations is beyond the current computational reach~\cite{si}.
\par
We now analyze capillary waves.  
As expected in chiral systems, these are clearly visible along domain 
edges~\cite{zuiden-2016,soni-2019,yang-2020,massana-cid-2021,beppu-2021,yashunsky-2022,%
caporusso-2024,langford-2025-chiral, hargus-2025-prl, hargus-2025-pre,caprini-2025,caprini-2025-thermodynamics-arxiv,%
pisegna-2025-chiral-arxiv,alsallom-2026-arxiv,wang-2026-arxiv,metzger-2026-arxiv}
as illustrated in Fig.~\ref{fig:amb*_waves-coarsening-phi0-0.0}a)--d) 
in the form of unidirectionally traveling waves that are damped by the surface tension.  
It was suggested that edge waves in chiral active systems 
may exhibit a dynamical exponent~$z=2$~\cite{langford-2025-chiral}.  
If this were the case in cAMB, one would expect a signature in the coarsening, 
at least at early times, which we do not observe numerically. 
To clarify this point, we thus derive the dynamics of capillary waves in cAMB.
Denoting by~$h_q$ the Fourier transform of the interface height fluctuations 
and using techniques introduced for other active systems~\cite{fausti-2021, %
caballero2024interface, burekovic-2026,ziethen2026}, we obtain~\cite{si}
\begin{align} 
    \partial_t h_q 
    = - 2  \sigma_{\mathrm{cw}} |q|^3 \tilde{A} h_q
    + \mathbbm{i} \zetac q|q|^2 \tilde{A} \tilde{c} h_q 
    + \eta(q,t) 
    \,,
    \label{eq:amb*_heighteq}
\end{align}
where $\tilde{A},\tilde{c}>0$ are constants and~$\eta$ a Gaussian noise~\cite{si}.  
In Eq.~\eqref{eq:amb*_heighteq},~$\sigma_{\mathrm{cw}}$ 
corresponds to the capillary-wave interfacial tension of AMB+~\cite{fausti-2021,cates-2025} which reduces, 
when $\lambda=\zeta=0$, to the interfacial tension of equilibrium Model~B, 
$\sigma_{\mathrm{cw}} = \sigma_{\mathrm{eq}} = \sqrt{-8 K_0 a^3/(9 b^2)}$.
Chirality generates propagating edge waves through the imaginary term proportional to~$\zetac$ 
and we find that the dynamical exponent associated with both the real 
and imaginary parts of the dispersion relation remain $z=3$ as in the achiral case. 
\par 
However, we now show that chirality has a profound effect on the dynamics of elongated droplets 
and can cause their splitting. 
(Exactly the same phenomenology happens for vapor bubbles given the symmetry 
$(\lambda,\zeta,\zetac,\phi)\to -(\lambda,\zeta,\zetac,\phi)$ of cAMB.) 
In achiral phase-separated systems in which the tension associated with the Ostwald process is positive, 
an initially elongated droplet whose vertical height is significantly larger than the interfacial width 
(so that the pinching of the two interfaces due to thermal fluctuations does not occur) 
evolves in time to become spherical. 
In contrast to this, we show that, in the presence of chirality, elongated droplets can split at their tips 
(see Fig.~\ref{fig:amb*_droplet-split}). 
Interestingly, an analogous phenomenon was experimentally observed 
in magnetically actuated chiral spinners suspended in water~\cite{soni-2019}, 
but the current theoretical description fundamentally hinges on hydrodynamic interactions~\cite{han-2021,neville-2026}. 
We find that droplet breakup also occurs in dry scalar systems. 
\par 
The effect of chirality is strongest near the droplet tips, where the interface curvature is largest, 
and the chiral current induces bending.  
The droplet breaks up when the chiral forcing overcomes the capillary restoring force, 
which opposes interfacial deformations. 
From dimensional analysis, this happens when
\begin{align}\label{eq:droplet-splitting}
    |\zetac| > c_{\mathrm{split}} \sigma_{\mathrm{cw}} \ell_y
    \,, 
\end{align}
where~$c_{\mathrm{split}}$ is a dimensionless prefactor 
that depends on the aspect ratio~$\Upsilon = \ell_x/\ell_y$, $\ell_x$ 
and~$\ell_y$ are the horizontal and vertical lengths of the droplet. 
We confirm the scaling relation~\eqref{eq:droplet-splitting} 
numerically for a range of~$\ell_y$ and~$\zetac$ values, 
as shown in Fig~\ref{fig:amb*_droplet-split}i. 
Furthermore, we have observed that~$c_{\mathrm{split}}$ decreases with increasing~$\Upsilon$ (data not shown). 

Performing simulations of cQSAPs starting from an initial configuration of an elongated droplet, 
we found that droplet splitting arises likewise in cQSAPs, 
see Figs.~\ref{fig:amb*_droplet-split}a--d and the corresponding movies in~\cite{si}. 
As is visible in the movies in~\cite{si}, 
it is likely that the splitting of elongated droplets contributes 
to the size selection of microphase-separated structures in cQSAPs, 
but a quantification of this effect goes beyond the scope of this Letter. 
\par
\inlinesection{Discussion}
We have shown that for chiral active matter, QS active systems can either undergo bulk or microphase separation 
depending on the strength of chirality and on the presence of translational diffusivity. 
This contrasts with achiral QS particles which undergo only bulk phase separation. 
Microphase separation in chiral QS particles occurs at both low and high mean densities, 
with the system forming dense clusters and vapor bubbles, respectively. 
In systems coarsening toward bulk phase separation, 
we have further shown the diffusive law~$t^{1/3}$ is unchanged by chirality.
We additionally identified a generic mechanism by which chirality breaks up elongated droplets, 
showing that a phenomenon previously observed and modeled in wet systems persists even in the dry limit. 
Finally, we introduced and analyzed a minimal continuum theory for chiral active phase separation, 
supporting the generality of our results. 
We have left for future work a description of the effect of chirality on
phase-separated states with negative interfacial tensions
induced by activity~\cite{cates-2025}. 
The generalization of cAMB to the regime in which chirality induces microphase separation 
may build on the model recently introduced in Ref.~\cite{toffenetti-2026-arxiv}, 
and this might enable the discovery of emergent network-like states in active chiral phase-separating systems.  
\par 
\inlinesection{Acknowledgments}
SB and CN thank M.E. Cates for fruitful discussion on the coarsening law. 
AM, SK and CN warmly thank A. Gareth and S. Ramaswamy for fruitful discussions 
on cAMB during the initial stages of this work.
AM and CN acknowledge the support of the ANR grant PSAM. 
CN acknowledges the support of the INP-IRP grant IFAM. 
AM acknowledges support from CY Cergy Paris Universit\'{e} through a TALENT fellowship.
SJK acknowledges the support provided by the Indo-French Centre 
for the Promotion of Advanced Research under the Raman Charpak Fellowship. 
The authors thank the Isaac Newton Institute for Mathematical Sciences for support and hospitality 
during the program ``Anti-diffusive dynamics: from sub-cellular to astrophysical scales'' 
during which a part of this work was undertaken. 
This research was supported in part by grant NSF PHY-2309135 to the Kavli Institute for Theoretical Physics (KITP).

\bibliography{refs}

\begin{thebibliography}{143}%
\makeatletter
\providecommand \@ifxundefined [1]{%
 \@ifx{#1\undefined}
}%
\providecommand \@ifnum [1]{%
 \ifnum #1\expandafter \@firstoftwo
 \else \expandafter \@secondoftwo
 \fi
}%
\providecommand \@ifx [1]{%
 \ifx #1\expandafter \@firstoftwo
 \else \expandafter \@secondoftwo
 \fi
}%
\providecommand \natexlab [1]{#1}%
\providecommand \enquote  [1]{``#1''}%
\providecommand \bibnamefont  [1]{#1}%
\providecommand \bibfnamefont [1]{#1}%
\providecommand \citenamefont [1]{#1}%
\providecommand \href@noop [0]{\@secondoftwo}%
\providecommand \href [0]{\begingroup \@sanitize@url \@href}%
\providecommand \@href[1]{\@@startlink{#1}\@@href}%
\providecommand \@@href[1]{\endgroup#1\@@endlink}%
\providecommand \@sanitize@url [0]{\catcode `\\12\catcode `\$12\catcode
  `\&12\catcode `\#12\catcode `\^12\catcode `\_12\catcode `\%12\relax}%
\providecommand \@@startlink[1]{}%
\providecommand \@@endlink[0]{}%
\providecommand \url  [0]{\begingroup\@sanitize@url \@url }%
\providecommand \@url [1]{\endgroup\@href {#1}{\urlprefix }}%
\providecommand \urlprefix  [0]{URL }%
\providecommand \Eprint [0]{\href }%
\providecommand \doibase [0]{https://doi.org/}%
\providecommand \selectlanguage [0]{\@gobble}%
\providecommand \bibinfo  [0]{\@secondoftwo}%
\providecommand \bibfield  [0]{\@secondoftwo}%
\providecommand \translation [1]{[#1]}%
\providecommand \BibitemOpen [0]{}%
\providecommand \bibitemStop [0]{}%
\providecommand \bibitemNoStop [0]{.\EOS\space}%
\providecommand \EOS [0]{\spacefactor3000\relax}%
\providecommand \BibitemShut  [1]{\csname bibitem#1\endcsname}%
\let\auto@bib@innerbib\@empty
\bibitem [{\citenamefont {van Teeffelen}\ and\ \citenamefont
  {L\"owen}(2008)}]{teeffelen-2008}%
  \BibitemOpen
  \bibfield  {author} {\bibinfo {author} {\bibfnamefont {S.}~\bibnamefont {van
  Teeffelen}}\ and\ \bibinfo {author} {\bibfnamefont {H.}~\bibnamefont
  {L\"owen}},\ }\bibfield  {title} {\bibinfo {title} {Dynamics of a {Brownian}
  circle swimmer},\ }\href {https://doi.org/10.1103/PhysRevE.78.020101}
  {\bibfield  {journal} {\bibinfo  {journal} {Phys. Rev. E}\ }\textbf {\bibinfo
  {volume} {78}},\ \bibinfo {pages} {020101} (\bibinfo {year}
  {2008})}\BibitemShut {NoStop}%
\bibitem [{\citenamefont {L{\"o}wen}(2016)}]{lowen-2016}%
  \BibitemOpen
  \bibfield  {author} {\bibinfo {author} {\bibfnamefont {H.}~\bibnamefont
  {L{\"o}wen}},\ }\bibfield  {title} {\bibinfo {title} {Chirality in
  microswimmer motion: From circle swimmers to active turbulence},\ }\href
  {https://doi.org/10.1140/epjst/e2016-60054-6} {\bibfield  {journal} {\bibinfo
   {journal} {Eur. Phys. J. Spec. Top.}\ }\textbf {\bibinfo {volume} {225}},\
  \bibinfo {pages} {2319} (\bibinfo {year} {2016})}\BibitemShut {NoStop}%
\bibitem [{\citenamefont {Liebchen}\ and\ \citenamefont
  {Levis}(2022)}]{liebchen-2022}%
  \BibitemOpen
  \bibfield  {author} {\bibinfo {author} {\bibfnamefont {B.}~\bibnamefont
  {Liebchen}}\ and\ \bibinfo {author} {\bibfnamefont {D.}~\bibnamefont
  {Levis}},\ }\bibfield  {title} {\bibinfo {title} {Chiral active matter},\
  }\href {https://doi.org/10.1209/0295-5075/ac8f69} {\bibfield  {journal}
  {\bibinfo  {journal} {EPL}\ }\textbf {\bibinfo {volume} {139}},\ \bibinfo
  {pages} {67001} (\bibinfo {year} {2022})}\BibitemShut {NoStop}%
\bibitem [{\citenamefont {{Mecke, J.}}\ \emph {et~al.}(2024)\citenamefont
  {{Mecke, J.}}, \citenamefont {{Nketsiah, J. O.}}, \citenamefont {{Li, R.}},\
  and\ \citenamefont {{Gao, Y.}}}]{mecke-2024}%
  \BibitemOpen
  \bibfield  {author} {\bibinfo {author} {\bibnamefont {{Mecke, J.}}}, \bibinfo
  {author} {\bibnamefont {{Nketsiah, J. O.}}}, \bibinfo {author} {\bibnamefont
  {{Li, R.}}},\ and\ \bibinfo {author} {\bibnamefont {{Gao, Y.}}},\ }\bibfield
  {title} {\bibinfo {title} {Emergent phenomena in chiral active matter},\
  }\href {https://doi.org/10.1360/nso/20230086} {\bibfield  {journal} {\bibinfo
   {journal} {Natl Sci Open}\ }\textbf {\bibinfo {volume} {3}},\ \bibinfo
  {pages} {20230086} (\bibinfo {year} {2024})}\BibitemShut {NoStop}%
\bibitem [{\citenamefont {DiLuzio}\ \emph {et~al.}(2005)\citenamefont
  {DiLuzio}, \citenamefont {Turner}, \citenamefont {Mayer}, \citenamefont
  {Garstecki}, \citenamefont {Weibel}, \citenamefont {Berg},\ and\
  \citenamefont {Whitesides}}]{diLuzio-2005}%
  \BibitemOpen
  \bibfield  {author} {\bibinfo {author} {\bibfnamefont {W.~R.}\ \bibnamefont
  {DiLuzio}}, \bibinfo {author} {\bibfnamefont {L.}~\bibnamefont {Turner}},
  \bibinfo {author} {\bibfnamefont {M.}~\bibnamefont {Mayer}}, \bibinfo
  {author} {\bibfnamefont {P.}~\bibnamefont {Garstecki}}, \bibinfo {author}
  {\bibfnamefont {D.~B.}\ \bibnamefont {Weibel}}, \bibinfo {author}
  {\bibfnamefont {H.~C.}\ \bibnamefont {Berg}},\ and\ \bibinfo {author}
  {\bibfnamefont {G.}~\bibnamefont {Whitesides}},\ }\bibfield  {title}
  {\bibinfo {title} {Escherichia coli swim on the right-hand side},\ }\href
  {https://doi.org/10.1038/nature03660} {\bibfield  {journal} {\bibinfo
  {journal} {Nature}\ }\textbf {\bibinfo {volume} {435}},\ \bibinfo {pages}
  {1271} (\bibinfo {year} {2005})}\BibitemShut {NoStop}%
\bibitem [{\citenamefont {Riedel}\ \emph {et~al.}(2005)\citenamefont {Riedel},
  \citenamefont {Kruse},\ and\ \citenamefont {Howard}}]{riedel-2005}%
  \BibitemOpen
  \bibfield  {author} {\bibinfo {author} {\bibfnamefont {I.~H.}\ \bibnamefont
  {Riedel}}, \bibinfo {author} {\bibfnamefont {K.}~\bibnamefont {Kruse}},\ and\
  \bibinfo {author} {\bibfnamefont {J.}~\bibnamefont {Howard}},\ }\bibfield
  {title} {\bibinfo {title} {A self-organized vortex array of hydrodynamically
  entrained sperm cells},\ }\href {https://doi.org/10.1126/science.1110329}
  {\bibfield  {journal} {\bibinfo  {journal} {Science}\ }\textbf {\bibinfo
  {volume} {309}},\ \bibinfo {pages} {300} (\bibinfo {year}
  {2005})}\BibitemShut {NoStop}%
\bibitem [{\citenamefont {Lauga}\ \emph {et~al.}(2006)\citenamefont {Lauga},
  \citenamefont {DiLuzio}, \citenamefont {Whitesides},\ and\ \citenamefont
  {Stone}}]{lauga-2006}%
  \BibitemOpen
  \bibfield  {author} {\bibinfo {author} {\bibfnamefont {E.}~\bibnamefont
  {Lauga}}, \bibinfo {author} {\bibfnamefont {W.~R.}\ \bibnamefont {DiLuzio}},
  \bibinfo {author} {\bibfnamefont {G.~M.}\ \bibnamefont {Whitesides}},\ and\
  \bibinfo {author} {\bibfnamefont {H.~A.}\ \bibnamefont {Stone}},\ }\bibfield
  {title} {\bibinfo {title} {Swimming in circles: Motion of bacteria near solid
  boundaries},\ }\href {https://doi.org/10.1529/biophysj.105.069401} {\bibfield
   {journal} {\bibinfo  {journal} {Biophys. J.}\ }\textbf {\bibinfo {volume}
  {90}},\ \bibinfo {pages} {400} (\bibinfo {year} {2006})}\BibitemShut
  {NoStop}%
\bibitem [{\citenamefont {Tan}\ \emph {et~al.}(2022)\citenamefont {Tan},
  \citenamefont {Mietke}, \citenamefont {Li}, \citenamefont {Chen},
  \citenamefont {Higinbotham}, \citenamefont {Foster}, \citenamefont {Gokhale},
  \citenamefont {Dunkel},\ and\ \citenamefont {Fakhri}}]{tan-2022}%
  \BibitemOpen
  \bibfield  {author} {\bibinfo {author} {\bibfnamefont {T.~H.}\ \bibnamefont
  {Tan}}, \bibinfo {author} {\bibfnamefont {A.}~\bibnamefont {Mietke}},
  \bibinfo {author} {\bibfnamefont {J.}~\bibnamefont {Li}}, \bibinfo {author}
  {\bibfnamefont {Y.}~\bibnamefont {Chen}}, \bibinfo {author} {\bibfnamefont
  {H.}~\bibnamefont {Higinbotham}}, \bibinfo {author} {\bibfnamefont {P.~J.}\
  \bibnamefont {Foster}}, \bibinfo {author} {\bibfnamefont {S.}~\bibnamefont
  {Gokhale}}, \bibinfo {author} {\bibfnamefont {J.}~\bibnamefont {Dunkel}},\
  and\ \bibinfo {author} {\bibfnamefont {N.}~\bibnamefont {Fakhri}},\
  }\bibfield  {title} {\bibinfo {title} {Odd dynamics of living chiral
  crystals},\ }\href {https://doi.org/10.1038/s41586-022-04889-6} {\bibfield
  {journal} {\bibinfo  {journal} {Nature}\ }\textbf {\bibinfo {volume} {607}},\
  \bibinfo {pages} {287} (\bibinfo {year} {2022})}\BibitemShut {NoStop}%
\bibitem [{\citenamefont {Yang}\ \emph {et~al.}(2020)\citenamefont {Yang},
  \citenamefont {Ren}, \citenamefont {Cheng},\ and\ \citenamefont
  {Zhang}}]{yang-2020}%
  \BibitemOpen
  \bibfield  {author} {\bibinfo {author} {\bibfnamefont {X.}~\bibnamefont
  {Yang}}, \bibinfo {author} {\bibfnamefont {C.}~\bibnamefont {Ren}}, \bibinfo
  {author} {\bibfnamefont {K.}~\bibnamefont {Cheng}},\ and\ \bibinfo {author}
  {\bibfnamefont {H.~P.}\ \bibnamefont {Zhang}},\ }\bibfield  {title} {\bibinfo
  {title} {Robust boundary flow in chiral active fluid},\ }\href
  {https://doi.org/10.1103/PhysRevE.101.022603} {\bibfield  {journal} {\bibinfo
   {journal} {Phys. Rev. E}\ }\textbf {\bibinfo {volume} {101}},\ \bibinfo
  {pages} {022603} (\bibinfo {year} {2020})}\BibitemShut {NoStop}%
\bibitem [{\citenamefont {K\"ummel}\ \emph {et~al.}(2013)\citenamefont
  {K\"ummel}, \citenamefont {ten Hagen}, \citenamefont {Wittkowski},
  \citenamefont {Buttinoni}, \citenamefont {Eichhorn}, \citenamefont {Volpe},
  \citenamefont {L\"owen},\ and\ \citenamefont {Bechinger}}]{kummel-2013}%
  \BibitemOpen
  \bibfield  {author} {\bibinfo {author} {\bibfnamefont {F.}~\bibnamefont
  {K\"ummel}}, \bibinfo {author} {\bibfnamefont {B.}~\bibnamefont {ten Hagen}},
  \bibinfo {author} {\bibfnamefont {R.}~\bibnamefont {Wittkowski}}, \bibinfo
  {author} {\bibfnamefont {I.}~\bibnamefont {Buttinoni}}, \bibinfo {author}
  {\bibfnamefont {R.}~\bibnamefont {Eichhorn}}, \bibinfo {author}
  {\bibfnamefont {G.}~\bibnamefont {Volpe}}, \bibinfo {author} {\bibfnamefont
  {H.}~\bibnamefont {L\"owen}},\ and\ \bibinfo {author} {\bibfnamefont
  {C.}~\bibnamefont {Bechinger}},\ }\bibfield  {title} {\bibinfo {title}
  {Circular motion of asymmetric self-propelling particles},\ }\href
  {https://doi.org/10.1103/PhysRevLett.110.198302} {\bibfield  {journal}
  {\bibinfo  {journal} {Phys. Rev. Lett.}\ }\textbf {\bibinfo {volume} {110}},\
  \bibinfo {pages} {198302} (\bibinfo {year} {2013})}\BibitemShut {NoStop}%
\bibitem [{\citenamefont {Soni}\ \emph {et~al.}(2019)\citenamefont {Soni},
  \citenamefont {Bililign}, \citenamefont {Magkiriadou}, \citenamefont
  {Sacanna}, \citenamefont {Bartolo}, \citenamefont {Shelley},\ and\
  \citenamefont {Irvine}}]{soni-2019}%
  \BibitemOpen
  \bibfield  {author} {\bibinfo {author} {\bibfnamefont {V.}~\bibnamefont
  {Soni}}, \bibinfo {author} {\bibfnamefont {E.~S.}\ \bibnamefont {Bililign}},
  \bibinfo {author} {\bibfnamefont {S.}~\bibnamefont {Magkiriadou}}, \bibinfo
  {author} {\bibfnamefont {S.}~\bibnamefont {Sacanna}}, \bibinfo {author}
  {\bibfnamefont {D.}~\bibnamefont {Bartolo}}, \bibinfo {author} {\bibfnamefont
  {M.~J.}\ \bibnamefont {Shelley}},\ and\ \bibinfo {author} {\bibfnamefont
  {W.~T.~M.}\ \bibnamefont {Irvine}},\ }\bibfield  {title} {\bibinfo {title}
  {The odd free surface flows of a colloidal chiral fluid},\ }\href
  {https://doi.org/10.1038/s41567-019-0603-8} {\bibfield  {journal} {\bibinfo
  {journal} {Nat. Phys.}\ }\textbf {\bibinfo {volume} {15}},\ \bibinfo {pages}
  {1188} (\bibinfo {year} {2019})}\BibitemShut {NoStop}%
\bibitem [{\citenamefont {Massana-Cid}\ \emph {et~al.}(2021)\citenamefont
  {Massana-Cid}, \citenamefont {Levis}, \citenamefont {Hern\'andez},
  \citenamefont {Pagonabarraga},\ and\ \citenamefont
  {Tierno}}]{massana-cid-2021}%
  \BibitemOpen
  \bibfield  {author} {\bibinfo {author} {\bibfnamefont {H.}~\bibnamefont
  {Massana-Cid}}, \bibinfo {author} {\bibfnamefont {D.}~\bibnamefont {Levis}},
  \bibinfo {author} {\bibfnamefont {R.~J.~H.}\ \bibnamefont {Hern\'andez}},
  \bibinfo {author} {\bibfnamefont {I.}~\bibnamefont {Pagonabarraga}},\ and\
  \bibinfo {author} {\bibfnamefont {P.}~\bibnamefont {Tierno}},\ }\bibfield
  {title} {\bibinfo {title} {Arrested phase separation in chiral fluids of
  colloidal spinners},\ }\href
  {https://doi.org/10.1103/PhysRevResearch.3.L042021} {\bibfield  {journal}
  {\bibinfo  {journal} {Phys. Rev. Res.}\ }\textbf {\bibinfo {volume} {3}},\
  \bibinfo {pages} {L042021} (\bibinfo {year} {2021})}\BibitemShut {NoStop}%
\bibitem [{\citenamefont {Hargus}\ \emph {et~al.}(2021)\citenamefont {Hargus},
  \citenamefont {Epstein},\ and\ \citenamefont {Mandadapu}}]{hargus-2021}%
  \BibitemOpen
  \bibfield  {author} {\bibinfo {author} {\bibfnamefont {C.}~\bibnamefont
  {Hargus}}, \bibinfo {author} {\bibfnamefont {J.~M.}\ \bibnamefont
  {Epstein}},\ and\ \bibinfo {author} {\bibfnamefont {K.~K.}\ \bibnamefont
  {Mandadapu}},\ }\bibfield  {title} {\bibinfo {title} {Odd diffusivity of
  chiral random motion},\ }\href
  {https://doi.org/10.1103/PhysRevLett.127.178001} {\bibfield  {journal}
  {\bibinfo  {journal} {Phys. Rev. Lett.}\ }\textbf {\bibinfo {volume} {127}},\
  \bibinfo {pages} {178001} (\bibinfo {year} {2021})}\BibitemShut {NoStop}%
\bibitem [{\citenamefont {Faedi}\ \emph {et~al.}(2026)\citenamefont {Faedi},
  \citenamefont {Kalz}, \citenamefont {Metzler},\ and\ \citenamefont
  {Sharma}}]{faedi-2026-arxiv}%
  \BibitemOpen
  \bibfield  {author} {\bibinfo {author} {\bibfnamefont {F.}~\bibnamefont
  {Faedi}}, \bibinfo {author} {\bibfnamefont {E.}~\bibnamefont {Kalz}},
  \bibinfo {author} {\bibfnamefont {R.}~\bibnamefont {Metzler}},\ and\ \bibinfo
  {author} {\bibfnamefont {A.}~\bibnamefont {Sharma}},\ }\bibfield  {title}
  {\bibinfo {title} {A mobility based approach to transport in chiral fluids}\
  }\href {https://doi.org/10.48550/arXiv.2602.18091}
  {10.48550/arXiv.2602.18091} (\bibinfo {year} {2026})\BibitemShut {NoStop}%
\bibitem [{\citenamefont {Maire}\ \emph {et~al.}(2026)\citenamefont {Maire},
  \citenamefont {Petrini}, \citenamefont {{Marini Bettolo Marconi}},\ and\
  \citenamefont {Caprini}}]{maire-2026-arxiv}%
  \BibitemOpen
  \bibfield  {author} {\bibinfo {author} {\bibfnamefont {R.}~\bibnamefont
  {Maire}}, \bibinfo {author} {\bibfnamefont {A.}~\bibnamefont {Petrini}},
  \bibinfo {author} {\bibfnamefont {U.}~\bibnamefont {{Marini Bettolo
  Marconi}}},\ and\ \bibinfo {author} {\bibfnamefont {L.}~\bibnamefont
  {Caprini}},\ }\bibfield  {title} {\bibinfo {title} {Kinetic theory of chiral
  active disks: Odd transport and torque density}\ }\href
  {https://doi.org/10.48550/arXiv.2603.04273} {10.48550/arXiv.2603.04273}
  (\bibinfo {year} {2026})\BibitemShut {NoStop}%
\bibitem [{\citenamefont {Avron}(1998)}]{avron-1998}%
  \BibitemOpen
  \bibfield  {author} {\bibinfo {author} {\bibfnamefont {J.~E.}\ \bibnamefont
  {Avron}},\ }\bibfield  {title} {\bibinfo {title} {Odd viscosity},\ }\href
  {https://doi.org/10.1023/A:1023084404080} {\bibfield  {journal} {\bibinfo
  {journal} {J. Stat. Phys.}\ }\textbf {\bibinfo {volume} {92}},\ \bibinfo
  {pages} {543} (\bibinfo {year} {1998})}\BibitemShut {NoStop}%
\bibitem [{\citenamefont {Banerjee}\ \emph {et~al.}(2017)\citenamefont
  {Banerjee}, \citenamefont {Souslov}, \citenamefont {Abanov},\ and\
  \citenamefont {Vitelli}}]{banerjee-2017}%
  \BibitemOpen
  \bibfield  {author} {\bibinfo {author} {\bibfnamefont {D.}~\bibnamefont
  {Banerjee}}, \bibinfo {author} {\bibfnamefont {A.}~\bibnamefont {Souslov}},
  \bibinfo {author} {\bibfnamefont {A.~G.}\ \bibnamefont {Abanov}},\ and\
  \bibinfo {author} {\bibfnamefont {V.}~\bibnamefont {Vitelli}},\ }\bibfield
  {title} {\bibinfo {title} {Odd viscosity in chiral active fluids},\ }\href
  {https://doi.org/10.1038/s41467-017-01378-7} {\bibfield  {journal} {\bibinfo
  {journal} {Nat. Commun.}\ }\textbf {\bibinfo {volume} {8}},\ \bibinfo {pages}
  {1573} (\bibinfo {year} {2017})}\BibitemShut {NoStop}%
\bibitem [{\citenamefont {Scheibner}\ \emph {et~al.}(2020)\citenamefont
  {Scheibner}, \citenamefont {Souslov}, \citenamefont {Banerjee}, \citenamefont
  {Sur{\'o}wka}, \citenamefont {Irvine},\ and\ \citenamefont
  {Vitelli}}]{scheibner-2020}%
  \BibitemOpen
  \bibfield  {author} {\bibinfo {author} {\bibfnamefont {C.}~\bibnamefont
  {Scheibner}}, \bibinfo {author} {\bibfnamefont {A.}~\bibnamefont {Souslov}},
  \bibinfo {author} {\bibfnamefont {D.}~\bibnamefont {Banerjee}}, \bibinfo
  {author} {\bibfnamefont {P.}~\bibnamefont {Sur{\'o}wka}}, \bibinfo {author}
  {\bibfnamefont {W.~T.~M.}\ \bibnamefont {Irvine}},\ and\ \bibinfo {author}
  {\bibfnamefont {V.}~\bibnamefont {Vitelli}},\ }\bibfield  {title} {\bibinfo
  {title} {Odd elasticity},\ }\href {https://doi.org/10.1038/s41567-020-0795-y}
  {\bibfield  {journal} {\bibinfo  {journal} {Nat. Phys.}\ }\textbf {\bibinfo
  {volume} {16}},\ \bibinfo {pages} {475} (\bibinfo {year} {2020})}\BibitemShut
  {NoStop}%
\bibitem [{\citenamefont {Maitra}\ \emph {et~al.}(2020)\citenamefont {Maitra},
  \citenamefont {Lenz},\ and\ \citenamefont {Voituriez}}]{maitra-2020}%
  \BibitemOpen
  \bibfield  {author} {\bibinfo {author} {\bibfnamefont {A.}~\bibnamefont
  {Maitra}}, \bibinfo {author} {\bibfnamefont {M.}~\bibnamefont {Lenz}},\ and\
  \bibinfo {author} {\bibfnamefont {R.}~\bibnamefont {Voituriez}},\ }\bibfield
  {title} {\bibinfo {title} {Chiral active hexatics: Giant number fluctuations,
  waves, and destruction of order},\ }\href
  {https://doi.org/10.1103/PhysRevLett.125.238005} {\bibfield  {journal}
  {\bibinfo  {journal} {Phys. Rev. Lett.}\ }\textbf {\bibinfo {volume} {125}},\
  \bibinfo {pages} {238005} (\bibinfo {year} {2020})}\BibitemShut {NoStop}%
\bibitem [{\citenamefont {Fruchart}\ \emph {et~al.}(2023)\citenamefont
  {Fruchart}, \citenamefont {Scheibner},\ and\ \citenamefont
  {Vitelli}}]{fruchart-2023}%
  \BibitemOpen
  \bibfield  {author} {\bibinfo {author} {\bibfnamefont {M.}~\bibnamefont
  {Fruchart}}, \bibinfo {author} {\bibfnamefont {C.}~\bibnamefont
  {Scheibner}},\ and\ \bibinfo {author} {\bibfnamefont {V.}~\bibnamefont
  {Vitelli}},\ }\bibfield  {title} {\bibinfo {title} {Odd viscosity and odd
  elasticity},\ }\href
  {https://doi.org/10.1146/annurev-conmatphys-040821-125506} {\bibfield
  {journal} {\bibinfo  {journal} {Annu. Rev. Condens. Matter Phys.}\ }\textbf
  {\bibinfo {volume} {14}},\ \bibinfo {pages} {471} (\bibinfo {year}
  {2023})}\BibitemShut {NoStop}%
\bibitem [{\citenamefont {van Zuiden}\ \emph {et~al.}(2016)\citenamefont {van
  Zuiden}, \citenamefont {Paulose}, \citenamefont {Irvine}, \citenamefont
  {Bartolo},\ and\ \citenamefont {Vitelli}}]{zuiden-2016}%
  \BibitemOpen
  \bibfield  {author} {\bibinfo {author} {\bibfnamefont {B.~C.}\ \bibnamefont
  {van Zuiden}}, \bibinfo {author} {\bibfnamefont {J.}~\bibnamefont {Paulose}},
  \bibinfo {author} {\bibfnamefont {W.~T.~M.}\ \bibnamefont {Irvine}}, \bibinfo
  {author} {\bibfnamefont {D.}~\bibnamefont {Bartolo}},\ and\ \bibinfo {author}
  {\bibfnamefont {V.}~\bibnamefont {Vitelli}},\ }\bibfield  {title} {\bibinfo
  {title} {Spatiotemporal order and emergent edge currents in active spinner
  materials},\ }\href {https://doi.org/10.1073/pnas.1609572113} {\bibfield
  {journal} {\bibinfo  {journal} {Proc. Natl. Acad. Sci.}\ }\textbf {\bibinfo
  {volume} {113}},\ \bibinfo {pages} {12919} (\bibinfo {year}
  {2016})}\BibitemShut {NoStop}%
\bibitem [{\citenamefont {Beppu}\ \emph {et~al.}(2021)\citenamefont {Beppu},
  \citenamefont {Izri}, \citenamefont {Sato}, \citenamefont {Yamanishi},
  \citenamefont {Sumino},\ and\ \citenamefont {Maeda}}]{beppu-2021}%
  \BibitemOpen
  \bibfield  {author} {\bibinfo {author} {\bibfnamefont {K.}~\bibnamefont
  {Beppu}}, \bibinfo {author} {\bibfnamefont {Z.}~\bibnamefont {Izri}},
  \bibinfo {author} {\bibfnamefont {T.}~\bibnamefont {Sato}}, \bibinfo {author}
  {\bibfnamefont {Y.}~\bibnamefont {Yamanishi}}, \bibinfo {author}
  {\bibfnamefont {Y.}~\bibnamefont {Sumino}},\ and\ \bibinfo {author}
  {\bibfnamefont {Y.~T.}\ \bibnamefont {Maeda}},\ }\bibfield  {title} {\bibinfo
  {title} {Edge current and pairing order transition in chiral bacterial
  vortices},\ }\href {https://doi.org/10.1073/pnas.2107461118} {\bibfield
  {journal} {\bibinfo  {journal} {Proc. Natl. Acad. Sci. U.S.A.}\ }\textbf
  {\bibinfo {volume} {118}},\ \bibinfo {pages} {e2107461118} (\bibinfo {year}
  {2021})}\BibitemShut {NoStop}%
\bibitem [{\citenamefont {Yashunsky}\ \emph {et~al.}(2022)\citenamefont
  {Yashunsky}, \citenamefont {Pearce}, \citenamefont {Blanch-Mercader},
  \citenamefont {Ascione}, \citenamefont {Silberzan},\ and\ \citenamefont
  {Giomi}}]{yashunsky-2022}%
  \BibitemOpen
  \bibfield  {author} {\bibinfo {author} {\bibfnamefont {V.}~\bibnamefont
  {Yashunsky}}, \bibinfo {author} {\bibfnamefont {D.~J.~G.}\ \bibnamefont
  {Pearce}}, \bibinfo {author} {\bibfnamefont {C.}~\bibnamefont
  {Blanch-Mercader}}, \bibinfo {author} {\bibfnamefont {F.}~\bibnamefont
  {Ascione}}, \bibinfo {author} {\bibfnamefont {P.}~\bibnamefont {Silberzan}},\
  and\ \bibinfo {author} {\bibfnamefont {L.}~\bibnamefont {Giomi}},\ }\bibfield
   {title} {\bibinfo {title} {Chiral edge current in nematic cell monolayers},\
  }\href {https://doi.org/10.1103/PhysRevX.12.041017} {\bibfield  {journal}
  {\bibinfo  {journal} {Phys. Rev. X}\ }\textbf {\bibinfo {volume} {12}},\
  \bibinfo {pages} {041017} (\bibinfo {year} {2022})}\BibitemShut {NoStop}%
\bibitem [{\citenamefont {Caporusso}\ \emph {et~al.}(2024)\citenamefont
  {Caporusso}, \citenamefont {Gonnella},\ and\ \citenamefont
  {Levis}}]{caporusso-2024}%
  \BibitemOpen
  \bibfield  {author} {\bibinfo {author} {\bibfnamefont {C.~B.}\ \bibnamefont
  {Caporusso}}, \bibinfo {author} {\bibfnamefont {G.}~\bibnamefont
  {Gonnella}},\ and\ \bibinfo {author} {\bibfnamefont {D.}~\bibnamefont
  {Levis}},\ }\bibfield  {title} {\bibinfo {title} {Phase coexistence and edge
  currents in the chiral {Lennard-Jones} fluid},\ }\href
  {https://doi.org/10.1103/PhysRevLett.132.168201} {\bibfield  {journal}
  {\bibinfo  {journal} {Phys. Rev. Lett.}\ }\textbf {\bibinfo {volume} {132}},\
  \bibinfo {pages} {168201} (\bibinfo {year} {2024})}\BibitemShut {NoStop}%
\bibitem [{\citenamefont {Langford}\ and\ \citenamefont
  {Omar}(2025)}]{langford-2025-chiral}%
  \BibitemOpen
  \bibfield  {author} {\bibinfo {author} {\bibfnamefont {L.}~\bibnamefont
  {Langford}}\ and\ \bibinfo {author} {\bibfnamefont {A.~K.}\ \bibnamefont
  {Omar}},\ }\bibfield  {title} {\bibinfo {title} {Phase separation,
  capillarity, and odd-surface flows in chiral active matter},\ }\href
  {https://doi.org/10.1103/PhysRevLett.134.068301} {\bibfield  {journal}
  {\bibinfo  {journal} {Phys. Rev. Lett.}\ }\textbf {\bibinfo {volume} {134}},\
  \bibinfo {pages} {068301} (\bibinfo {year} {2025})}\BibitemShut {NoStop}%
\bibitem [{\citenamefont {Hargus}\ \emph
  {et~al.}(2025{\natexlab{a}})\citenamefont {Hargus}, \citenamefont {Ghimenti},
  \citenamefont {Tailleur},\ and\ \citenamefont {van
  Wijland}}]{hargus-2025-prl}%
  \BibitemOpen
  \bibfield  {author} {\bibinfo {author} {\bibfnamefont {C.}~\bibnamefont
  {Hargus}}, \bibinfo {author} {\bibfnamefont {F.}~\bibnamefont {Ghimenti}},
  \bibinfo {author} {\bibfnamefont {J.}~\bibnamefont {Tailleur}},\ and\
  \bibinfo {author} {\bibfnamefont {F.}~\bibnamefont {van Wijland}},\
  }\bibfield  {title} {\bibinfo {title} {Odd dynamics of passive objects in a
  chiral active bath},\ }\href {https://doi.org/10.1103/pdpf-sd9q} {\bibfield
  {journal} {\bibinfo  {journal} {Phys. Rev. Lett.}\ }\textbf {\bibinfo
  {volume} {135}},\ \bibinfo {pages} {167102} (\bibinfo {year}
  {2025}{\natexlab{a}})}\BibitemShut {NoStop}%
\bibitem [{\citenamefont {Hargus}\ \emph
  {et~al.}(2025{\natexlab{b}})\citenamefont {Hargus}, \citenamefont {Ghimenti},
  \citenamefont {Tailleur},\ and\ \citenamefont {van
  Wijland}}]{hargus-2025-pre}%
  \BibitemOpen
  \bibfield  {author} {\bibinfo {author} {\bibfnamefont {C.}~\bibnamefont
  {Hargus}}, \bibinfo {author} {\bibfnamefont {F.}~\bibnamefont {Ghimenti}},
  \bibinfo {author} {\bibfnamefont {J.}~\bibnamefont {Tailleur}},\ and\
  \bibinfo {author} {\bibfnamefont {F.}~\bibnamefont {van Wijland}},\
  }\bibfield  {title} {\bibinfo {title} {Passive objects in a chiral active
  bath: From microscopic to macroscopic},\ }\href
  {https://doi.org/10.1103/qb5j-y254} {\bibfield  {journal} {\bibinfo
  {journal} {Phys. Rev. E}\ }\textbf {\bibinfo {volume} {112}},\ \bibinfo
  {pages} {044127} (\bibinfo {year} {2025}{\natexlab{b}})}\BibitemShut
  {NoStop}%
\bibitem [{\citenamefont {Caprini}\ and\ \citenamefont {Marini
  Bettolo~Marconi}(2025)}]{caprini-2025}%
  \BibitemOpen
  \bibfield  {author} {\bibinfo {author} {\bibfnamefont {L.}~\bibnamefont
  {Caprini}}\ and\ \bibinfo {author} {\bibfnamefont {U.}~\bibnamefont {Marini
  Bettolo~Marconi}},\ }\bibfield  {title} {\bibinfo {title} {Bubble phase
  induced by odd interactions in chiral systems},\ }\href
  {https://doi.org/10.1063/5.0262594} {\bibfield  {journal} {\bibinfo
  {journal} {J. Chem. Phys.}\ }\textbf {\bibinfo {volume} {162}},\ \bibinfo
  {pages} {161101} (\bibinfo {year} {2025})}\BibitemShut {NoStop}%
\bibitem [{\citenamefont {Caprini}\ \emph {et~al.}(2025)\citenamefont
  {Caprini}, \citenamefont {Marconi}, \citenamefont {Liebchen},\ and\
  \citenamefont {Löwen}}]{caprini-2025-thermodynamics-arxiv}%
  \BibitemOpen
  \bibfield  {author} {\bibinfo {author} {\bibfnamefont {L.}~\bibnamefont
  {Caprini}}, \bibinfo {author} {\bibfnamefont {U.~M.~B.}\ \bibnamefont
  {Marconi}}, \bibinfo {author} {\bibfnamefont {B.}~\bibnamefont {Liebchen}},\
  and\ \bibinfo {author} {\bibfnamefont {H.}~\bibnamefont {Löwen}},\
  }\bibfield  {title} {\bibinfo {title} {Active thermodynamics of inertial
  chiral active gases: equation of state and edge currents},\ }\bibfield
  {journal} {\bibinfo  {journal} {arXiv preprint}\ }\href
  {https://doi.org/10.48550/arXiv.2509.05053} {10.48550/arXiv.2509.05053}
  (\bibinfo {year} {2025}),\ \Eprint {https://arxiv.org/abs/2509.05053}
  {arXiv:2509.05053 [cond-mat.soft]} \BibitemShut {NoStop}%
\bibitem [{\citenamefont {Pisegna}\ and\ \citenamefont
  {Saha}(2025)}]{pisegna-2025-chiral-arxiv}%
  \BibitemOpen
  \bibfield  {author} {\bibinfo {author} {\bibfnamefont {G.}~\bibnamefont
  {Pisegna}}\ and\ \bibinfo {author} {\bibfnamefont {S.}~\bibnamefont {Saha}},\
  }\bibfield  {title} {\bibinfo {title} {Spinning mixtures: nonreciprocity
  transfers chirality across scales in scalar densities},\ }\href@noop {} {\
  (\bibinfo {year} {2025})},\ \Eprint {https://arxiv.org/abs/2509.07630}
  {arXiv:2509.07630 [cond-mat.stat-mech]} \BibitemShut {NoStop}%
\bibitem [{\citenamefont {Alsallom}\ and\ \citenamefont
  {Limmer}(2026)}]{alsallom-2026-arxiv}%
  \BibitemOpen
  \bibfield  {author} {\bibinfo {author} {\bibfnamefont {F.}~\bibnamefont
  {Alsallom}}\ and\ \bibinfo {author} {\bibfnamefont {D.~T.}\ \bibnamefont
  {Limmer}},\ }\bibfield  {title} {\bibinfo {title} {Origin of edge currents in
  chiral active liquids},\ }\bibfield  {journal} {\bibinfo  {journal} {arXiv
  preprint}\ }\href {https://doi.org/10.48550/arXiv.2603.18159}
  {10.48550/arXiv.2603.18159} (\bibinfo {year} {2026}),\ \Eprint
  {https://arxiv.org/abs/2603.18159} {arXiv:2603.18159 [cond-mat.soft]}
  \BibitemShut {NoStop}%
\bibitem [{\citenamefont {Wang}\ \emph {et~al.}(2026)\citenamefont {Wang},
  \citenamefont {Pietzonka},\ and\ \citenamefont
  {Jülicher}}]{wang-2026-arxiv}%
  \BibitemOpen
  \bibfield  {author} {\bibinfo {author} {\bibfnamefont {B.}~\bibnamefont
  {Wang}}, \bibinfo {author} {\bibfnamefont {P.}~\bibnamefont {Pietzonka}},\
  and\ \bibinfo {author} {\bibfnamefont {F.}~\bibnamefont {Jülicher}},\
  }\bibfield  {title} {\bibinfo {title} {Edge currents shape condensates in
  chiral active matter}\ }\href {https://doi.org/10.48550/arXiv.2603.20064}
  {10.48550/arXiv.2603.20064} (\bibinfo {year} {2026})\BibitemShut {NoStop}%
\bibitem [{\citenamefont {Metzger}\ \emph {et~al.}(2026)\citenamefont
  {Metzger}, \citenamefont {Hargus}, \citenamefont {Tailleur},\ and\
  \citenamefont {van Wijland}}]{metzger-2026-arxiv}%
  \BibitemOpen
  \bibfield  {author} {\bibinfo {author} {\bibfnamefont {J.}~\bibnamefont
  {Metzger}}, \bibinfo {author} {\bibfnamefont {C.}~\bibnamefont {Hargus}},
  \bibinfo {author} {\bibfnamefont {J.}~\bibnamefont {Tailleur}},\ and\
  \bibinfo {author} {\bibfnamefont {F.}~\bibnamefont {van Wijland}},\
  }\bibfield  {title} {\bibinfo {title} {Equation of state for the edge flow of
  chiral colloidal fluids},\ }\href@noop {} {\bibfield  {journal} {\bibinfo
  {journal} {arXiv preprint}\ } (\bibinfo {year} {2026})},\ \Eprint
  {https://arxiv.org/abs/2604.18708} {arXiv:2604.18708 [cond-mat.soft]}
  \BibitemShut {NoStop}%
\bibitem [{\citenamefont {Liebchen}\ and\ \citenamefont
  {Levis}(2017)}]{liebchen-2017}%
  \BibitemOpen
  \bibfield  {author} {\bibinfo {author} {\bibfnamefont {B.}~\bibnamefont
  {Liebchen}}\ and\ \bibinfo {author} {\bibfnamefont {D.}~\bibnamefont
  {Levis}},\ }\bibfield  {title} {\bibinfo {title} {Collective behavior of
  chiral active matter: Pattern formation and enhanced flocking},\ }\href
  {https://doi.org/10.1103/PhysRevLett.119.058002} {\bibfield  {journal}
  {\bibinfo  {journal} {Phys. Rev. Lett.}\ }\textbf {\bibinfo {volume} {119}},\
  \bibinfo {pages} {058002} (\bibinfo {year} {2017})}\BibitemShut {NoStop}%
\bibitem [{\citenamefont {Maitra}(2025)}]{maitra-2025}%
  \BibitemOpen
  \bibfield  {author} {\bibinfo {author} {\bibfnamefont {A.}~\bibnamefont
  {Maitra}},\ }\bibfield  {title} {\bibinfo {title} {Activity unmasks chirality
  in liquid-crystalline matter},\ }\href
  {https://doi.org/10.1146/annurev-conmatphys-032922-101439} {\bibfield
  {journal} {\bibinfo  {journal} {Annu. Rev. Condens. Matter Phys.}\ }\textbf
  {\bibinfo {volume} {16}},\ \bibinfo {pages} {275} (\bibinfo {year}
  {2025})}\BibitemShut {NoStop}%
\bibitem [{\citenamefont {Cates}\ and\ \citenamefont
  {Nardini}(2025)}]{cates-2025}%
  \BibitemOpen
  \bibfield  {author} {\bibinfo {author} {\bibfnamefont {M.~E.}\ \bibnamefont
  {Cates}}\ and\ \bibinfo {author} {\bibfnamefont {C.}~\bibnamefont
  {Nardini}},\ }\bibfield  {title} {\bibinfo {title} {Active phase separation:
  new phenomenology from non-equilibrium physics},\ }\href
  {https://doi.org/10.1088/1361-6633/add278} {\bibfield  {journal} {\bibinfo
  {journal} {Rep. Prog. Phys.}\ }\textbf {\bibinfo {volume} {88}},\ \bibinfo
  {pages} {056601} (\bibinfo {year} {2025})}\BibitemShut {NoStop}%
\bibitem [{\citenamefont {Tailleur}\ and\ \citenamefont
  {Cates}(2008)}]{tailleur-2008}%
  \BibitemOpen
  \bibfield  {author} {\bibinfo {author} {\bibfnamefont {J.}~\bibnamefont
  {Tailleur}}\ and\ \bibinfo {author} {\bibfnamefont {M.~E.}\ \bibnamefont
  {Cates}},\ }\bibfield  {title} {\bibinfo {title} {Statistical mechanics of
  interacting run-and-tumble bacteria},\ }\href
  {https://doi.org/10.1103/PhysRevLett.100.218103} {\bibfield  {journal}
  {\bibinfo  {journal} {Phys. Rev. Lett.}\ }\textbf {\bibinfo {volume} {100}},\
  \bibinfo {pages} {218103} (\bibinfo {year} {2008})}\BibitemShut {NoStop}%
\bibitem [{\citenamefont {Cates}\ and\ \citenamefont
  {Tailleur}(2015)}]{cates-2015}%
  \BibitemOpen
  \bibfield  {author} {\bibinfo {author} {\bibfnamefont {M.~E.}\ \bibnamefont
  {Cates}}\ and\ \bibinfo {author} {\bibfnamefont {J.}~\bibnamefont
  {Tailleur}},\ }\bibfield  {title} {\bibinfo {title} {{Motility-Induced Phase
  Separation}},\ }\href
  {https://doi.org/10.1146/annurev-conmatphys-031214-014710} {\bibfield
  {journal} {\bibinfo  {journal} {Annu. Rev. Condens. Matter Phys.}\ }\textbf
  {\bibinfo {volume} {6}},\ \bibinfo {pages} {219} (\bibinfo {year}
  {2015})}\BibitemShut {NoStop}%
\bibitem [{\citenamefont {Fily}\ and\ \citenamefont
  {Marchetti}(2012)}]{fily-2012}%
  \BibitemOpen
  \bibfield  {author} {\bibinfo {author} {\bibfnamefont {Y.}~\bibnamefont
  {Fily}}\ and\ \bibinfo {author} {\bibfnamefont {M.~C.}\ \bibnamefont
  {Marchetti}},\ }\bibfield  {title} {\bibinfo {title} {Athermal phase
  separation of self-propelled particles with no alignment},\ }\href
  {https://doi.org/10.1103/PhysRevLett.108.235702} {\bibfield  {journal}
  {\bibinfo  {journal} {Phys. Rev. Lett.}\ }\textbf {\bibinfo {volume} {108}},\
  \bibinfo {pages} {235702} (\bibinfo {year} {2012})}\BibitemShut {NoStop}%
\bibitem [{\citenamefont {Redner}\ \emph
  {et~al.}(2013{\natexlab{a}})\citenamefont {Redner}, \citenamefont {Hagan},\
  and\ \citenamefont {Baskaran}}]{redner-2013a}%
  \BibitemOpen
  \bibfield  {author} {\bibinfo {author} {\bibfnamefont {G.~S.}\ \bibnamefont
  {Redner}}, \bibinfo {author} {\bibfnamefont {M.~F.}\ \bibnamefont {Hagan}},\
  and\ \bibinfo {author} {\bibfnamefont {A.}~\bibnamefont {Baskaran}},\
  }\bibfield  {title} {\bibinfo {title} {Structure and dynamics of a
  phase-separating active colloidal fluid},\ }\href
  {https://doi.org/10.1103/PhysRevLett.110.055701} {\bibfield  {journal}
  {\bibinfo  {journal} {Phys. Rev. Lett.}\ }\textbf {\bibinfo {volume} {110}},\
  \bibinfo {pages} {055701} (\bibinfo {year} {2013}{\natexlab{a}})}\BibitemShut
  {NoStop}%
\bibitem [{\citenamefont {Stenhammar}\ \emph {et~al.}(2014)\citenamefont
  {Stenhammar}, \citenamefont {Marenduzzo}, \citenamefont {Allen},\ and\
  \citenamefont {Cates}}]{stenhammar-2014}%
  \BibitemOpen
  \bibfield  {author} {\bibinfo {author} {\bibfnamefont {J.}~\bibnamefont
  {Stenhammar}}, \bibinfo {author} {\bibfnamefont {D.}~\bibnamefont
  {Marenduzzo}}, \bibinfo {author} {\bibfnamefont {R.~J.}\ \bibnamefont
  {Allen}},\ and\ \bibinfo {author} {\bibfnamefont {M.~E.}\ \bibnamefont
  {Cates}},\ }\bibfield  {title} {\bibinfo {title} {Phase behaviour of active
  {Brownian} particles: the role of dimensionality},\ }\href
  {https://doi.org/10.1039/C3SM52813H} {\bibfield  {journal} {\bibinfo
  {journal} {Soft Matter}\ }\textbf {\bibinfo {volume} {10}},\ \bibinfo {pages}
  {1489} (\bibinfo {year} {2014})}\BibitemShut {NoStop}%
\bibitem [{\citenamefont {Stenhammar}(2021)}]{stenhammar-2021}%
  \BibitemOpen
  \bibfield  {author} {\bibinfo {author} {\bibfnamefont {J.}~\bibnamefont
  {Stenhammar}},\ }\bibfield  {title} {\bibinfo {title} {An introduction to
  motility-induced phase separation}\ }\href
  {https://doi.org/10.48550/arXiv.2112.05024} {10.48550/arXiv.2112.05024}
  (\bibinfo {year} {2021})\BibitemShut {NoStop}%
\bibitem [{\citenamefont {O'Byrne}\ \emph {et~al.}(2023)\citenamefont
  {O'Byrne}, \citenamefont {Solon}, \citenamefont {Tailleur},\ and\
  \citenamefont {Zhao}}]{obyrne-2023-mips}%
  \BibitemOpen
  \bibfield  {author} {\bibinfo {author} {\bibfnamefont {J.}~\bibnamefont
  {O'Byrne}}, \bibinfo {author} {\bibfnamefont {A.}~\bibnamefont {Solon}},
  \bibinfo {author} {\bibfnamefont {J.}~\bibnamefont {Tailleur}},\ and\
  \bibinfo {author} {\bibfnamefont {Y.}~\bibnamefont {Zhao}},\ }\bibfield
  {title} {\bibinfo {title} {An introduction to motility-induced phase
  separation},\ }in\ \href {https://doi.org/10.1039/9781839169465-00107} {\emph
  {\bibinfo {booktitle} {Out-of-equilibrium Soft Matter}}},\ \bibinfo {editor}
  {edited by\ \bibinfo {editor} {\bibfnamefont {C.}~\bibnamefont {Kurzthaler}},
  \bibinfo {editor} {\bibfnamefont {L.}~\bibnamefont {Gentile}},\ and\ \bibinfo
  {editor} {\bibfnamefont {H.~A.}\ \bibnamefont {Stone}}}\ (\bibinfo
  {publisher} {The Royal Society of Chemistry},\ \bibinfo {address} {Cambridge,
  UK},\ \bibinfo {year} {2023})\ pp.\ \bibinfo {pages} {107--150}\BibitemShut
  {NoStop}%
\bibitem [{\citenamefont {Bárdfalvy}\ \emph {et~al.}(2024)\citenamefont
  {Bárdfalvy}, \citenamefont {Škultéty}, \citenamefont {Nardini},
  \citenamefont {Morozov},\ and\ \citenamefont
  {Stenhammar}}]{bardfalvy2024collective}%
  \BibitemOpen
  \bibfield  {author} {\bibinfo {author} {\bibfnamefont {D.}~\bibnamefont
  {Bárdfalvy}}, \bibinfo {author} {\bibfnamefont {V.}~\bibnamefont
  {Škultéty}}, \bibinfo {author} {\bibfnamefont {C.}~\bibnamefont {Nardini}},
  \bibinfo {author} {\bibfnamefont {A.}~\bibnamefont {Morozov}},\ and\ \bibinfo
  {author} {\bibfnamefont {J.}~\bibnamefont {Stenhammar}},\ }\bibfield  {title}
  {\bibinfo {title} {Collective motion in a sheet of microswimmers},\ }\href
  {https://doi.org/10.1038/s42005-024-01587-9} {\bibfield  {journal} {\bibinfo
  {journal} {Commun. Phys.}\ }\textbf {\bibinfo {volume} {7}},\ \bibinfo
  {pages} {93} (\bibinfo {year} {2024})}\BibitemShut {NoStop}%
\bibitem [{\citenamefont {Thutupalli}\ \emph {et~al.}(2018)\citenamefont
  {Thutupalli}, \citenamefont {Geyer}, \citenamefont {Singh}, \citenamefont
  {Adhikari},\ and\ \citenamefont {Stone}}]{thutupalli2018flow}%
  \BibitemOpen
  \bibfield  {author} {\bibinfo {author} {\bibfnamefont {S.}~\bibnamefont
  {Thutupalli}}, \bibinfo {author} {\bibfnamefont {D.}~\bibnamefont {Geyer}},
  \bibinfo {author} {\bibfnamefont {R.}~\bibnamefont {Singh}}, \bibinfo
  {author} {\bibfnamefont {R.}~\bibnamefont {Adhikari}},\ and\ \bibinfo
  {author} {\bibfnamefont {H.~A.}\ \bibnamefont {Stone}},\ }\bibfield  {title}
  {\bibinfo {title} {Flow-induced phase separation of active particles is
  controlled by boundary conditions},\ }\href
  {https://doi.org/10.1073/pnas.1718807115} {\bibfield  {journal} {\bibinfo
  {journal} {Proc. Natl. Acad. Sci. U.S.A.}\ }\textbf {\bibinfo {volume}
  {115}},\ \bibinfo {pages} {5403} (\bibinfo {year} {2018})}\BibitemShut
  {NoStop}%
\bibitem [{\citenamefont {Hupe}\ \emph {et~al.}(2026)\citenamefont {Hupe},
  \citenamefont {Materska}, \citenamefont {Zwicker}, \citenamefont
  {Golestanian}, \citenamefont {Waclaw},\ and\ \citenamefont
  {Bittihn}}]{hupe-2026}%
  \BibitemOpen
  \bibfield  {author} {\bibinfo {author} {\bibfnamefont {L.}~\bibnamefont
  {Hupe}}, \bibinfo {author} {\bibfnamefont {J.~M.}\ \bibnamefont {Materska}},
  \bibinfo {author} {\bibfnamefont {D.}~\bibnamefont {Zwicker}}, \bibinfo
  {author} {\bibfnamefont {R.}~\bibnamefont {Golestanian}}, \bibinfo {author}
  {\bibfnamefont {B.}~\bibnamefont {Waclaw}},\ and\ \bibinfo {author}
  {\bibfnamefont {P.}~\bibnamefont {Bittihn}},\ }\bibfield  {title} {\bibinfo
  {title} {Phase separation in a mixture of proliferating and motile active
  matter},\ }\href {https://doi.org/10.1103/9pns-h5ll} {\bibfield  {journal}
  {\bibinfo  {journal} {Phys. Rev. Res.}\ }\textbf {\bibinfo {volume} {8}},\
  \bibinfo {pages} {L022012} (\bibinfo {year} {2026})}\BibitemShut {NoStop}%
\bibitem [{\citenamefont {Weber}\ \emph {et~al.}(2016)\citenamefont {Weber},
  \citenamefont {Weber},\ and\ \citenamefont {Frey}}]{weber2016binary}%
  \BibitemOpen
  \bibfield  {author} {\bibinfo {author} {\bibfnamefont {S.~N.}\ \bibnamefont
  {Weber}}, \bibinfo {author} {\bibfnamefont {C.~A.}\ \bibnamefont {Weber}},\
  and\ \bibinfo {author} {\bibfnamefont {E.}~\bibnamefont {Frey}},\ }\bibfield
  {title} {\bibinfo {title} {Binary mixtures of particles with different
  diffusivities demix},\ }\href
  {https://doi.org/10.1103/PhysRevLett.116.058301} {\bibfield  {journal}
  {\bibinfo  {journal} {Phys. Rev. Lett.}\ }\textbf {\bibinfo {volume} {116}},\
  \bibinfo {pages} {058301} (\bibinfo {year} {2016})}\BibitemShut {NoStop}%
\bibitem [{\citenamefont {Grosberg}\ and\ \citenamefont
  {Joanny}(2015)}]{grosberg2015nonequilibrium}%
  \BibitemOpen
  \bibfield  {author} {\bibinfo {author} {\bibfnamefont {A.~Y.}\ \bibnamefont
  {Grosberg}}\ and\ \bibinfo {author} {\bibfnamefont {J.-F.}\ \bibnamefont
  {Joanny}},\ }\bibfield  {title} {\bibinfo {title} {Nonequilibrium statistical
  mechanics of mixtures of particles in contact with different thermostats},\
  }\href {https://doi.org/10.1103/PhysRevE.92.032118} {\bibfield  {journal}
  {\bibinfo  {journal} {Phys. Rev. E}\ }\textbf {\bibinfo {volume} {92}},\
  \bibinfo {pages} {032118} (\bibinfo {year} {2015})}\BibitemShut {NoStop}%
\bibitem [{\citenamefont {Ilker}\ and\ \citenamefont
  {Joanny}(2020)}]{ilker2020phase}%
  \BibitemOpen
  \bibfield  {author} {\bibinfo {author} {\bibfnamefont {E.}~\bibnamefont
  {Ilker}}\ and\ \bibinfo {author} {\bibfnamefont {J.-F.}\ \bibnamefont
  {Joanny}},\ }\bibfield  {title} {\bibinfo {title} {Phase separation and
  nucleation in mixtures of particles with different temperatures},\ }\href
  {https://doi.org/10.1103/PhysRevResearch.2.023200} {\bibfield  {journal}
  {\bibinfo  {journal} {Phys. Rev. Res.}\ }\textbf {\bibinfo {volume} {2}},\
  \bibinfo {pages} {023200} (\bibinfo {year} {2020})}\BibitemShut {NoStop}%
\bibitem [{\citenamefont {Smrek}\ and\ \citenamefont
  {Kremer}(2017)}]{smrek2017small}%
  \BibitemOpen
  \bibfield  {author} {\bibinfo {author} {\bibfnamefont {J.}~\bibnamefont
  {Smrek}}\ and\ \bibinfo {author} {\bibfnamefont {K.}~\bibnamefont {Kremer}},\
  }\bibfield  {title} {\bibinfo {title} {Small activity differences drive phase
  separation in active-passive polymer mixtures},\ }\href
  {https://doi.org/10.1103/PhysRevLett.118.098002} {\bibfield  {journal}
  {\bibinfo  {journal} {Phys. Rev. Lett.}\ }\textbf {\bibinfo {volume} {118}},\
  \bibinfo {pages} {098002} (\bibinfo {year} {2017})}\BibitemShut {NoStop}%
\bibitem [{\citenamefont {McCarthy}\ \emph {et~al.}(2024)\citenamefont
  {McCarthy}, \citenamefont {Manna}, \citenamefont {Damavandi},\ and\
  \citenamefont {Manning}}]{mccarthy2023demixing}%
  \BibitemOpen
  \bibfield  {author} {\bibinfo {author} {\bibfnamefont {E.}~\bibnamefont
  {McCarthy}}, \bibinfo {author} {\bibfnamefont {R.~K.}\ \bibnamefont {Manna}},
  \bibinfo {author} {\bibfnamefont {O.}~\bibnamefont {Damavandi}},\ and\
  \bibinfo {author} {\bibfnamefont {M.~L.}\ \bibnamefont {Manning}},\
  }\bibfield  {title} {\bibinfo {title} {Demixing in binary mixtures with
  differential diffusivity at high density},\ }\href
  {https://doi.org/10.1103/PhysRevLett.132.098301} {\bibfield  {journal}
  {\bibinfo  {journal} {Phys. Rev. Lett.}\ }\textbf {\bibinfo {volume} {132}},\
  \bibinfo {pages} {098301} (\bibinfo {year} {2024})}\BibitemShut {NoStop}%
\bibitem [{\citenamefont {Schnitzer}(1993)}]{schnitzer-1993}%
  \BibitemOpen
  \bibfield  {author} {\bibinfo {author} {\bibfnamefont {M.~J.}\ \bibnamefont
  {Schnitzer}},\ }\bibfield  {title} {\bibinfo {title} {Theory of continuum
  random walks and application to chemotaxis},\ }\href
  {https://doi.org/10.1103/PhysRevE.48.2553} {\bibfield  {journal} {\bibinfo
  {journal} {Phys. Rev. E}\ }\textbf {\bibinfo {volume} {48}},\ \bibinfo
  {pages} {2553} (\bibinfo {year} {1993})}\BibitemShut {NoStop}%
\bibitem [{\citenamefont {Miller}\ and\ \citenamefont
  {Bassler}(2001)}]{miller-2001}%
  \BibitemOpen
  \bibfield  {author} {\bibinfo {author} {\bibfnamefont {M.~B.}\ \bibnamefont
  {Miller}}\ and\ \bibinfo {author} {\bibfnamefont {B.~L.}\ \bibnamefont
  {Bassler}},\ }\bibfield  {title} {\bibinfo {title} {Quorum sensing in
  bacteria},\ }\href {https://doi.org/10.1146/annurev.micro.55.1.165}
  {\bibfield  {journal} {\bibinfo  {journal} {Annu. Rev. Microbiol.}\ }\textbf
  {\bibinfo {volume} {55}},\ \bibinfo {pages} {165} (\bibinfo {year}
  {2001})}\BibitemShut {NoStop}%
\bibitem [{\citenamefont {Liu}\ \emph {et~al.}(2011)\citenamefont {Liu},
  \citenamefont {Fu}, \citenamefont {Liu}, \citenamefont {Ren}, \citenamefont
  {Chau}, \citenamefont {Li}, \citenamefont {Xiang}, \citenamefont {Zeng},
  \citenamefont {Chen}, \citenamefont {Tang}, \citenamefont {Lenz},
  \citenamefont {Cui}, \citenamefont {Huang}, \citenamefont {Hwa},\ and\
  \citenamefont {Huang}}]{liu-2011}%
  \BibitemOpen
  \bibfield  {author} {\bibinfo {author} {\bibfnamefont {C.}~\bibnamefont
  {Liu}}, \bibinfo {author} {\bibfnamefont {X.}~\bibnamefont {Fu}}, \bibinfo
  {author} {\bibfnamefont {L.}~\bibnamefont {Liu}}, \bibinfo {author}
  {\bibfnamefont {X.}~\bibnamefont {Ren}}, \bibinfo {author} {\bibfnamefont
  {C.~K.}\ \bibnamefont {Chau}}, \bibinfo {author} {\bibfnamefont
  {S.}~\bibnamefont {Li}}, \bibinfo {author} {\bibfnamefont {L.}~\bibnamefont
  {Xiang}}, \bibinfo {author} {\bibfnamefont {H.}~\bibnamefont {Zeng}},
  \bibinfo {author} {\bibfnamefont {G.}~\bibnamefont {Chen}}, \bibinfo {author}
  {\bibfnamefont {L.-H.}\ \bibnamefont {Tang}}, \bibinfo {author}
  {\bibfnamefont {P.}~\bibnamefont {Lenz}}, \bibinfo {author} {\bibfnamefont
  {X.}~\bibnamefont {Cui}}, \bibinfo {author} {\bibfnamefont {W.}~\bibnamefont
  {Huang}}, \bibinfo {author} {\bibfnamefont {T.}~\bibnamefont {Hwa}},\ and\
  \bibinfo {author} {\bibfnamefont {J.-D.}\ \bibnamefont {Huang}},\ }\bibfield
  {title} {\bibinfo {title} {Sequential establishment of stripe patterns in an
  expanding cell population},\ }\href {https://doi.org/10.1126/science.1209042}
  {\bibfield  {journal} {\bibinfo  {journal} {Science}\ }\textbf {\bibinfo
  {volume} {334}},\ \bibinfo {pages} {238} (\bibinfo {year}
  {2011})}\BibitemShut {NoStop}%
\bibitem [{\citenamefont {Fu}\ \emph {et~al.}(2012)\citenamefont {Fu},
  \citenamefont {Tang}, \citenamefont {Liu}, \citenamefont {Huang},
  \citenamefont {Hwa},\ and\ \citenamefont {Lenz}}]{fu-2012}%
  \BibitemOpen
  \bibfield  {author} {\bibinfo {author} {\bibfnamefont {X.}~\bibnamefont
  {Fu}}, \bibinfo {author} {\bibfnamefont {L.-H.}\ \bibnamefont {Tang}},
  \bibinfo {author} {\bibfnamefont {C.}~\bibnamefont {Liu}}, \bibinfo {author}
  {\bibfnamefont {J.-D.}\ \bibnamefont {Huang}}, \bibinfo {author}
  {\bibfnamefont {T.}~\bibnamefont {Hwa}},\ and\ \bibinfo {author}
  {\bibfnamefont {P.}~\bibnamefont {Lenz}},\ }\bibfield  {title} {\bibinfo
  {title} {Stripe formation in bacterial systems with density-suppressed
  motility},\ }\href {https://doi.org/10.1103/PhysRevLett.108.198102}
  {\bibfield  {journal} {\bibinfo  {journal} {Phys. Rev. Lett.}\ }\textbf
  {\bibinfo {volume} {108}},\ \bibinfo {pages} {198102} (\bibinfo {year}
  {2012})}\BibitemShut {NoStop}%
\bibitem [{\citenamefont {Mukherjee}\ and\ \citenamefont
  {Bassler}(2019)}]{mukherjee-2019}%
  \BibitemOpen
  \bibfield  {author} {\bibinfo {author} {\bibfnamefont {S.}~\bibnamefont
  {Mukherjee}}\ and\ \bibinfo {author} {\bibfnamefont {B.~L.}\ \bibnamefont
  {Bassler}},\ }\bibfield  {title} {\bibinfo {title} {Bacterial quorum sensing
  in complex and dynamically changing environments},\ }\href
  {https://doi.org/10.1038/s41579-019-0186-5} {\bibfield  {journal} {\bibinfo
  {journal} {Nat. Rev. Microbiol.}\ }\textbf {\bibinfo {volume} {17}},\
  \bibinfo {pages} {371} (\bibinfo {year} {2019})}\BibitemShut {NoStop}%
\bibitem [{\citenamefont {Curatolo}\ \emph {et~al.}(2020)\citenamefont
  {Curatolo}, \citenamefont {Zhou}, \citenamefont {Zhao}, \citenamefont {Liu},
  \citenamefont {Daerr}, \citenamefont {Tailleur},\ and\ \citenamefont
  {Huang}}]{curatolo-2020}%
  \BibitemOpen
  \bibfield  {author} {\bibinfo {author} {\bibfnamefont {A.~I.}\ \bibnamefont
  {Curatolo}}, \bibinfo {author} {\bibfnamefont {N.}~\bibnamefont {Zhou}},
  \bibinfo {author} {\bibfnamefont {Y.}~\bibnamefont {Zhao}}, \bibinfo {author}
  {\bibfnamefont {C.}~\bibnamefont {Liu}}, \bibinfo {author} {\bibfnamefont
  {A.}~\bibnamefont {Daerr}}, \bibinfo {author} {\bibfnamefont
  {J.}~\bibnamefont {Tailleur}},\ and\ \bibinfo {author} {\bibfnamefont
  {J.}~\bibnamefont {Huang}},\ }\bibfield  {title} {\bibinfo {title}
  {Cooperative pattern formation in multi-component bacterial systems through
  reciprocal motility regulation},\ }\href
  {https://doi.org/10.1038/s41567-020-0964-z} {\bibfield  {journal} {\bibinfo
  {journal} {Nat. Phys.}\ }\textbf {\bibinfo {volume} {16}},\ \bibinfo {pages}
  {1152} (\bibinfo {year} {2020})}\BibitemShut {NoStop}%
\bibitem [{\citenamefont {Dinelli}\ \emph {et~al.}(2023)\citenamefont
  {Dinelli}, \citenamefont {O'Byrne}, \citenamefont {Curatolo}, \citenamefont
  {Zhao}, \citenamefont {Sollich},\ and\ \citenamefont
  {Tailleur}}]{dinelli-2023}%
  \BibitemOpen
  \bibfield  {author} {\bibinfo {author} {\bibfnamefont {A.}~\bibnamefont
  {Dinelli}}, \bibinfo {author} {\bibfnamefont {J.}~\bibnamefont {O'Byrne}},
  \bibinfo {author} {\bibfnamefont {A.}~\bibnamefont {Curatolo}}, \bibinfo
  {author} {\bibfnamefont {Y.}~\bibnamefont {Zhao}}, \bibinfo {author}
  {\bibfnamefont {P.}~\bibnamefont {Sollich}},\ and\ \bibinfo {author}
  {\bibfnamefont {J.}~\bibnamefont {Tailleur}},\ }\bibfield  {title} {\bibinfo
  {title} {Non-reciprocity across scales in active mixtures},\ }\href
  {https://doi.org/10.1038/s41467-023-42713-5} {\bibfield  {journal} {\bibinfo
  {journal} {Nat. Commun.}\ }\textbf {\bibinfo {volume} {14}},\ \bibinfo
  {pages} {7035} (\bibinfo {year} {2023})}\BibitemShut {NoStop}%
\bibitem [{\citenamefont {Ridgway}\ \emph {et~al.}(2023)\citenamefont
  {Ridgway}, \citenamefont {Dalwadi}, \citenamefont {Pearce},\ and\
  \citenamefont {Chapman}}]{ridgway-2023}%
  \BibitemOpen
  \bibfield  {author} {\bibinfo {author} {\bibfnamefont {W.~J.~M.}\
  \bibnamefont {Ridgway}}, \bibinfo {author} {\bibfnamefont {M.~P.}\
  \bibnamefont {Dalwadi}}, \bibinfo {author} {\bibfnamefont {P.}~\bibnamefont
  {Pearce}},\ and\ \bibinfo {author} {\bibfnamefont {S.~J.}\ \bibnamefont
  {Chapman}},\ }\bibfield  {title} {\bibinfo {title} {Motility-induced phase
  separation mediated by bacterial quorum sensing},\ }\href
  {https://doi.org/10.1103/PhysRevLett.131.228302} {\bibfield  {journal}
  {\bibinfo  {journal} {Phys. Rev. Lett.}\ }\textbf {\bibinfo {volume} {131}},\
  \bibinfo {pages} {228302} (\bibinfo {year} {2023})}\BibitemShut {NoStop}%
\bibitem [{\citenamefont {Li}\ \emph {et~al.}(2024)\citenamefont {Li},
  \citenamefont {Chat\'e}, \citenamefont {Sano}, \citenamefont {Shi},\ and\
  \citenamefont {Zhang}}]{li-2024}%
  \BibitemOpen
  \bibfield  {author} {\bibinfo {author} {\bibfnamefont {H.}~\bibnamefont
  {Li}}, \bibinfo {author} {\bibfnamefont {H.}~\bibnamefont {Chat\'e}},
  \bibinfo {author} {\bibfnamefont {M.}~\bibnamefont {Sano}}, \bibinfo {author}
  {\bibfnamefont {X.-q.}\ \bibnamefont {Shi}},\ and\ \bibinfo {author}
  {\bibfnamefont {H.~P.}\ \bibnamefont {Zhang}},\ }\bibfield  {title} {\bibinfo
  {title} {Robust edge flows in swarming bacterial colonies},\ }\href
  {https://doi.org/10.1103/PhysRevX.14.041006} {\bibfield  {journal} {\bibinfo
  {journal} {Phys. Rev. X}\ }\textbf {\bibinfo {volume} {14}},\ \bibinfo
  {pages} {041006} (\bibinfo {year} {2024})}\BibitemShut {NoStop}%
\bibitem [{\citenamefont {He}\ \emph {et~al.}(2025)\citenamefont {He},
  \citenamefont {Yin}, \citenamefont {Liang}, \citenamefont {Chang},\ and\
  \citenamefont {Xu}}]{he-2025}%
  \BibitemOpen
  \bibfield  {author} {\bibinfo {author} {\bibfnamefont {S.-Q.}\ \bibnamefont
  {He}}, \bibinfo {author} {\bibfnamefont {X.}~\bibnamefont {Yin}}, \bibinfo
  {author} {\bibfnamefont {D.}~\bibnamefont {Liang}}, \bibinfo {author}
  {\bibfnamefont {Z.}~\bibnamefont {Chang}},\ and\ \bibinfo {author}
  {\bibfnamefont {G.-K.}\ \bibnamefont {Xu}},\ }\bibfield  {title} {\bibinfo
  {title} {Spontaneous oscillation in collective microswimmers: Insights from a
  chiral self-propelled rod model},\ }\href
  {https://doi.org/10.1103/PhysRevE.111.014411} {\bibfield  {journal} {\bibinfo
   {journal} {Phys. Rev. E}\ }\textbf {\bibinfo {volume} {111}},\ \bibinfo
  {pages} {014411} (\bibinfo {year} {2025})}\BibitemShut {NoStop}%
\bibitem [{\citenamefont {Dinelli}\ \emph {et~al.}(2026)\citenamefont
  {Dinelli}, \citenamefont {Altieri},\ and\ \citenamefont
  {Tailleur}}]{dinelli-2026-random}%
  \BibitemOpen
  \bibfield  {author} {\bibinfo {author} {\bibfnamefont {A.}~\bibnamefont
  {Dinelli}}, \bibinfo {author} {\bibfnamefont {A.}~\bibnamefont {Altieri}},\
  and\ \bibinfo {author} {\bibfnamefont {J.}~\bibnamefont {Tailleur}},\
  }\bibfield  {title} {\bibinfo {title} {Random motility regulation as a
  generic mechanism of community formation},\ }\href
  {https://doi.org/10.1103/cl1f-mfgp} {\bibfield  {journal} {\bibinfo
  {journal} {Phys. Rev. E}\ }\textbf {\bibinfo {volume} {114}},\ \bibinfo
  {pages} {014404} (\bibinfo {year} {2026})}\BibitemShut {NoStop}%
\bibitem [{\citenamefont {van~de Koppel}\ \emph {et~al.}(2008)\citenamefont
  {van~de Koppel}, \citenamefont {Gascoigne}, \citenamefont {Theraulaz},
  \citenamefont {Rietkerk}, \citenamefont {Mooij},\ and\ \citenamefont
  {Herman}}]{van-de-koppel-2008}%
  \BibitemOpen
  \bibfield  {author} {\bibinfo {author} {\bibfnamefont {J.}~\bibnamefont
  {van~de Koppel}}, \bibinfo {author} {\bibfnamefont {J.~C.}\ \bibnamefont
  {Gascoigne}}, \bibinfo {author} {\bibfnamefont {G.}~\bibnamefont
  {Theraulaz}}, \bibinfo {author} {\bibfnamefont {M.}~\bibnamefont {Rietkerk}},
  \bibinfo {author} {\bibfnamefont {W.~M.}\ \bibnamefont {Mooij}},\ and\
  \bibinfo {author} {\bibfnamefont {P.~M.~J.}\ \bibnamefont {Herman}},\
  }\bibfield  {title} {\bibinfo {title} {Experimental evidence for spatial
  self-organization and its emergent effects in mussel bed ecosystems},\ }\href
  {https://doi.org/10.1126/science.1163952} {\bibfield  {journal} {\bibinfo
  {journal} {Science}\ }\textbf {\bibinfo {volume} {322}},\ \bibinfo {pages}
  {739} (\bibinfo {year} {2008})}\BibitemShut {NoStop}%
\bibitem [{\citenamefont {Liu}\ \emph {et~al.}(2013)\citenamefont {Liu},
  \citenamefont {Doelman}, \citenamefont {Rottschäfer}, \citenamefont
  {de~Jager}, \citenamefont {Herman}, \citenamefont {Rietkerk},\ and\
  \citenamefont {van~de Koppel}}]{liu-2013}%
  \BibitemOpen
  \bibfield  {author} {\bibinfo {author} {\bibfnamefont {Q.-X.}\ \bibnamefont
  {Liu}}, \bibinfo {author} {\bibfnamefont {A.}~\bibnamefont {Doelman}},
  \bibinfo {author} {\bibfnamefont {V.}~\bibnamefont {Rottschäfer}}, \bibinfo
  {author} {\bibfnamefont {M.}~\bibnamefont {de~Jager}}, \bibinfo {author}
  {\bibfnamefont {P.~M.~J.}\ \bibnamefont {Herman}}, \bibinfo {author}
  {\bibfnamefont {M.}~\bibnamefont {Rietkerk}},\ and\ \bibinfo {author}
  {\bibfnamefont {J.}~\bibnamefont {van~de Koppel}},\ }\bibfield  {title}
  {\bibinfo {title} {Phase separation explains a new class of self-organized
  spatial patterns in ecological systems},\ }\href
  {https://doi.org/10.1073/pnas.1222339110} {\bibfield  {journal} {\bibinfo
  {journal} {Proc. Natl. Acad. Sci.}\ }\textbf {\bibinfo {volume} {110}},\
  \bibinfo {pages} {11905} (\bibinfo {year} {2013})}\BibitemShut {NoStop}%
\bibitem [{\citenamefont {de~Souza}\ \emph {et~al.}(2025)\citenamefont
  {de~Souza}, \citenamefont {Teixeira}, \citenamefont {Viswanathan},
  \citenamefont {Sollich},\ and\ \citenamefont {de~Castro}}]{desouza-2025}%
  \BibitemOpen
  \bibfield  {author} {\bibinfo {author} {\bibfnamefont {L.}~\bibnamefont
  {de~Souza}}, \bibinfo {author} {\bibfnamefont {E.~F.}\ \bibnamefont
  {Teixeira}}, \bibinfo {author} {\bibfnamefont {G.~M.}\ \bibnamefont
  {Viswanathan}}, \bibinfo {author} {\bibfnamefont {P.}~\bibnamefont
  {Sollich}},\ and\ \bibinfo {author} {\bibfnamefont {P.}~\bibnamefont
  {de~Castro}},\ }\bibfield  {title} {\bibinfo {title} {How quorum sensing
  shapes clustering in active matter}\ }\href
  {https://doi.org/10.48550/arXiv.2512.02935} {10.48550/arXiv.2512.02935}
  (\bibinfo {year} {2025})\BibitemShut {NoStop}%
\bibitem [{\citenamefont {B{\"a}uerle}\ \emph {et~al.}(2018)\citenamefont
  {B{\"a}uerle}, \citenamefont {Fischer}, \citenamefont {Speck},\ and\
  \citenamefont {Bechinger}}]{bauerle-2018}%
  \BibitemOpen
  \bibfield  {author} {\bibinfo {author} {\bibfnamefont {T.}~\bibnamefont
  {B{\"a}uerle}}, \bibinfo {author} {\bibfnamefont {A.}~\bibnamefont
  {Fischer}}, \bibinfo {author} {\bibfnamefont {T.}~\bibnamefont {Speck}},\
  and\ \bibinfo {author} {\bibfnamefont {C.}~\bibnamefont {Bechinger}},\
  }\bibfield  {title} {\bibinfo {title} {Self-organization of active particles
  by quorum sensing rules},\ }\href
  {https://doi.org/10.1038/s41467-018-05675-7} {\bibfield  {journal} {\bibinfo
  {journal} {Nat. Commun.}\ }\textbf {\bibinfo {volume} {9}},\ \bibinfo {pages}
  {3232} (\bibinfo {year} {2018})}\BibitemShut {NoStop}%
\bibitem [{\citenamefont {Lefranc}\ \emph {et~al.}(2025)\citenamefont
  {Lefranc}, \citenamefont {Dinelli}, \citenamefont {Fern\'andez-Rico},
  \citenamefont {Dullens}, \citenamefont {Tailleur},\ and\ \citenamefont
  {Bartolo}}]{lefranc-2025}%
  \BibitemOpen
  \bibfield  {author} {\bibinfo {author} {\bibfnamefont {T.}~\bibnamefont
  {Lefranc}}, \bibinfo {author} {\bibfnamefont {A.}~\bibnamefont {Dinelli}},
  \bibinfo {author} {\bibfnamefont {C.}~\bibnamefont {Fern\'andez-Rico}},
  \bibinfo {author} {\bibfnamefont {R.~P.~A.}\ \bibnamefont {Dullens}},
  \bibinfo {author} {\bibfnamefont {J.}~\bibnamefont {Tailleur}},\ and\
  \bibinfo {author} {\bibfnamefont {D.}~\bibnamefont {Bartolo}},\ }\bibfield
  {title} {\bibinfo {title} {Synthetic quorum sensing and absorbing phase
  transitions in colloidal active matter},\ }\href
  {https://doi.org/10.1103/8csn-71jk} {\bibfield  {journal} {\bibinfo
  {journal} {Phys. Rev. X}\ }\textbf {\bibinfo {volume} {15}},\ \bibinfo
  {pages} {031050} (\bibinfo {year} {2025})}\BibitemShut {NoStop}%
\bibitem [{\citenamefont {De~Luca}\ \emph {et~al.}(2026)\citenamefont
  {De~Luca}, \citenamefont {Cates},\ and\ \citenamefont
  {Nardini}}]{de2026generic}%
  \BibitemOpen
  \bibfield  {author} {\bibinfo {author} {\bibfnamefont {F.}~\bibnamefont
  {De~Luca}}, \bibinfo {author} {\bibfnamefont {M.~E.}\ \bibnamefont {Cates}},\
  and\ \bibinfo {author} {\bibfnamefont {C.}~\bibnamefont {Nardini}},\
  }\bibfield  {title} {\bibinfo {title} {Generic nonlocal statistics of the
  stationary measure in conserved active systems},\ }\href@noop {} {\bibfield
  {journal} {\bibinfo  {journal} {arXiv preprint arXiv:2606.14483}\ } (\bibinfo
  {year} {2026})}\BibitemShut {NoStop}%
\bibitem [{\citenamefont {Tjhung}\ \emph {et~al.}(2018)\citenamefont {Tjhung},
  \citenamefont {Nardini},\ and\ \citenamefont {Cates}}]{tjhung-2018}%
  \BibitemOpen
  \bibfield  {author} {\bibinfo {author} {\bibfnamefont {E.}~\bibnamefont
  {Tjhung}}, \bibinfo {author} {\bibfnamefont {C.}~\bibnamefont {Nardini}},\
  and\ \bibinfo {author} {\bibfnamefont {M.~E.}\ \bibnamefont {Cates}},\
  }\bibfield  {title} {\bibinfo {title} {Cluster phases and bubbly phase
  separation in active fluids: Reversal of the {Ostwald} process},\ }\href
  {https://doi.org/10.1103/PhysRevX.8.031080} {\bibfield  {journal} {\bibinfo
  {journal} {Phys. Rev. X}\ }\textbf {\bibinfo {volume} {8}},\ \bibinfo {pages}
  {031080} (\bibinfo {year} {2018})}\BibitemShut {NoStop}%
\bibitem [{\citenamefont {Fausti}\ \emph {et~al.}(2024)\citenamefont {Fausti},
  \citenamefont {Cates},\ and\ \citenamefont {Nardini}}]{fausti-2024}%
  \BibitemOpen
  \bibfield  {author} {\bibinfo {author} {\bibfnamefont {G.}~\bibnamefont
  {Fausti}}, \bibinfo {author} {\bibfnamefont {M.~E.}\ \bibnamefont {Cates}},\
  and\ \bibinfo {author} {\bibfnamefont {C.}~\bibnamefont {Nardini}},\
  }\bibfield  {title} {\bibinfo {title} {Statistical properties of microphase
  and bubbly phase-separated active fluids},\ }\href
  {https://doi.org/10.1103/PhysRevE.110.L042103} {\bibfield  {journal}
  {\bibinfo  {journal} {Phys. Rev. E}\ }\textbf {\bibinfo {volume} {110}},\
  \bibinfo {pages} {L042103} (\bibinfo {year} {2024})}\BibitemShut {NoStop}%
\bibitem [{\citenamefont {Li}\ and\ \citenamefont
  {Cates}(2021)}]{li2021hierarchical}%
  \BibitemOpen
  \bibfield  {author} {\bibinfo {author} {\bibfnamefont {Y.~I.}\ \bibnamefont
  {Li}}\ and\ \bibinfo {author} {\bibfnamefont {M.~E.}\ \bibnamefont {Cates}},\
  }\bibfield  {title} {\bibinfo {title} {Hierarchical microphase separation in
  non-conserved active mixtures},\ }\href
  {https://doi.org/10.1140/epje/s10189-021-00113-x} {\bibfield  {journal}
  {\bibinfo  {journal} {Eur. Phys. J. E}\ }\textbf {\bibinfo {volume} {44}},\
  \bibinfo {pages} {119} (\bibinfo {year} {2021})}\BibitemShut {NoStop}%
\bibitem [{\citenamefont {Singh}\ and\ \citenamefont
  {Cates}(2019)}]{singh2019hydrodynamically}%
  \BibitemOpen
  \bibfield  {author} {\bibinfo {author} {\bibfnamefont {R.}~\bibnamefont
  {Singh}}\ and\ \bibinfo {author} {\bibfnamefont {M.}~\bibnamefont {Cates}},\
  }\bibfield  {title} {\bibinfo {title} {Hydrodynamically interrupted droplet
  growth in scalar active matter},\ }\href
  {https://doi.org/10.1103/PhysRevLett.123.148005} {\bibfield  {journal}
  {\bibinfo  {journal} {Phys. Rev. Lett.}\ }\textbf {\bibinfo {volume} {123}},\
  \bibinfo {pages} {148005} (\bibinfo {year} {2019})}\BibitemShut {NoStop}%
\bibitem [{\citenamefont {Zwicker}\ \emph {et~al.}(2017)\citenamefont
  {Zwicker}, \citenamefont {Seyboldt}, \citenamefont {Weber}, \citenamefont
  {Hyman},\ and\ \citenamefont {J{\"u}licher}}]{zwicker2017growth}%
  \BibitemOpen
  \bibfield  {author} {\bibinfo {author} {\bibfnamefont {D.}~\bibnamefont
  {Zwicker}}, \bibinfo {author} {\bibfnamefont {R.}~\bibnamefont {Seyboldt}},
  \bibinfo {author} {\bibfnamefont {C.~A.}\ \bibnamefont {Weber}}, \bibinfo
  {author} {\bibfnamefont {A.~A.}\ \bibnamefont {Hyman}},\ and\ \bibinfo
  {author} {\bibfnamefont {F.}~\bibnamefont {J{\"u}licher}},\ }\bibfield
  {title} {\bibinfo {title} {Growth and division of active droplets provides a
  model for protocells},\ }\href {https://doi.org/10.1038/nphys3984} {\bibfield
   {journal} {\bibinfo  {journal} {Nat. Phys.}\ }\textbf {\bibinfo {volume}
  {13}},\ \bibinfo {pages} {408} (\bibinfo {year} {2017})}\BibitemShut
  {NoStop}%
\bibitem [{\citenamefont {Caporusso}\ \emph {et~al.}(2020)\citenamefont
  {Caporusso}, \citenamefont {Digregorio}, \citenamefont {Levis}, \citenamefont
  {Cugliandolo},\ and\ \citenamefont {Gonnella}}]{caporusso2020micro}%
  \BibitemOpen
  \bibfield  {author} {\bibinfo {author} {\bibfnamefont {C.~B.}\ \bibnamefont
  {Caporusso}}, \bibinfo {author} {\bibfnamefont {P.}~\bibnamefont
  {Digregorio}}, \bibinfo {author} {\bibfnamefont {D.}~\bibnamefont {Levis}},
  \bibinfo {author} {\bibfnamefont {L.~F.}\ \bibnamefont {Cugliandolo}},\ and\
  \bibinfo {author} {\bibfnamefont {G.}~\bibnamefont {Gonnella}},\ }\bibfield
  {title} {\bibinfo {title} {Motility-induced microphase and macrophase
  separation in a two-dimensional active {Brownian} particle system},\ }\href
  {https://doi.org/10.1103/PhysRevLett.125.178004} {\bibfield  {journal}
  {\bibinfo  {journal} {Phys. Rev. Lett.}\ }\textbf {\bibinfo {volume} {125}},\
  \bibinfo {pages} {178004} (\bibinfo {year} {2020})}\BibitemShut {NoStop}%
\bibitem [{\citenamefont {Langford}\ and\ \citenamefont
  {Omar}(2026)}]{langford2026hexatic}%
  \BibitemOpen
  \bibfield  {author} {\bibinfo {author} {\bibfnamefont {L.}~\bibnamefont
  {Langford}}\ and\ \bibinfo {author} {\bibfnamefont {A.~K.}\ \bibnamefont
  {Omar}},\ }\bibfield  {title} {\bibinfo {title} {Hexatic order coupled with
  thermal noise produces bubbles in two-dimensional active matter},\
  }\href@noop {} {\bibfield  {journal} {\bibinfo  {journal} {arXiv preprint
  arXiv:2603.17320}\ } (\bibinfo {year} {2026})}\BibitemShut {NoStop}%
\bibitem [{\citenamefont {Shi}\ \emph {et~al.}(2020)\citenamefont {Shi},
  \citenamefont {Fausti}, \citenamefont {Chat\'e}, \citenamefont {Nardini},\
  and\ \citenamefont {Solon}}]{shi-2020}%
  \BibitemOpen
  \bibfield  {author} {\bibinfo {author} {\bibfnamefont {X.-q.}\ \bibnamefont
  {Shi}}, \bibinfo {author} {\bibfnamefont {G.}~\bibnamefont {Fausti}},
  \bibinfo {author} {\bibfnamefont {H.}~\bibnamefont {Chat\'e}}, \bibinfo
  {author} {\bibfnamefont {C.}~\bibnamefont {Nardini}},\ and\ \bibinfo {author}
  {\bibfnamefont {A.}~\bibnamefont {Solon}},\ }\bibfield  {title} {\bibinfo
  {title} {Self-organized critical coexistence phase in repulsive active
  particles},\ }\href {https://doi.org/10.1103/PhysRevLett.125.168001}
  {\bibfield  {journal} {\bibinfo  {journal} {Phys. Rev. Lett.}\ }\textbf
  {\bibinfo {volume} {125}},\ \bibinfo {pages} {168001} (\bibinfo {year}
  {2020})}\BibitemShut {NoStop}%
\bibitem [{\citenamefont {Redner}\ \emph
  {et~al.}(2013{\natexlab{b}})\citenamefont {Redner}, \citenamefont
  {Baskaran},\ and\ \citenamefont {Hagan}}]{redner-2013b}%
  \BibitemOpen
  \bibfield  {author} {\bibinfo {author} {\bibfnamefont {G.~S.}\ \bibnamefont
  {Redner}}, \bibinfo {author} {\bibfnamefont {A.}~\bibnamefont {Baskaran}},\
  and\ \bibinfo {author} {\bibfnamefont {M.~F.}\ \bibnamefont {Hagan}},\
  }\bibfield  {title} {\bibinfo {title} {Reentrant phase behavior in active
  colloids with attraction},\ }\href
  {https://doi.org/10.1103/PhysRevE.88.012305} {\bibfield  {journal} {\bibinfo
  {journal} {Phys. Rev. E}\ }\textbf {\bibinfo {volume} {88}},\ \bibinfo
  {pages} {012305} (\bibinfo {year} {2013}{\natexlab{b}})}\BibitemShut
  {NoStop}%
\bibitem [{\citenamefont {Prymidis}\ \emph {et~al.}(2015)\citenamefont
  {Prymidis}, \citenamefont {Sielcken},\ and\ \citenamefont
  {Filion}}]{prymidis2015self}%
  \BibitemOpen
  \bibfield  {author} {\bibinfo {author} {\bibfnamefont {V.}~\bibnamefont
  {Prymidis}}, \bibinfo {author} {\bibfnamefont {H.}~\bibnamefont {Sielcken}},\
  and\ \bibinfo {author} {\bibfnamefont {L.}~\bibnamefont {Filion}},\
  }\bibfield  {title} {\bibinfo {title} {Self-assembly of active attractive
  spheres},\ }\href {https://doi.org/10.1039/C5SM00127G} {\bibfield  {journal}
  {\bibinfo  {journal} {Soft Matter}\ }\textbf {\bibinfo {volume} {11}},\
  \bibinfo {pages} {4158} (\bibinfo {year} {2015})}\BibitemShut {NoStop}%
\bibitem [{\citenamefont {Bialk{\'e}}\ \emph {et~al.}(2015)\citenamefont
  {Bialk{\'e}}, \citenamefont {Siebert}, \citenamefont {L{\"o}wen},\ and\
  \citenamefont {Speck}}]{bialke-2015}%
  \BibitemOpen
  \bibfield  {author} {\bibinfo {author} {\bibfnamefont {J.}~\bibnamefont
  {Bialk{\'e}}}, \bibinfo {author} {\bibfnamefont {J.~T.}\ \bibnamefont
  {Siebert}}, \bibinfo {author} {\bibfnamefont {H.}~\bibnamefont {L{\"o}wen}},\
  and\ \bibinfo {author} {\bibfnamefont {T.}~\bibnamefont {Speck}},\ }\bibfield
   {title} {\bibinfo {title} {Negative interfacial tension in phase-separated
  active {Brownian} particles},\ }\href
  {https://doi.org/10.1103/PhysRevLett.115.098301} {\bibfield  {journal}
  {\bibinfo  {journal} {Phys. Rev. Lett.}\ }\textbf {\bibinfo {volume} {115}},\
  \bibinfo {pages} {098301} (\bibinfo {year} {2015})}\BibitemShut {NoStop}%
\bibitem [{\citenamefont {Fausti}\ \emph {et~al.}(2021)\citenamefont {Fausti},
  \citenamefont {Tjhung}, \citenamefont {Cates},\ and\ \citenamefont
  {Nardini}}]{fausti-2021}%
  \BibitemOpen
  \bibfield  {author} {\bibinfo {author} {\bibfnamefont {G.}~\bibnamefont
  {Fausti}}, \bibinfo {author} {\bibfnamefont {E.}~\bibnamefont {Tjhung}},
  \bibinfo {author} {\bibfnamefont {M.~E.}\ \bibnamefont {Cates}},\ and\
  \bibinfo {author} {\bibfnamefont {C.}~\bibnamefont {Nardini}},\ }\bibfield
  {title} {\bibinfo {title} {Capillary interfacial tension in active phase
  separation},\ }\href {https://doi.org/10.1103/PhysRevLett.127.068001}
  {\bibfield  {journal} {\bibinfo  {journal} {Phys. Rev. Lett.}\ }\textbf
  {\bibinfo {volume} {127}},\ \bibinfo {pages} {068001} (\bibinfo {year}
  {2021})}\BibitemShut {NoStop}%
\bibitem [{\citenamefont {Gulati}\ \emph {et~al.}(2025)\citenamefont {Gulati},
  \citenamefont {Caballero},\ and\ \citenamefont
  {Cristina~Marchetti}}]{gulati2025active}%
  \BibitemOpen
  \bibfield  {author} {\bibinfo {author} {\bibfnamefont {P.}~\bibnamefont
  {Gulati}}, \bibinfo {author} {\bibfnamefont {F.}~\bibnamefont {Caballero}},\
  and\ \bibinfo {author} {\bibfnamefont {M.}~\bibnamefont
  {Cristina~Marchetti}},\ }\bibfield  {title} {\bibinfo {title} {Active fluids
  form system-spanning filamentary networks},\ }\href
  {https://doi.org/10.1103/PhysRevLett.134.138301} {\bibfield  {journal}
  {\bibinfo  {journal} {Phys. Rev. Lett.}\ }\textbf {\bibinfo {volume} {134}},\
  \bibinfo {pages} {138301} (\bibinfo {year} {2025})}\BibitemShut {NoStop}%
\bibitem [{\citenamefont {Toffenetti}\ \emph {et~al.}(2026)\citenamefont
  {Toffenetti}, \citenamefont {Nettuno}, \citenamefont {Weyer},\ and\
  \citenamefont {Frey}}]{toffenetti-2026-arxiv}%
  \BibitemOpen
  \bibfield  {author} {\bibinfo {author} {\bibfnamefont {D.}~\bibnamefont
  {Toffenetti}}, \bibinfo {author} {\bibfnamefont {B.}~\bibnamefont {Nettuno}},
  \bibinfo {author} {\bibfnamefont {H.}~\bibnamefont {Weyer}},\ and\ \bibinfo
  {author} {\bibfnamefont {E.}~\bibnamefont {Frey}},\ }\bibfield  {title}
  {\bibinfo {title} {Active model b$^-$ from mass-conserving reaction-diffusion
  systems},\ }\href@noop {} {\bibfield  {journal} {\bibinfo  {journal} {arXiv
  preprint}\ } (\bibinfo {year} {2026})},\ \Eprint
  {https://arxiv.org/abs/2605.15903} {arXiv:2605.15903 [cond-mat.soft]}
  \BibitemShut {NoStop}%
\bibitem [{\citenamefont {Prymidis}\ \emph {et~al.}(2016)\citenamefont
  {Prymidis}, \citenamefont {Paliwal}, \citenamefont {Dijkstra},\ and\
  \citenamefont {Filion}}]{prymidis2016vapour}%
  \BibitemOpen
  \bibfield  {author} {\bibinfo {author} {\bibfnamefont {V.}~\bibnamefont
  {Prymidis}}, \bibinfo {author} {\bibfnamefont {S.}~\bibnamefont {Paliwal}},
  \bibinfo {author} {\bibfnamefont {M.}~\bibnamefont {Dijkstra}},\ and\
  \bibinfo {author} {\bibfnamefont {L.}~\bibnamefont {Filion}},\ }\bibfield
  {title} {\bibinfo {title} {Vapour-liquid coexistence of an active
  lennard-jones fluid},\ }\href {https://doi.org/10.1063/1.4963191} {\bibfield
  {journal} {\bibinfo  {journal} {J. Chem. Phys.}\ }\textbf {\bibinfo {volume}
  {145}},\ \bibinfo {pages} {124904} (\bibinfo {year} {2016})}\BibitemShut
  {NoStop}%
\bibitem [{\citenamefont {Saha}\ \emph {et~al.}(2020)\citenamefont {Saha},
  \citenamefont {Agudo-Canalejo},\ and\ \citenamefont
  {Golestanian}}]{saha-2020}%
  \BibitemOpen
  \bibfield  {author} {\bibinfo {author} {\bibfnamefont {S.}~\bibnamefont
  {Saha}}, \bibinfo {author} {\bibfnamefont {J.}~\bibnamefont
  {Agudo-Canalejo}},\ and\ \bibinfo {author} {\bibfnamefont {R.}~\bibnamefont
  {Golestanian}},\ }\bibfield  {title} {\bibinfo {title} {Scalar active
  mixtures: The nonreciprocal {Cahn-Hilliard} model},\ }\href
  {https://doi.org/10.1103/PhysRevX.10.041009} {\bibfield  {journal} {\bibinfo
  {journal} {Phys. Rev. X}\ }\textbf {\bibinfo {volume} {10}},\ \bibinfo
  {pages} {041009} (\bibinfo {year} {2020})}\BibitemShut {NoStop}%
\bibitem [{\citenamefont {You}\ \emph {et~al.}(2020)\citenamefont {You},
  \citenamefont {Baskaran},\ and\ \citenamefont
  {Marchetti}}]{you2020nonreciprocity}%
  \BibitemOpen
  \bibfield  {author} {\bibinfo {author} {\bibfnamefont {Z.}~\bibnamefont
  {You}}, \bibinfo {author} {\bibfnamefont {A.}~\bibnamefont {Baskaran}},\ and\
  \bibinfo {author} {\bibfnamefont {M.~C.}\ \bibnamefont {Marchetti}},\
  }\bibfield  {title} {\bibinfo {title} {Nonreciprocity as a generic route to
  traveling states},\ }\href {https://doi.org/10.1073/pnas.2010318117}
  {\bibfield  {journal} {\bibinfo  {journal} {Proc. Natl. Acad. Sci. U.S.A.}\
  }\textbf {\bibinfo {volume} {117}},\ \bibinfo {pages} {19767} (\bibinfo
  {year} {2020})}\BibitemShut {NoStop}%
\bibitem [{\citenamefont {Frohoff-H{\"u}lsmann}\ \emph
  {et~al.}(2023)\citenamefont {Frohoff-H{\"u}lsmann}, \citenamefont {Thiele},\
  and\ \citenamefont {Pismen}}]{frohoff2023non}%
  \BibitemOpen
  \bibfield  {author} {\bibinfo {author} {\bibfnamefont {T.}~\bibnamefont
  {Frohoff-H{\"u}lsmann}}, \bibinfo {author} {\bibfnamefont {U.}~\bibnamefont
  {Thiele}},\ and\ \bibinfo {author} {\bibfnamefont {L.~M.}\ \bibnamefont
  {Pismen}},\ }\bibfield  {title} {\bibinfo {title} {Non-reciprocity induces
  resonances in a two-field {Cahn--Hilliard} model},\ }\href
  {https://doi.org/10.1098/rsta.2022.0087} {\bibfield  {journal} {\bibinfo
  {journal} {Philos. Trans. R. Soc. A}\ }\textbf {\bibinfo {volume} {381}},\
  \bibinfo {pages} {20220087} (\bibinfo {year} {2023})}\BibitemShut {NoStop}%
\bibitem [{\citenamefont {Frohoff-H{\"u}lsmann}\ \emph
  {et~al.}(2021)\citenamefont {Frohoff-H{\"u}lsmann}, \citenamefont {Wrembel},\
  and\ \citenamefont {Thiele}}]{frohoff2021suppression}%
  \BibitemOpen
  \bibfield  {author} {\bibinfo {author} {\bibfnamefont {T.}~\bibnamefont
  {Frohoff-H{\"u}lsmann}}, \bibinfo {author} {\bibfnamefont {J.}~\bibnamefont
  {Wrembel}},\ and\ \bibinfo {author} {\bibfnamefont {U.}~\bibnamefont
  {Thiele}},\ }\bibfield  {title} {\bibinfo {title} {Suppression of coarsening
  and emergence of oscillatory behavior in a {Cahn-Hilliard} model with
  nonvariational coupling},\ }\href
  {https://doi.org/10.1103/PhysRevE.103.042602} {\bibfield  {journal} {\bibinfo
   {journal} {Phys. Rev. E}\ }\textbf {\bibinfo {volume} {103}},\ \bibinfo
  {pages} {042602} (\bibinfo {year} {2021})}\BibitemShut {NoStop}%
\bibitem [{\citenamefont {Brauns}\ and\ \citenamefont
  {Marchetti}(2024)}]{brauns-2024}%
  \BibitemOpen
  \bibfield  {author} {\bibinfo {author} {\bibfnamefont {F.}~\bibnamefont
  {Brauns}}\ and\ \bibinfo {author} {\bibfnamefont {M.~C.}\ \bibnamefont
  {Marchetti}},\ }\bibfield  {title} {\bibinfo {title} {Nonreciprocal pattern
  formation of conserved fields},\ }\href
  {https://doi.org/10.1103/PhysRevX.14.021014} {\bibfield  {journal} {\bibinfo
  {journal} {Phys. Rev. X}\ }\textbf {\bibinfo {volume} {14}},\ \bibinfo
  {pages} {021014} (\bibinfo {year} {2024})}\BibitemShut {NoStop}%
\bibitem [{\citenamefont {Solon}\ \emph
  {et~al.}(2018{\natexlab{a}})\citenamefont {Solon}, \citenamefont
  {Stenhammar}, \citenamefont {Cates}, \citenamefont {Kafri},\ and\
  \citenamefont {Tailleur}}]{solon-2018}%
  \BibitemOpen
  \bibfield  {author} {\bibinfo {author} {\bibfnamefont {A.~P.}\ \bibnamefont
  {Solon}}, \bibinfo {author} {\bibfnamefont {J.}~\bibnamefont {Stenhammar}},
  \bibinfo {author} {\bibfnamefont {M.~E.}\ \bibnamefont {Cates}}, \bibinfo
  {author} {\bibfnamefont {Y.}~\bibnamefont {Kafri}},\ and\ \bibinfo {author}
  {\bibfnamefont {J.}~\bibnamefont {Tailleur}},\ }\bibfield  {title} {\bibinfo
  {title} {Generalized thermodynamics of phase equilibria in scalar active
  matter},\ }\href {https://doi.org/10.1103/PhysRevE.97.020602} {\bibfield
  {journal} {\bibinfo  {journal} {Phys. Rev. E}\ }\textbf {\bibinfo {volume}
  {97}},\ \bibinfo {pages} {020602} (\bibinfo {year}
  {2018}{\natexlab{a}})}\BibitemShut {NoStop}%
\bibitem [{\citenamefont {Solon}\ \emph
  {et~al.}(2018{\natexlab{b}})\citenamefont {Solon}, \citenamefont
  {Stenhammar}, \citenamefont {Cates}, \citenamefont {Kafri},\ and\
  \citenamefont {Tailleur}}]{solon-2018-njp}%
  \BibitemOpen
  \bibfield  {author} {\bibinfo {author} {\bibfnamefont {A.~P.}\ \bibnamefont
  {Solon}}, \bibinfo {author} {\bibfnamefont {J.}~\bibnamefont {Stenhammar}},
  \bibinfo {author} {\bibfnamefont {M.~E.}\ \bibnamefont {Cates}}, \bibinfo
  {author} {\bibfnamefont {Y.}~\bibnamefont {Kafri}},\ and\ \bibinfo {author}
  {\bibfnamefont {J.}~\bibnamefont {Tailleur}},\ }\bibfield  {title} {\bibinfo
  {title} {Generalized thermodynamics of motility-induced phase separation:
  phase equilibria, {Laplace} pressure, and change of ensembles},\ }\href
  {https://doi.org/10.1088/1367-2630/aaccdd} {\bibfield  {journal} {\bibinfo
  {journal} {New J. Phys.}\ }\textbf {\bibinfo {volume} {20}},\ \bibinfo
  {pages} {075001} (\bibinfo {year} {2018}{\natexlab{b}})}\BibitemShut
  {NoStop}%
\bibitem [{\citenamefont {Dinelli}\ \emph {et~al.}(2024)\citenamefont
  {Dinelli}, \citenamefont {O’Byrne},\ and\ \citenamefont
  {Tailleur}}]{dinelli-2024}%
  \BibitemOpen
  \bibfield  {author} {\bibinfo {author} {\bibfnamefont {A.}~\bibnamefont
  {Dinelli}}, \bibinfo {author} {\bibfnamefont {J.}~\bibnamefont {O’Byrne}},\
  and\ \bibinfo {author} {\bibfnamefont {J.}~\bibnamefont {Tailleur}},\
  }\bibfield  {title} {\bibinfo {title} {Fluctuating hydrodynamics of active
  particles interacting via taxis and quorum sensing: static and dynamics},\
  }\href {https://doi.org/10.1088/1751-8121/ad72bc} {\bibfield  {journal}
  {\bibinfo  {journal} {J. Phys. A Math. Theor.}\ }\textbf {\bibinfo {volume}
  {57}},\ \bibinfo {pages} {395002} (\bibinfo {year} {2024})}\BibitemShut
  {NoStop}%
\bibitem [{\citenamefont {Burekovi\'{c}}\ \emph {et~al.}(2026)\citenamefont
  {Burekovi\'{c}}, \citenamefont {De~Luca}, \citenamefont {Cates},\ and\
  \citenamefont {Nardini}}]{burekovic-2026}%
  \BibitemOpen
  \bibfield  {author} {\bibinfo {author} {\bibfnamefont {S.}~\bibnamefont
  {Burekovi\'{c}}}, \bibinfo {author} {\bibfnamefont {F.}~\bibnamefont
  {De~Luca}}, \bibinfo {author} {\bibfnamefont {M.~E.}\ \bibnamefont {Cates}},\
  and\ \bibinfo {author} {\bibfnamefont {C.}~\bibnamefont {Nardini}},\
  }\bibfield  {title} {\bibinfo {title} {Active {Cahn--Hilliard} theory for
  non-equilibrium phase separation: quantitative macroscopic predictions and a
  microscopic derivation}\ }\href {https://doi.org/10.48550/arXiv.2601.16539}
  {10.48550/arXiv.2601.16539} (\bibinfo {year} {2026})\BibitemShut {NoStop}%
\bibitem [{\citenamefont {Stenhammar}\ \emph {et~al.}(2013)\citenamefont
  {Stenhammar}, \citenamefont {Tiribocchi}, \citenamefont {Allen},
  \citenamefont {Marenduzzo},\ and\ \citenamefont {Cates}}]{stenhammar-2013}%
  \BibitemOpen
  \bibfield  {author} {\bibinfo {author} {\bibfnamefont {J.}~\bibnamefont
  {Stenhammar}}, \bibinfo {author} {\bibfnamefont {A.}~\bibnamefont
  {Tiribocchi}}, \bibinfo {author} {\bibfnamefont {R.~J.}\ \bibnamefont
  {Allen}}, \bibinfo {author} {\bibfnamefont {D.}~\bibnamefont {Marenduzzo}},\
  and\ \bibinfo {author} {\bibfnamefont {M.~E.}\ \bibnamefont {Cates}},\
  }\bibfield  {title} {\bibinfo {title} {Continuum theory of phase separation
  kinetics for active brownian particles},\ }\href
  {https://doi.org/10.1103/PhysRevLett.111.145702} {\bibfield  {journal}
  {\bibinfo  {journal} {Phys. Rev. Lett.}\ }\textbf {\bibinfo {volume} {111}},\
  \bibinfo {pages} {145702} (\bibinfo {year} {2013})}\BibitemShut {NoStop}%
\bibitem [{\citenamefont {Speck}\ \emph {et~al.}(2014)\citenamefont {Speck},
  \citenamefont {Bialk\'e}, \citenamefont {Menzel},\ and\ \citenamefont
  {L\"owen}}]{speck-2014-weaknl-prl}%
  \BibitemOpen
  \bibfield  {author} {\bibinfo {author} {\bibfnamefont {T.}~\bibnamefont
  {Speck}}, \bibinfo {author} {\bibfnamefont {J.}~\bibnamefont {Bialk\'e}},
  \bibinfo {author} {\bibfnamefont {A.~M.}\ \bibnamefont {Menzel}},\ and\
  \bibinfo {author} {\bibfnamefont {H.}~\bibnamefont {L\"owen}},\ }\bibfield
  {title} {\bibinfo {title} {Effective {Cahn-Hilliard} equation for the phase
  separation of active {Brownian} particles},\ }\href
  {https://doi.org/10.1103/PhysRevLett.112.218304} {\bibfield  {journal}
  {\bibinfo  {journal} {Phys. Rev. Lett.}\ }\textbf {\bibinfo {volume} {112}},\
  \bibinfo {pages} {218304} (\bibinfo {year} {2014})}\BibitemShut {NoStop}%
\bibitem [{\citenamefont {te~Vrugt}\ \emph {et~al.}(2023)\citenamefont
  {te~Vrugt}, \citenamefont {Bickmann},\ and\ \citenamefont
  {Wittkowski}}]{vrugt-2023}%
  \BibitemOpen
  \bibfield  {author} {\bibinfo {author} {\bibfnamefont {M.}~\bibnamefont
  {te~Vrugt}}, \bibinfo {author} {\bibfnamefont {J.}~\bibnamefont {Bickmann}},\
  and\ \bibinfo {author} {\bibfnamefont {R.}~\bibnamefont {Wittkowski}},\
  }\bibfield  {title} {\bibinfo {title} {How to derive a predictive field
  theory for active {Brownian} particles: a step-by-step tutorial},\ }\href
  {https://doi.org/10.1088/1361-648X/acc440} {\bibfield  {journal} {\bibinfo
  {journal} {J. Phys. Condens. Matter}\ }\textbf {\bibinfo {volume} {35}},\
  \bibinfo {pages} {313001} (\bibinfo {year} {2023})}\BibitemShut {NoStop}%
\bibitem [{\citenamefont {Liao}\ and\ \citenamefont {Klapp}(2018)}]{liao-2018}%
  \BibitemOpen
  \bibfield  {author} {\bibinfo {author} {\bibfnamefont {G.-J.}\ \bibnamefont
  {Liao}}\ and\ \bibinfo {author} {\bibfnamefont {S.~H.~L.}\ \bibnamefont
  {Klapp}},\ }\bibfield  {title} {\bibinfo {title} {Clustering and phase
  separation of circle swimmers dispersed in a monolayer},\ }\href
  {https://doi.org/10.1039/C8SM01366G} {\bibfield  {journal} {\bibinfo
  {journal} {Soft Matter}\ }\textbf {\bibinfo {volume} {14}},\ \bibinfo {pages}
  {7873} (\bibinfo {year} {2018})}\BibitemShut {NoStop}%
\bibitem [{\citenamefont {Bickmann}\ \emph {et~al.}(2022)\citenamefont
  {Bickmann}, \citenamefont {Bröker}, \citenamefont {Jeggle},\ and\
  \citenamefont {Wittkowski}}]{bickmann-2022}%
  \BibitemOpen
  \bibfield  {author} {\bibinfo {author} {\bibfnamefont {J.}~\bibnamefont
  {Bickmann}}, \bibinfo {author} {\bibfnamefont {S.}~\bibnamefont {Bröker}},
  \bibinfo {author} {\bibfnamefont {J.}~\bibnamefont {Jeggle}},\ and\ \bibinfo
  {author} {\bibfnamefont {R.}~\bibnamefont {Wittkowski}},\ }\bibfield  {title}
  {\bibinfo {title} {{Analytical approach to chiral active systems: Suppressed
  phase separation of interacting {Brownian} circle swimmers}},\ }\href
  {https://doi.org/10.1063/5.0085122} {\bibfield  {journal} {\bibinfo
  {journal} {J. Chem. Phys.}\ }\textbf {\bibinfo {volume} {156}},\ \bibinfo
  {pages} {194904} (\bibinfo {year} {2022})}\BibitemShut {NoStop}%
\bibitem [{\citenamefont {Kalz}\ \emph {et~al.}(2024)\citenamefont {Kalz},
  \citenamefont {Sharma},\ and\ \citenamefont {Metzler}}]{kalz-2024}%
  \BibitemOpen
  \bibfield  {author} {\bibinfo {author} {\bibfnamefont {E.}~\bibnamefont
  {Kalz}}, \bibinfo {author} {\bibfnamefont {A.}~\bibnamefont {Sharma}},\ and\
  \bibinfo {author} {\bibfnamefont {R.}~\bibnamefont {Metzler}},\ }\bibfield
  {title} {\bibinfo {title} {Field theory of active chiral hard disks: a
  first-principles approach to steric interactions},\ }\href
  {https://doi.org/10.1088/1751-8121/ad5089} {\bibfield  {journal} {\bibinfo
  {journal} {J. Phys. A Math. Theor.}\ }\textbf {\bibinfo {volume} {57}},\
  \bibinfo {pages} {265002} (\bibinfo {year} {2024})}\BibitemShut {NoStop}%
\bibitem [{\citenamefont {Sesé-Sansa}\ \emph {et~al.}(2022)\citenamefont
  {Sesé-Sansa}, \citenamefont {Levis},\ and\ \citenamefont
  {Pagonabarraga}}]{sese-sansa-2022}%
  \BibitemOpen
  \bibfield  {author} {\bibinfo {author} {\bibfnamefont {E.}~\bibnamefont
  {Sesé-Sansa}}, \bibinfo {author} {\bibfnamefont {D.}~\bibnamefont {Levis}},\
  and\ \bibinfo {author} {\bibfnamefont {I.}~\bibnamefont {Pagonabarraga}},\
  }\bibfield  {title} {\bibinfo {title} {{Microscopic field theory for
  structure formation in systems of self-propelled particles with generic
  torques}},\ }\href {https://doi.org/10.1063/5.0123680} {\bibfield  {journal}
  {\bibinfo  {journal} {J. Chem. Phys.}\ }\textbf {\bibinfo {volume} {157}},\
  \bibinfo {pages} {224905} (\bibinfo {year} {2022})}\BibitemShut {NoStop}%
\bibitem [{\citenamefont {Ma}\ and\ \citenamefont {Ni}(2022)}]{ma-2022}%
  \BibitemOpen
  \bibfield  {author} {\bibinfo {author} {\bibfnamefont {Z.}~\bibnamefont
  {Ma}}\ and\ \bibinfo {author} {\bibfnamefont {R.}~\bibnamefont {Ni}},\
  }\bibfield  {title} {\bibinfo {title} {{Dynamical clustering interrupts
  motility-induced phase separation in chiral active {Brownian} particles}},\
  }\href {https://doi.org/10.1063/5.0077389} {\bibfield  {journal} {\bibinfo
  {journal} {J. Chem. Phys.}\ }\textbf {\bibinfo {volume} {156}},\ \bibinfo
  {pages} {021102} (\bibinfo {year} {2022})}\BibitemShut {NoStop}%
\bibitem [{\citenamefont {Semwal}\ \emph {et~al.}(2024)\citenamefont {Semwal},
  \citenamefont {Joshi}, \citenamefont {Dikshit},\ and\ \citenamefont
  {Mishra}}]{semwal-2024}%
  \BibitemOpen
  \bibfield  {author} {\bibinfo {author} {\bibfnamefont {V.}~\bibnamefont
  {Semwal}}, \bibinfo {author} {\bibfnamefont {J.}~\bibnamefont {Joshi}},
  \bibinfo {author} {\bibfnamefont {S.}~\bibnamefont {Dikshit}},\ and\ \bibinfo
  {author} {\bibfnamefont {S.}~\bibnamefont {Mishra}},\ }\bibfield  {title}
  {\bibinfo {title} {Macro to micro phase separation of chiral active
  swimmers},\ }\href {https://doi.org/10.1016/j.physa.2023.129435} {\bibfield
  {journal} {\bibinfo  {journal} {Physica A}\ }\textbf {\bibinfo {volume}
  {634}},\ \bibinfo {pages} {129435} (\bibinfo {year} {2024})}\BibitemShut
  {NoStop}%
\bibitem [{\citenamefont {Hohenberg}\ and\ \citenamefont
  {Halperin}(1977)}]{hohenberg-1977}%
  \BibitemOpen
  \bibfield  {author} {\bibinfo {author} {\bibfnamefont {P.~C.}\ \bibnamefont
  {Hohenberg}}\ and\ \bibinfo {author} {\bibfnamefont {B.~I.}\ \bibnamefont
  {Halperin}},\ }\bibfield  {title} {\bibinfo {title} {Theory of dynamic
  critical phenomena},\ }\href {https://doi.org/10.1103/RevModPhys.49.435}
  {\bibfield  {journal} {\bibinfo  {journal} {Rev. Mod. Phys.}\ }\textbf
  {\bibinfo {volume} {49}},\ \bibinfo {pages} {435} (\bibinfo {year}
  {1977})}\BibitemShut {NoStop}%
\bibitem [{\citenamefont {Chaikin}\ and\ \citenamefont
  {Lubensky}(2000)}]{chaikin2000principles}%
  \BibitemOpen
  \bibfield  {author} {\bibinfo {author} {\bibfnamefont {P.~M.}\ \bibnamefont
  {Chaikin}}\ and\ \bibinfo {author} {\bibfnamefont {T.~C.}\ \bibnamefont
  {Lubensky}},\ }\href@noop {} {\emph {\bibinfo {title} {Principles of
  condensed matter physics}}},\ Vol.~\bibinfo {volume} {1}\ (\bibinfo
  {publisher} {Cambridge University Press},\ \bibinfo {address} {Cambridge,
  UK},\ \bibinfo {year} {2000})\BibitemShut {NoStop}%
\bibitem [{\citenamefont {Cates}\ and\ \citenamefont
  {Tailleur}(2013)}]{cates-2013}%
  \BibitemOpen
  \bibfield  {author} {\bibinfo {author} {\bibfnamefont {M.~E.}\ \bibnamefont
  {Cates}}\ and\ \bibinfo {author} {\bibfnamefont {J.}~\bibnamefont
  {Tailleur}},\ }\bibfield  {title} {\bibinfo {title} {When are active
  {Brownian} particles and run-and-tumble particles equivalent? {C}onsequences
  for motility-induced phase separation},\ }\href
  {https://doi.org/10.1209/0295-5075/101/20010} {\bibfield  {journal} {\bibinfo
   {journal} {Europhys. Lett.}\ }\textbf {\bibinfo {volume} {101}},\ \bibinfo
  {pages} {20010} (\bibinfo {year} {2013})}\BibitemShut {NoStop}%
\bibitem [{\citenamefont {Solon}\ \emph {et~al.}(2015)\citenamefont {Solon},
  \citenamefont {Cates},\ and\ \citenamefont {Tailleur}}]{solon-2015}%
  \BibitemOpen
  \bibfield  {author} {\bibinfo {author} {\bibfnamefont {A.~P.}\ \bibnamefont
  {Solon}}, \bibinfo {author} {\bibfnamefont {M.~E.}\ \bibnamefont {Cates}},\
  and\ \bibinfo {author} {\bibfnamefont {J.}~\bibnamefont {Tailleur}},\
  }\bibfield  {title} {\bibinfo {title} {Active brownian particles and
  run-and-tumble particles: A comparative study},\ }\href
  {https://doi.org/10.1140/epjst/e2015-02457-0} {\bibfield  {journal} {\bibinfo
   {journal} {Eur. Phys. J. Spec. Top.}\ }\textbf {\bibinfo {volume} {224}},\
  \bibinfo {pages} {1231} (\bibinfo {year} {2015})}\BibitemShut {NoStop}%
\bibitem [{\citenamefont {Dinelli}\ and\ \citenamefont
  {Muzzeddu}(2026)}]{dinelli-2026-long}%
  \BibitemOpen
  \bibfield  {author} {\bibinfo {author} {\bibfnamefont {A.}~\bibnamefont
  {Dinelli}}\ and\ \bibinfo {author} {\bibfnamefont {P.}~\bibnamefont
  {Muzzeddu}},\ }\bibfield  {title} {\bibinfo {title} {Multiscale perturbative
  approach to active matter with motility regulation}\ }\href
  {https://doi.org/10.48550/arXiv.2604.09453} {10.48550/arXiv.2604.09453}
  (\bibinfo {year} {2026})\BibitemShut {NoStop}%
\bibitem [{\citenamefont {Martin}\ \emph {et~al.}(2021)\citenamefont {Martin},
  \citenamefont {O'Byrne}, \citenamefont {Cates}, \citenamefont {Fodor},
  \citenamefont {Nardini}, \citenamefont {Tailleur},\ and\ \citenamefont {van
  Wijland}}]{martin-2021}%
  \BibitemOpen
  \bibfield  {author} {\bibinfo {author} {\bibfnamefont {D.}~\bibnamefont
  {Martin}}, \bibinfo {author} {\bibfnamefont {J.}~\bibnamefont {O'Byrne}},
  \bibinfo {author} {\bibfnamefont {M.~E.}\ \bibnamefont {Cates}}, \bibinfo
  {author} {\bibfnamefont {E.}~\bibnamefont {Fodor}}, \bibinfo {author}
  {\bibfnamefont {C.}~\bibnamefont {Nardini}}, \bibinfo {author} {\bibfnamefont
  {J.}~\bibnamefont {Tailleur}},\ and\ \bibinfo {author} {\bibfnamefont
  {F.}~\bibnamefont {van Wijland}},\ }\bibfield  {title} {\bibinfo {title}
  {Statistical mechanics of active {Ornstein-Uhlenbeck} particles},\ }\href
  {https://doi.org/10.1103/PhysRevE.103.032607} {\bibfield  {journal} {\bibinfo
   {journal} {Phys. Rev. E}\ }\textbf {\bibinfo {volume} {103}},\ \bibinfo
  {pages} {032607} (\bibinfo {year} {2021})}\BibitemShut {NoStop}%
\bibitem [{si()}]{si}%
  \BibitemOpen
  \href@noop {} {}\bibinfo {note} {See Supplemental Material at [URL will be
  inserted by publisher], which also includes
  Refs.~\onlinecite{ascher-1997,besse-2023,burns-2020,cahn-1958,cates-2018,deluca-2024,kevorkian-1996,pavliotis-2008,chatzittofi2026nonequilibrium},
  for details on the chiral QS particle model and chiral active Model~B,
  coarse-graining derivations, linear stability analyses, numerical methods,
  simulation snapshots and movies.}\BibitemShut {Stop}%
\bibitem [{\citenamefont {Ebbens}\ \emph {et~al.}(2010)\citenamefont {Ebbens},
  \citenamefont {Jones}, \citenamefont {Ryan}, \citenamefont {Golestanian},\
  and\ \citenamefont {Howse}}]{ebbens-2010}%
  \BibitemOpen
  \bibfield  {author} {\bibinfo {author} {\bibfnamefont {S.}~\bibnamefont
  {Ebbens}}, \bibinfo {author} {\bibfnamefont {R.~A.~L.}\ \bibnamefont
  {Jones}}, \bibinfo {author} {\bibfnamefont {A.~J.}\ \bibnamefont {Ryan}},
  \bibinfo {author} {\bibfnamefont {R.}~\bibnamefont {Golestanian}},\ and\
  \bibinfo {author} {\bibfnamefont {J.~R.}\ \bibnamefont {Howse}},\ }\bibfield
  {title} {\bibinfo {title} {Self-assembled autonomous runners and tumblers},\
  }\href {https://doi.org/10.1103/PhysRevE.82.015304} {\bibfield  {journal}
  {\bibinfo  {journal} {Phys. Rev. E}\ }\textbf {\bibinfo {volume} {82}},\
  \bibinfo {pages} {015304} (\bibinfo {year} {2010})}\BibitemShut {NoStop}%
\bibitem [{\citenamefont {Marine}\ \emph {et~al.}(2013)\citenamefont {Marine},
  \citenamefont {Wheat}, \citenamefont {Ault},\ and\ \citenamefont
  {Posner}}]{marine-2013}%
  \BibitemOpen
  \bibfield  {author} {\bibinfo {author} {\bibfnamefont {N.~A.}\ \bibnamefont
  {Marine}}, \bibinfo {author} {\bibfnamefont {P.~M.}\ \bibnamefont {Wheat}},
  \bibinfo {author} {\bibfnamefont {J.}~\bibnamefont {Ault}},\ and\ \bibinfo
  {author} {\bibfnamefont {J.~D.}\ \bibnamefont {Posner}},\ }\bibfield  {title}
  {\bibinfo {title} {Diffusive behaviors of circle-swimming motors},\ }\href
  {https://doi.org/10.1103/PhysRevE.87.052305} {\bibfield  {journal} {\bibinfo
  {journal} {Phys. Rev. E}\ }\textbf {\bibinfo {volume} {87}},\ \bibinfo
  {pages} {052305} (\bibinfo {year} {2013})}\BibitemShut {NoStop}%
\bibitem [{\citenamefont {Sevilla}(2016)}]{sevilla-2016}%
  \BibitemOpen
  \bibfield  {author} {\bibinfo {author} {\bibfnamefont {F.~J.}\ \bibnamefont
  {Sevilla}},\ }\bibfield  {title} {\bibinfo {title} {Diffusion of active
  chiral particles},\ }\href {https://doi.org/10.1103/PhysRevE.94.062120}
  {\bibfield  {journal} {\bibinfo  {journal} {Phys. Rev. E}\ }\textbf {\bibinfo
  {volume} {94}},\ \bibinfo {pages} {062120} (\bibinfo {year}
  {2016})}\BibitemShut {NoStop}%
\bibitem [{\citenamefont {Caprini}\ and\ \citenamefont {Marini
  Bettolo~Marconi}(2019)}]{caprini-2019}%
  \BibitemOpen
  \bibfield  {author} {\bibinfo {author} {\bibfnamefont {L.}~\bibnamefont
  {Caprini}}\ and\ \bibinfo {author} {\bibfnamefont {U.}~\bibnamefont {Marini
  Bettolo~Marconi}},\ }\bibfield  {title} {\bibinfo {title} {Active chiral
  particles under confinement: surface currents and bulk accumulation
  phenomena},\ }\href {https://doi.org/10.1039/C8SM02492H} {\bibfield
  {journal} {\bibinfo  {journal} {Soft Matter}\ }\textbf {\bibinfo {volume}
  {15}},\ \bibinfo {pages} {2627} (\bibinfo {year} {2019})}\BibitemShut
  {NoStop}%
\bibitem [{\citenamefont {Cates}\ and\ \citenamefont
  {Nardini}(2023)}]{cates-2023}%
  \BibitemOpen
  \bibfield  {author} {\bibinfo {author} {\bibfnamefont {M.~E.}\ \bibnamefont
  {Cates}}\ and\ \bibinfo {author} {\bibfnamefont {C.}~\bibnamefont
  {Nardini}},\ }\bibfield  {title} {\bibinfo {title} {{Classical Nucleation
  Theory} for active fluid phase separation},\ }\href
  {https://doi.org/10.1103/PhysRevLett.130.098203} {\bibfield  {journal}
  {\bibinfo  {journal} {Phys. Rev. Lett.}\ }\textbf {\bibinfo {volume} {130}},\
  \bibinfo {pages} {098203} (\bibinfo {year} {2023})}\BibitemShut {NoStop}%
\bibitem [{\citenamefont {Wittkowski}\ \emph {et~al.}(2014)\citenamefont
  {Wittkowski}, \citenamefont {Tiribocchi}, \citenamefont {Stenhammar},
  \citenamefont {Allen}, \citenamefont {Marenduzzo},\ and\ \citenamefont
  {Cates}}]{wittkowski-2014}%
  \BibitemOpen
  \bibfield  {author} {\bibinfo {author} {\bibfnamefont {R.}~\bibnamefont
  {Wittkowski}}, \bibinfo {author} {\bibfnamefont {A.}~\bibnamefont
  {Tiribocchi}}, \bibinfo {author} {\bibfnamefont {J.}~\bibnamefont
  {Stenhammar}}, \bibinfo {author} {\bibfnamefont {R.~J.}\ \bibnamefont
  {Allen}}, \bibinfo {author} {\bibfnamefont {D.}~\bibnamefont {Marenduzzo}},\
  and\ \bibinfo {author} {\bibfnamefont {M.~E.}\ \bibnamefont {Cates}},\
  }\bibfield  {title} {\bibinfo {title} {Scalar {$\varphi^4$} field theory for
  active-particle phase separation},\ }\href
  {https://doi.org/10.1038/ncomms5351} {\bibfield  {journal} {\bibinfo
  {journal} {Nat. Commun.}\ }\textbf {\bibinfo {volume} {5}},\ \bibinfo {pages}
  {4351} (\bibinfo {year} {2014})}\BibitemShut {NoStop}%
\bibitem [{\citenamefont {Omar}\ \emph {et~al.}(2023)\citenamefont {Omar},
  \citenamefont {Row}, \citenamefont {Mallory},\ and\ \citenamefont
  {Brady}}]{omar-2023}%
  \BibitemOpen
  \bibfield  {author} {\bibinfo {author} {\bibfnamefont {A.~K.}\ \bibnamefont
  {Omar}}, \bibinfo {author} {\bibfnamefont {H.}~\bibnamefont {Row}}, \bibinfo
  {author} {\bibfnamefont {S.~A.}\ \bibnamefont {Mallory}},\ and\ \bibinfo
  {author} {\bibfnamefont {J.~F.}\ \bibnamefont {Brady}},\ }\bibfield  {title}
  {\bibinfo {title} {Mechanical theory of nonequilibrium coexistence and
  motility-induced phase separation},\ }\href
  {https://doi.org/10.1073/pnas.2219900120} {\bibfield  {journal} {\bibinfo
  {journal} {Proc. Natl. Acad. Sci. U.S.A.}\ }\textbf {\bibinfo {volume}
  {120}},\ \bibinfo {pages} {e2219900120} (\bibinfo {year} {2023})}\BibitemShut
  {NoStop}%
\bibitem [{\citenamefont {Emmerich}\ \emph {et~al.}(2012)\citenamefont
  {Emmerich}, \citenamefont {Löwen}, \citenamefont {Wittkowski}, \citenamefont
  {Gruhn}, \citenamefont {Tóth}, \citenamefont {Tegze},\ and\ \citenamefont
  {Gránásy}}]{emmerich2012phase}%
  \BibitemOpen
  \bibfield  {author} {\bibinfo {author} {\bibfnamefont {H.}~\bibnamefont
  {Emmerich}}, \bibinfo {author} {\bibfnamefont {H.}~\bibnamefont {Löwen}},
  \bibinfo {author} {\bibfnamefont {R.}~\bibnamefont {Wittkowski}}, \bibinfo
  {author} {\bibfnamefont {T.}~\bibnamefont {Gruhn}}, \bibinfo {author}
  {\bibfnamefont {G.~I.}\ \bibnamefont {Tóth}}, \bibinfo {author}
  {\bibfnamefont {G.}~\bibnamefont {Tegze}},\ and\ \bibinfo {author}
  {\bibfnamefont {L.}~\bibnamefont {Gránásy}},\ }\bibfield  {title} {\bibinfo
  {title} {Phase-field-crystal models for condensed matter dynamics on atomic
  length and diffusive time scales: an overview},\ }\href
  {https://doi.org/10.1080/00018732.2012.737555} {\bibfield  {journal}
  {\bibinfo  {journal} {Adv. Phys.}\ }\textbf {\bibinfo {volume} {61}},\
  \bibinfo {pages} {665} (\bibinfo {year} {2012})}\BibitemShut {NoStop}%
\bibitem [{\citenamefont {Matthews}\ and\ \citenamefont
  {Cox}(2000)}]{matthews-2000}%
  \BibitemOpen
  \bibfield  {author} {\bibinfo {author} {\bibfnamefont {P.~C.}\ \bibnamefont
  {Matthews}}\ and\ \bibinfo {author} {\bibfnamefont {S.~M.}\ \bibnamefont
  {Cox}},\ }\bibfield  {title} {\bibinfo {title} {Pattern formation with a
  conservation law},\ }\href {https://doi.org/10.1088/0951-7715/13/4/317}
  {\bibfield  {journal} {\bibinfo  {journal} {Nonlinearity}\ }\textbf {\bibinfo
  {volume} {13}},\ \bibinfo {pages} {1293} (\bibinfo {year}
  {2000})}\BibitemShut {NoStop}%
\bibitem [{\citenamefont {Elder}\ \emph {et~al.}(2002)\citenamefont {Elder},
  \citenamefont {Katakowski}, \citenamefont {Haataja},\ and\ \citenamefont
  {Grant}}]{elder2002modeling}%
  \BibitemOpen
  \bibfield  {author} {\bibinfo {author} {\bibfnamefont {K.~R.}\ \bibnamefont
  {Elder}}, \bibinfo {author} {\bibfnamefont {M.}~\bibnamefont {Katakowski}},
  \bibinfo {author} {\bibfnamefont {M.}~\bibnamefont {Haataja}},\ and\ \bibinfo
  {author} {\bibfnamefont {M.}~\bibnamefont {Grant}},\ }\bibfield  {title}
  {\bibinfo {title} {Modeling elasticity in crystal growth},\ }\href
  {https://doi.org/10.1103/PhysRevLett.88.245701} {\bibfield  {journal}
  {\bibinfo  {journal} {Phys. Rev. Lett.}\ }\textbf {\bibinfo {volume} {88}},\
  \bibinfo {pages} {245701} (\bibinfo {year} {2002})}\BibitemShut {NoStop}%
\bibitem [{\citenamefont {Thiele}\ \emph {et~al.}(2013)\citenamefont {Thiele},
  \citenamefont {Archer}, \citenamefont {Robbins}, \citenamefont {Gomez},\ and\
  \citenamefont {Knobloch}}]{thiele2013localized}%
  \BibitemOpen
  \bibfield  {author} {\bibinfo {author} {\bibfnamefont {U.}~\bibnamefont
  {Thiele}}, \bibinfo {author} {\bibfnamefont {A.~J.}\ \bibnamefont {Archer}},
  \bibinfo {author} {\bibfnamefont {M.~J.}\ \bibnamefont {Robbins}}, \bibinfo
  {author} {\bibfnamefont {H.}~\bibnamefont {Gomez}},\ and\ \bibinfo {author}
  {\bibfnamefont {E.}~\bibnamefont {Knobloch}},\ }\bibfield  {title} {\bibinfo
  {title} {Localized states in the conserved {Swift-Hohenberg} equation with
  cubic nonlinearity},\ }\href {https://doi.org/10.1103/PhysRevE.87.042915}
  {\bibfield  {journal} {\bibinfo  {journal} {Phys. Rev. E}\ }\textbf {\bibinfo
  {volume} {87}},\ \bibinfo {pages} {042915} (\bibinfo {year}
  {2013})}\BibitemShut {NoStop}%
\bibitem [{\citenamefont {Caprini}\ \emph {et~al.}(2026)\citenamefont
  {Caprini}, \citenamefont {Petrini},\ and\ \citenamefont {Marini
  Bettolo~Marconi}}]{caprini-2026-modelingchiral}%
  \BibitemOpen
  \bibfield  {author} {\bibinfo {author} {\bibfnamefont {L.}~\bibnamefont
  {Caprini}}, \bibinfo {author} {\bibfnamefont {A.}~\bibnamefont {Petrini}},\
  and\ \bibinfo {author} {\bibfnamefont {U.}~\bibnamefont {Marini
  Bettolo~Marconi}},\ }\bibfield  {title} {\bibinfo {title} {Modeling chiral
  active particles: from circular motion to odd interactions},\ }\href
  {https://doi.org/10.1088/1742-5468/ae3d28} {\bibfield  {journal} {\bibinfo
  {journal} {J. Stat. Mech.: Theory Exp.}\ }\textbf {\bibinfo {volume}
  {2026}}\bibinfo  {number} { (2)},\ \bibinfo {pages} {024001}}\BibitemShut
  {NoStop}%
\bibitem [{\citenamefont {Marini Bettolo~Marconi}\ \emph
  {et~al.}(2026)\citenamefont {Marini Bettolo~Marconi}, \citenamefont
  {Petrini}, \citenamefont {Maire},\ and\ \citenamefont
  {Caprini}}]{marconi-2026}%
  \BibitemOpen
\bibfield  {number} {  }\bibfield  {author} {\bibinfo {author} {\bibfnamefont
  {U.}~\bibnamefont {Marini Bettolo~Marconi}}, \bibinfo {author} {\bibfnamefont
  {A.}~\bibnamefont {Petrini}}, \bibinfo {author} {\bibfnamefont
  {R.}~\bibnamefont {Maire}},\ and\ \bibinfo {author} {\bibfnamefont
  {L.}~\bibnamefont {Caprini}},\ }\bibfield  {title} {\bibinfo {title}
  {Emergent hydrodynamics of chiral active fluids: vortices, bubbles and odd
  diffusion},\ }\href {https://doi.org/10.1088/1367-2630/ae6c55} {\bibfield
  {journal} {\bibinfo  {journal} {New J. Phys.}\ }\textbf {\bibinfo {volume}
  {28}},\ \bibinfo {pages} {064401} (\bibinfo {year} {2026})}\BibitemShut
  {NoStop}%
\bibitem [{\citenamefont {Petrini}\ \emph {et~al.}(2026)\citenamefont
  {Petrini}, \citenamefont {Maire}, \citenamefont {Marconi},\ and\
  \citenamefont {Caprini}}]{petrini-2026-arxiv}%
  \BibitemOpen
  \bibfield  {author} {\bibinfo {author} {\bibfnamefont {A.}~\bibnamefont
  {Petrini}}, \bibinfo {author} {\bibfnamefont {R.}~\bibnamefont {Maire}},
  \bibinfo {author} {\bibfnamefont {U.~M.~B.}\ \bibnamefont {Marconi}},\ and\
  \bibinfo {author} {\bibfnamefont {L.}~\bibnamefont {Caprini}},\ }\bibfield
  {title} {\bibinfo {title} {Sparkling bubbles in chiral active fluids},\
  }\bibfield  {journal} {\bibinfo  {journal} {arXiv preprint}\ }\href
  {https://doi.org/10.48550/arXiv.2605.03404} {10.48550/arXiv.2605.03404}
  (\bibinfo {year} {2026}),\ \Eprint {https://arxiv.org/abs/2605.03404}
  {arXiv:2605.03404 [cond-mat.soft]} \BibitemShut {NoStop}%
\bibitem [{\citenamefont {Digregorio}\ \emph {et~al.}(2026)\citenamefont
  {Digregorio}, \citenamefont {Pagonabarraga},\ and\ \citenamefont
  {Vega~Reyes}}]{digregorio-2026}%
  \BibitemOpen
  \bibfield  {author} {\bibinfo {author} {\bibfnamefont {P.}~\bibnamefont
  {Digregorio}}, \bibinfo {author} {\bibfnamefont {I.}~\bibnamefont
  {Pagonabarraga}},\ and\ \bibinfo {author} {\bibfnamefont {F.}~\bibnamefont
  {Vega~Reyes}},\ }\bibfield  {title} {\bibinfo {title} {Phase separation in a
  chiral active fluid of inertial self-spinning disks},\ }\href
  {https://doi.org/10.1103/dbgp-pqsh} {\bibfield  {journal} {\bibinfo
  {journal} {Phys. Rev. Lett.}\ }\textbf {\bibinfo {volume} {136}},\ \bibinfo
  {pages} {218301} (\bibinfo {year} {2026})}\BibitemShut {NoStop}%
\bibitem [{\citenamefont {Shen}\ and\ \citenamefont
  {Lintuvuori}(2023)}]{shen-2023}%
  \BibitemOpen
  \bibfield  {author} {\bibinfo {author} {\bibfnamefont {Z.}~\bibnamefont
  {Shen}}\ and\ \bibinfo {author} {\bibfnamefont {J.~S.}\ \bibnamefont
  {Lintuvuori}},\ }\bibfield  {title} {\bibinfo {title} {Collective flows drive
  cavitation in spinner monolayers},\ }\href
  {https://doi.org/10.1103/PhysRevLett.130.188202} {\bibfield  {journal}
  {\bibinfo  {journal} {Phys. Rev. Lett.}\ }\textbf {\bibinfo {volume} {130}},\
  \bibinfo {pages} {188202} (\bibinfo {year} {2023})}\BibitemShut {NoStop}%
\bibitem [{mod()}]{model-footnote}%
  \BibitemOpen
  \href@noop {} {}\bibinfo {note} {{cAMB} is distinct from the model recently
  introduced in Ref.~\cite{wang-2026-arxiv} that focuses on anisotropic
  effects.}\BibitemShut {Stop}%
\bibitem [{\citenamefont {Bray}(1994)}]{bray-1994}%
  \BibitemOpen
  \bibfield  {author} {\bibinfo {author} {\bibfnamefont {A.}~\bibnamefont
  {Bray}},\ }\bibfield  {title} {\bibinfo {title} {Theory of phase-ordering
  kinetics},\ }\href {https://doi.org/10.1080/00018739400101505} {\bibfield
  {journal} {\bibinfo  {journal} {Adv. Phys.}\ }\textbf {\bibinfo {volume}
  {43}},\ \bibinfo {pages} {357} (\bibinfo {year} {1994})}\BibitemShut
  {NoStop}%
\bibitem [{\citenamefont {Pattanayak}\ \emph {et~al.}(2021)\citenamefont
  {Pattanayak}, \citenamefont {Mishra},\ and\ \citenamefont
  {Puri}}]{pattanayak-2021}%
  \BibitemOpen
  \bibfield  {author} {\bibinfo {author} {\bibfnamefont {S.}~\bibnamefont
  {Pattanayak}}, \bibinfo {author} {\bibfnamefont {S.}~\bibnamefont {Mishra}},\
  and\ \bibinfo {author} {\bibfnamefont {S.}~\bibnamefont {Puri}},\ }\bibfield
  {title} {\bibinfo {title} {Ordering kinetics in the active {model $B$}},\
  }\href {https://doi.org/10.1103/PhysRevE.104.014606} {\bibfield  {journal}
  {\bibinfo  {journal} {Phys. Rev. E}\ }\textbf {\bibinfo {volume} {104}},\
  \bibinfo {pages} {014606} (\bibinfo {year} {2021})}\BibitemShut {NoStop}%
\bibitem [{\citenamefont {Dikshit}\ \emph {et~al.}(2024)\citenamefont
  {Dikshit}, \citenamefont {Pattanayak}, \citenamefont {Mishra},\ and\
  \citenamefont {Puri}}]{dikshit-2024-arxiv}%
  \BibitemOpen
  \bibfield  {author} {\bibinfo {author} {\bibfnamefont {S.}~\bibnamefont
  {Dikshit}}, \bibinfo {author} {\bibfnamefont {S.}~\bibnamefont {Pattanayak}},
  \bibinfo {author} {\bibfnamefont {S.}~\bibnamefont {Mishra}},\ and\ \bibinfo
  {author} {\bibfnamefont {S.}~\bibnamefont {Puri}},\ }\bibfield  {title}
  {\bibinfo {title} {Domain growth kinetics of active {model B} with thermal
  fluctuations},\ }\bibfield  {journal} {\bibinfo  {journal} {arXiv preprint}\
  }\href {https://doi.org/10.48550/arXiv.2402.18977}
  {10.48550/arXiv.2402.18977} (\bibinfo {year} {2024}),\ \Eprint
  {https://arxiv.org/abs/2402.18977} {arXiv:2402.18977 [cond-mat.soft]}
  \BibitemShut {NoStop}%
\bibitem [{\citenamefont {Yadav}\ \emph {et~al.}(2025)\citenamefont {Yadav},
  \citenamefont {Mishra},\ and\ \citenamefont {Puri}}]{yadav-2025}%
  \BibitemOpen
  \bibfield  {author} {\bibinfo {author} {\bibfnamefont {P.~K.}\ \bibnamefont
  {Yadav}}, \bibinfo {author} {\bibfnamefont {S.}~\bibnamefont {Mishra}},\ and\
  \bibinfo {author} {\bibfnamefont {S.}~\bibnamefont {Puri}},\ }\bibfield
  {title} {\bibinfo {title} {Coarsening kinetics in active model $\mathrm{B}+$:
  Macroscale and microscale phase separation},\ }\href
  {https://doi.org/10.1103/kb9k-w7jr} {\bibfield  {journal} {\bibinfo
  {journal} {Phys. Rev. E}\ }\textbf {\bibinfo {volume} {112}},\ \bibinfo
  {pages} {035412} (\bibinfo {year} {2025})}\BibitemShut {NoStop}%
\bibitem [{\citenamefont {Bhowmick}\ and\ \citenamefont
  {Mohanty}(2026)}]{bhowmick2026critical}%
  \BibitemOpen
  \bibfield  {author} {\bibinfo {author} {\bibfnamefont {A.}~\bibnamefont
  {Bhowmick}}\ and\ \bibinfo {author} {\bibfnamefont {P.}~\bibnamefont
  {Mohanty}},\ }\bibfield  {title} {\bibinfo {title} {Critical scaling and
  supercritical coarsening in {Active Model B+}},\ }\href@noop {} {\bibfield
  {journal} {\bibinfo  {journal} {arXiv preprint}\ } (\bibinfo {year}
  {2026})},\ \Eprint {https://arxiv.org/abs/2604.07247} {arXiv:2604.07247
  [cond-mat.soft]} \BibitemShut {NoStop}%
\bibitem [{\citenamefont {Caballero}\ \emph {et~al.}(2025)\citenamefont
  {Caballero}, \citenamefont {Maitra},\ and\ \citenamefont
  {Nardini}}]{caballero2024interface}%
  \BibitemOpen
  \bibfield  {author} {\bibinfo {author} {\bibfnamefont {F.}~\bibnamefont
  {Caballero}}, \bibinfo {author} {\bibfnamefont {A.}~\bibnamefont {Maitra}},\
  and\ \bibinfo {author} {\bibfnamefont {C.}~\bibnamefont {Nardini}},\
  }\bibfield  {title} {\bibinfo {title} {Interface dynamics of wet active
  systems},\ }\href {https://doi.org/10.1103/PhysRevLett.134.087105} {\bibfield
   {journal} {\bibinfo  {journal} {Phys. Rev. Lett.}\ }\textbf {\bibinfo
  {volume} {134}},\ \bibinfo {pages} {087105} (\bibinfo {year}
  {2025})}\BibitemShut {NoStop}%
\bibitem [{\citenamefont {Ziethen}\ \emph {et~al.}(2026)\citenamefont
  {Ziethen}, \citenamefont {Chatzittofi}, \citenamefont {Cates},\ and\
  \citenamefont {Nardini}}]{ziethen2026}%
  \BibitemOpen
  \bibfield  {author} {\bibinfo {author} {\bibfnamefont {N.}~\bibnamefont
  {Ziethen}}, \bibinfo {author} {\bibfnamefont {M.}~\bibnamefont
  {Chatzittofi}}, \bibinfo {author} {\bibfnamefont {M.~E.}\ \bibnamefont
  {Cates}},\ and\ \bibinfo {author} {\bibfnamefont {C.}~\bibnamefont
  {Nardini}},\ }\bibfield  {title} {\bibinfo {title} {Nucleation and
  time-reversal symmetry breaking in nonconserved scalar field theories},\
  }\href@noop {} {\bibfield  {journal} {\bibinfo  {journal} {arXiv preprint
  arXiv:2607.05194}\ } (\bibinfo {year} {2026})}\BibitemShut {NoStop}%
\bibitem [{\citenamefont {Han}\ \emph {et~al.}(2021)\citenamefont {Han},
  \citenamefont {Fruchart}, \citenamefont {Scheibner}, \citenamefont
  {Vaikuntanathan}, \citenamefont {de~Pablo},\ and\ \citenamefont
  {Vitelli}}]{han-2021}%
  \BibitemOpen
  \bibfield  {author} {\bibinfo {author} {\bibfnamefont {M.}~\bibnamefont
  {Han}}, \bibinfo {author} {\bibfnamefont {M.}~\bibnamefont {Fruchart}},
  \bibinfo {author} {\bibfnamefont {C.}~\bibnamefont {Scheibner}}, \bibinfo
  {author} {\bibfnamefont {S.}~\bibnamefont {Vaikuntanathan}}, \bibinfo
  {author} {\bibfnamefont {J.~J.}\ \bibnamefont {de~Pablo}},\ and\ \bibinfo
  {author} {\bibfnamefont {V.}~\bibnamefont {Vitelli}},\ }\bibfield  {title}
  {\bibinfo {title} {Fluctuating hydrodynamics of chiral active fluids},\
  }\href {https://doi.org/10.1038/s41567-021-01360-7} {\bibfield  {journal}
  {\bibinfo  {journal} {Nat. Phys.}\ }\textbf {\bibinfo {volume} {17}},\
  \bibinfo {pages} {1260} (\bibinfo {year} {2021})}\BibitemShut {NoStop}%
\bibitem [{\citenamefont {Neville}\ \emph {et~al.}(2026)\citenamefont
  {Neville}, \citenamefont {Eggers},\ and\ \citenamefont
  {Liverpool}}]{neville-2026}%
  \BibitemOpen
  \bibfield  {author} {\bibinfo {author} {\bibfnamefont {L.}~\bibnamefont
  {Neville}}, \bibinfo {author} {\bibfnamefont {J.}~\bibnamefont {Eggers}},\
  and\ \bibinfo {author} {\bibfnamefont {T.~B.}\ \bibnamefont {Liverpool}},\
  }\bibfield  {title} {\bibinfo {title} {Breakup of an active chiral fluid}\
  }\href {https://doi.org/10.48550/arXiv.2506.10534}
  {10.48550/arXiv.2506.10534} (\bibinfo {year} {2026})\BibitemShut {NoStop}%
\bibitem [{\citenamefont {Ascher}\ \emph {et~al.}(1997)\citenamefont {Ascher},
  \citenamefont {Ruuth},\ and\ \citenamefont {Spiteri}}]{ascher-1997}%
  \BibitemOpen
  \bibfield  {author} {\bibinfo {author} {\bibfnamefont {U.~M.}\ \bibnamefont
  {Ascher}}, \bibinfo {author} {\bibfnamefont {S.~J.}\ \bibnamefont {Ruuth}},\
  and\ \bibinfo {author} {\bibfnamefont {R.~J.}\ \bibnamefont {Spiteri}},\
  }\bibfield  {title} {\bibinfo {title} {Implicit-explicit {Runge-Kutta}
  methods for time-dependent partial differential equations},\ }\href
  {https://doi.org/10.1016/S0168-9274(97)00056-1} {\bibfield  {journal}
  {\bibinfo  {journal} {Appl. Numer. Math.}\ }\textbf {\bibinfo {volume}
  {25}},\ \bibinfo {pages} {151} (\bibinfo {year} {1997})},\ \bibinfo {note}
  {special Issue on Time Integration}\BibitemShut {NoStop}%
\bibitem [{\citenamefont {Besse}\ \emph {et~al.}(2023)\citenamefont {Besse},
  \citenamefont {Fausti}, \citenamefont {Cates}, \citenamefont {Delamotte},\
  and\ \citenamefont {Nardini}}]{besse-2023}%
  \BibitemOpen
  \bibfield  {author} {\bibinfo {author} {\bibfnamefont {M.}~\bibnamefont
  {Besse}}, \bibinfo {author} {\bibfnamefont {G.}~\bibnamefont {Fausti}},
  \bibinfo {author} {\bibfnamefont {M.~E.}\ \bibnamefont {Cates}}, \bibinfo
  {author} {\bibfnamefont {B.}~\bibnamefont {Delamotte}},\ and\ \bibinfo
  {author} {\bibfnamefont {C.}~\bibnamefont {Nardini}},\ }\bibfield  {title}
  {\bibinfo {title} {Interface roughening in nonequilibrium phase-separated
  systems},\ }\href {https://doi.org/10.1103/PhysRevLett.130.187102} {\bibfield
   {journal} {\bibinfo  {journal} {Phys. Rev. Lett.}\ }\textbf {\bibinfo
  {volume} {130}},\ \bibinfo {pages} {187102} (\bibinfo {year}
  {2023})}\BibitemShut {NoStop}%
\bibitem [{\citenamefont {Burns}\ \emph {et~al.}(2020)\citenamefont {Burns},
  \citenamefont {Vasil}, \citenamefont {Oishi}, \citenamefont {Lecoanet},\ and\
  \citenamefont {Brown}}]{burns-2020}%
  \BibitemOpen
  \bibfield  {author} {\bibinfo {author} {\bibfnamefont {K.~J.}\ \bibnamefont
  {Burns}}, \bibinfo {author} {\bibfnamefont {G.~M.}\ \bibnamefont {Vasil}},
  \bibinfo {author} {\bibfnamefont {J.~S.}\ \bibnamefont {Oishi}}, \bibinfo
  {author} {\bibfnamefont {D.}~\bibnamefont {Lecoanet}},\ and\ \bibinfo
  {author} {\bibfnamefont {B.~P.}\ \bibnamefont {Brown}},\ }\bibfield  {title}
  {\bibinfo {title} {Dedalus: A flexible framework for numerical simulations
  with spectral methods},\ }\href
  {https://doi.org/10.1103/PhysRevResearch.2.023068} {\bibfield  {journal}
  {\bibinfo  {journal} {Phys. Rev. Res.}\ }\textbf {\bibinfo {volume} {2}},\
  \bibinfo {pages} {023068} (\bibinfo {year} {2020})}\BibitemShut {NoStop}%
\bibitem [{\citenamefont {Cahn}\ and\ \citenamefont
  {Hilliard}(1958)}]{cahn-1958}%
  \BibitemOpen
  \bibfield  {author} {\bibinfo {author} {\bibfnamefont {J.~W.}\ \bibnamefont
  {Cahn}}\ and\ \bibinfo {author} {\bibfnamefont {J.~E.}\ \bibnamefont
  {Hilliard}},\ }\bibfield  {title} {\bibinfo {title} {Free energy of a
  nonuniform system. {I}. {Interfacial} free energy},\ }\href
  {https://doi.org/10.1063/1.1744102} {\bibfield  {journal} {\bibinfo
  {journal} {J. Chem. Phys.}\ }\textbf {\bibinfo {volume} {28}},\ \bibinfo
  {pages} {258} (\bibinfo {year} {1958})}\BibitemShut {NoStop}%
\bibitem [{\citenamefont {Cates}\ and\ \citenamefont
  {Tjhung}(2018)}]{cates-2018}%
  \BibitemOpen
  \bibfield  {author} {\bibinfo {author} {\bibfnamefont {M.~E.}\ \bibnamefont
  {Cates}}\ and\ \bibinfo {author} {\bibfnamefont {E.}~\bibnamefont {Tjhung}},\
  }\bibfield  {title} {\bibinfo {title} {Theories of binary fluid mixtures:
  from phase-separation kinetics to active emulsions},\ }\href
  {https://doi.org/10.1017/jfm.2017.832} {\bibfield  {journal} {\bibinfo
  {journal} {J. Fluid Mech.}\ }\textbf {\bibinfo {volume} {836}},\ \bibinfo
  {pages} {P1} (\bibinfo {year} {2018})}\BibitemShut {NoStop}%
\bibitem [{\citenamefont {{De Luca}}\ \emph {et~al.}(2024)\citenamefont {{De
  Luca}}, \citenamefont {Ma}, \citenamefont {Nardini},\ and\ \citenamefont
  {Cates}}]{deluca-2024}%
  \BibitemOpen
  \bibfield  {author} {\bibinfo {author} {\bibfnamefont {F.}~\bibnamefont {{De
  Luca}}}, \bibinfo {author} {\bibfnamefont {X.}~\bibnamefont {Ma}}, \bibinfo
  {author} {\bibfnamefont {C.}~\bibnamefont {Nardini}},\ and\ \bibinfo {author}
  {\bibfnamefont {M.~E.}\ \bibnamefont {Cates}},\ }\bibfield  {title} {\bibinfo
  {title} {Hyperuniformity in phase ordering: The roles of activity, noise, and
  non-constant mobility},\ }\href {https://doi.org/10.1088/1361-648X/ad5b45}
  {\bibfield  {journal} {\bibinfo  {journal} {J. Phys. Condens. Matter}\
  }\textbf {\bibinfo {volume} {36}},\ \bibinfo {pages} {405101} (\bibinfo
  {year} {2024})}\BibitemShut {NoStop}%
\bibitem [{\citenamefont {Kevorkian}\ and\ \citenamefont
  {Cole}(1996)}]{kevorkian-1996}%
  \BibitemOpen
  \bibfield  {author} {\bibinfo {author} {\bibfnamefont {J.}~\bibnamefont
  {Kevorkian}}\ and\ \bibinfo {author} {\bibfnamefont {J.}~\bibnamefont
  {Cole}},\ }\href {https://doi.org/10.1007/978-1-4612-3968-0} {\emph {\bibinfo
  {title} {Multiple Scale and Singular Perturbation Methods}}},\ Applied
  Mathematical Sciences\ (\bibinfo  {publisher} {Springer New York},\ \bibinfo
  {year} {1996})\BibitemShut {NoStop}%
\bibitem [{\citenamefont {Pavliotis}\ and\ \citenamefont
  {Stuart}(2008)}]{pavliotis-2008}%
  \BibitemOpen
  \bibfield  {author} {\bibinfo {author} {\bibfnamefont {G.~A.}\ \bibnamefont
  {Pavliotis}}\ and\ \bibinfo {author} {\bibfnamefont {A.~M.}\ \bibnamefont
  {Stuart}},\ }\href {https://doi.org/10.1007/978-0-387-73829-1} {\emph
  {\bibinfo {title} {Multiscale Methods: Averaging and Homogenization}}},\
  \bibinfo {edition} {1st}\ ed.\ (\bibinfo  {publisher} {Springer},\ \bibinfo
  {address} {New York, NY},\ \bibinfo {year} {2008})\BibitemShut {NoStop}%
\bibitem [{\citenamefont {Chatzittofi}\ \emph {et~al.}(2026)\citenamefont
  {Chatzittofi}, \citenamefont {Ziethen}, \citenamefont {Nardini},\ and\
  \citenamefont {Cates}}]{chatzittofi2026nonequilibrium}%
  \BibitemOpen
  \bibfield  {author} {\bibinfo {author} {\bibfnamefont {M.}~\bibnamefont
  {Chatzittofi}}, \bibinfo {author} {\bibfnamefont {N.}~\bibnamefont
  {Ziethen}}, \bibinfo {author} {\bibfnamefont {C.}~\bibnamefont {Nardini}},\
  and\ \bibinfo {author} {\bibfnamefont {M.~E.}\ \bibnamefont {Cates}},\
  }\bibfield  {title} {\bibinfo {title} {Nonequilibrium nucleation theory for
  nonconserved fields: from active matter to population dynamics},\ }\href@noop
  {} {\bibfield  {journal} {\bibinfo  {journal} {arXiv preprint
  arXiv:2606.18911}\ } (\bibinfo {year} {2026})}\BibitemShut {NoStop}%
\end{thebibliography}%

\end{document}